\documentclass[11pt,a4paper]{article}
\pdfoutput=1

\usepackage{jheppub-nosort,amsthm}

\usepackage{amsmath,amssymb,amsthm,amscd}
\usepackage[dvipsnames]{xcolor}
\usepackage{physics,stmaryrd}
\usepackage{tikz,float}
\tikzset{node distance=2cm, auto}
\usepackage{epsfig}
\usepackage{psfrag,comment}
\usepackage{graphicx}
\usepackage{caption}
\usepackage{subcaption}
\usepackage{mathrsfs}
\usepackage[utf8]{inputenc}  
\usepackage{xcolor}
\usepackage[normalem]{ulem}
\usepackage{upgreek}
\usepackage{rotating}
\usepackage{mathtools}
\usepackage[bottom]{footmisc}

\usepackage[textsize=footnotesize]{todonotes}
\presetkeys%
    {todonotes}%
    {color=Apricot}{}%

\usepackage{url}
\newcommand\doilink[1]{\href{http://dx.doi.org/#1}{#1}}
\newcommand\arxivlink[1]{\href{http://arxiv.org/abs/#1}{#1}}

\newcommand{\CB}{{\mathcal B}}
\newcommand{\CC}{{\mathcal C}}

\newcommand{\CF}{{\mathcal F}}

\newcommand{\CI}{{\mathcal I}}

\newcommand{\CM}{{\mathcal M}}

\newcommand{\CR}{{\mathcal R}}
\newcommand{\CS}{{\mathcal S}}

\newcommand{\CV}{{\mathcal V}}

\newcommand{\CZ}{{\mathcal Z}}

\newcommand{\NCL}{{\mathscr L}}

\newcommand{\NCO}{{\mathscr O}}

\def\BN{{\mathbb N}}
\def\BZ{{\mathbb Z}}
\def\BR{{\mathbb R}}
\def\BC{{\mathbb C}}
\def\BP{{\mathbb P}}

\def\BS{{\mathbb S}}

\newcommand{\be}{\begin{equation}}
\newcommand{\ee}{\end{equation}}
\newcommand{\ba}{\begin{aligned}}
\newcommand{\ea}{\end{aligned}}
\newcommand{\bea}{\begin{eqnarray}}
\newcommand{\eea}{\end{eqnarray}}
\newcommand{\bean}{\begin{eqnarray*}}
\newcommand{\eean}{\end{eqnarray*}}

\def\r{\right\rangle}

\def\1{\mathbf{1}}
\def\0{|\1\r}
\def\im{{\mathbb{I}}{\mathrm{m}}}
\def\re{{\mathbb{R}}{\mathrm{e}}}
\def\Tr{{\mathbb{T}}{\mathrm{r}}}

\newcommand{\rme}{{\mathrm{e}}}
\newcommand{\rmi}{{\mathrm{i}}}
\newcommand{\rmd}{{\mathrm{d}}}

\def\XXint#1#2#3{{\setbox0=\hbox{$#1{#2#3}{\int}$}
     \vcenter{\hbox{$#2#3$}}\kern-.5\wd0}}

\theoremstyle{remark}

\makeatletter
\newsavebox\myboxA
\newsavebox\myboxB
\newlength\mylenA

\newcommand*\widebar[2][0.75]{%
    \sbox{\myboxA}{$\m@th#2$}%
    \setbox\myboxB\null
    \ht\myboxB=\ht\myboxA%
    \dp\myboxB=\dp\myboxA%
    \wd\myboxB=#1\wd\myboxA
    \sbox\myboxB{$\m@th\overline{\copy\myboxB}$}
    \setlength\mylenA{\the\wd\myboxA}
    \addtolength\mylenA{-\the\wd\myboxB}%
    \ifdim\wd\myboxB<\wd\myboxA%
       \rlap{\hskip 0.8\mylenA\usebox\myboxB}{\usebox\myboxA}%
    \else
        \hskip -0.5\mylenA\rlap{\usebox\myboxA}{\hskip 0.5\mylenA\usebox\myboxB}%
    \fi}
\makeatother

\newdimen\tableauside\tableauside=1.0ex
\newdimen\tableaurule\tableaurule=0.4pt
\newdimen\tableaustep
\def\phantomhrule#1{\hbox{\vbox to0pt{\hrule height\tableaurule width#1\vss}}}
\def\phantomvrule#1{\vbox{\hbox to0pt{\vrule width\tableaurule height#1\hss}}}
\def\sqr{\vbox{%
  \phantomhrule\tableaustep
  \hbox{\phantomvrule\tableaustep\kern\tableaustep\phantomvrule\tableaustep}%
  \hbox{\vbox{\phantomhrule\tableauside}\kern-\tableaurule}}}
\def\squares#1{\hbox{\count0=#1\noindent\loop\sqr
  \advance\count0 by-1 \ifnum\count0>0\repeat}}
\def\tableau#1{\vcenter{\offinterlineskip
  \tableaustep=\tableauside\advance\tableaustep by-\tableaurule
  \kern\normallineskip\hbox
    {\kern\normallineskip\vbox
      {\gettableau#1 0 }%
     \kern\normallineskip\kern\tableaurule}%
  \kern\normallineskip\kern\tableaurule}}
\def\gettableau#1{\ifnum#1=0\let\next=\null\else
\squares{#1}\let\next=\gettableau\fi\next}
\usepackage{pdftexcmds}

\DeclareFontFamily{U}{cbgreek}{}
\DeclareFontShape{U}{cbgreek}{m}{n}{
        <-6>    grmn0500
        <6-7>   grmn0600
        <7-8>   grmn0700
        <8-9>   grmn0800
        <9-10>  grmn0900
        <10-12> grmn1000
        <12-17> grmn1200
        <17->   grmn1728
      }{}
\DeclareFontShape{U}{cbgreek}{bx}{n}{
        <-6>    grxn0500
        <6-7>   grxn0600
        <7-8>   grxn0700
        <8-9>   grxn0800
        <9-10>  grxn0900
        <10-12> grxn1000
        <12-17> grxn1200
        <17->   grxn1728
      }{}

\makeatletter
\newcommand{\normalorbold}{%
  \ifnum\pdf@strcmp{\math@version}{bold}=\z@ bx\else m\fi
}
\makeatother

\subheader{\hfill
{\small{\textsf{CERN-TH-2021-097}}}
}

\title{Resurgent Asymptotics of Jackiw--Teitelboim Gravity and the Nonperturbative Topological Recursion}

\author[a]{Bertrand~Eynard,}
\emailAdd{bertrand.eynard@ipht.fr}
\affiliation[a]{Institut de Physique Th\'eorique (IPhT), Commissariat \`a l'\'Energie Atomique (CEA), CNRS,\\ Universit\'e Paris--Saclay, 91191 Gif-sur-Yvette, France}

\author[b]{Elba~Garcia-Failde,}
\emailAdd{elba.garcia-failde@imj-prg.fr}
\affiliation[b]{Sorbonne Universit\'e, IMJ-PRG,\\ Place Jussieu, 75252 Paris Cedex 05, France}

\author[a,c]{Paolo~Gregori,}
\emailAdd{paolo.gregori@ipht.fr}
\affiliation[c]{CAMGSD, Departamento de Matem\'atica, Instituto Superior T\'ecnico,\\ Universidade de Lisboa, 1049-001 Lisboa, Portugal}

\author[d,e]{Danilo~Lewa\'{n}ski,}
\emailAdd{danilo.lewanski@unige.ch}
\affiliation[d]{Section de Math\'ematiques, Universit\'e de Gen\`eve,\\ Rue de Conseil-G\'en\'eral 7, 1206 Gen\`eve, Switzerland}
\affiliation[e]{Dipartimento di Matematica e Geoscienze, Universit\`a di Trieste,\\ Via Valerio 12/1, 34127 Trieste, Italy\\}

\author[c]{Ricardo~Schiappa\,}
\emailAdd{ricardo.schiappa@tecnico.ulisboa.pt}

\abstract{
Jackiw--Teitelboim dilaton-quantum-gravity localizes on a double-scaled random-matrix model, whose perturbative free energy is an asymptotic series. Understanding the resurgent properties of this asymptotic series, including its completion into a full transseries, requires understanding the nonperturbative instanton sectors of the matrix model for Jackiw--Teitelboim gravity. The present work addresses this question by setting-up instanton calculus associated to eigenvalue tunneling (or ZZ-brane contributions), directly in the matrix model. In order to systematize such calculations, a nonperturbative extension of the topological recursion formalism is required---which is herein both constructed and applied to the present problem. Large-order tests of the perturbative genus expansion validate the resurgent nature of Jackiw--Teitelboim gravity, both for its free energy and for its (multi-resolvent) correlation functions. Both ZZ and FZZT nonperturbative effects are required by resurgence, and they further display resonance upon the Borel plane. Finally, the resurgence properties of the multi-resolvent correlation functions yield new and improved resurgence formulae for the large-genus growth of Weil--Petersson volumes.}


\arxivnumber{2305.16940}

\begin{document}

\maketitle

\vfill

\eject

\allowdisplaybreaks

\section{Introduction and Summary}
\label{sec:intro}

Jackiw--Teitelboim two-dimensional dilaton gravity \cite{t83, j84}---henceforth ``JT gravity''---has in recent years been shown to describe universal features of the near-horizon geometry of higher-dimensional near-extremal black-holes, at full quantum mechanical level \cite{ap14, ms16, j16, msy16, emv16, cghps4t16, sw17} (see, \textit{e.g.}, \cite{mt22} for a recent review and a list of references on JT gravity). It may in fact be one of the simplest (solvable) toy models describing quantum black holes; hence its great interest.

One of the many difficulties pertaining to the study of quantum black holes is that their physics is maximally chaotic \cite{ss13, mss15}. This has subtle consequences in regard to what the euclidean gravitational path-integral can actually compute, at least concerning observables involving black hole states \cite{sw22}: for these, all it seems to produce are averages over the aforementioned chaotic contributions (the same not occurring for any other, non-black-hole observables). For JT gravity in particular, this implies that the gravitational bulk theory ends up localizing on a (double scaled) random matrix ensemble; as had been earlier shown by Saad, Shenker, and Stanford in their seminal work \cite{sss19}. In this setting, the continuous spectrum\footnote{To be contrasted with an expected discrete spectrum, from generic AdS/CFT considerations.} found from the boundary JT disk-partition-function is interpreted not as any failure of AdS/CFT duality, but as the natural outcome of random-matrix ensemble-averaging \cite{sw22}. There has been enormous interest in studying this double-scaled matrix model; the main focus of our present work.

One immediate result obtainable from the matrix model in \cite{sss19} is the computation of local correlation-functions of multiple JT-gravity partition-function insertions, expressible via (topological) genus expansions. As further discussed\footnote{At least explicitly for the matrix-model free-energy, and for the one-point correlator of a JT partition function.} in \cite{sss19}, these expansions are in fact \textit{asymptotic}. What we want to address in the present work is whether these expansions are also \textit{resurgent} \cite{e81, e93}---in which case, opening the door to obtaining the full nonperturbative multi-instanton content of JT gravity. In the resurgence context, observables are described by transseries including both perturbative and (multi-instanton) nonperturbative sectors (alongside their negative-tension counterparts when \textit{resonant} \cite{mss22, sst23}). Then, hiding deep in the asymptotics of any such sector, lies the ``resurgence'' of any \textit{other} sector---which is to say, the resurgence of the full nonperturbative content in the transseries; see, \textit{e.g.}, \cite{abs18}. Because JT gravity may also be obtained as a specific large-central-charge limit of minimal string theory \cite{sss19, os19, t:mt20, h:mt20, km20, tuw20, gs21}, in principle this complete nonperturbative content will be made out of minimal-string-theoretic ZZ branes \cite{zz01}, FZZT branes \cite{fzz00, t00}, and their negative-tension counterparts \cite{mss22, sst23} (the latter being associated to the resonant nature of the resurgent transseries of JT gravity; shown in \cite{gs21}). We shall verify all this carefully; also with very explicit tests of the large-genus asymptotics of correlation functions. Establishing this resurgent nature of the JT-gravity matrix-model will eventually allow for further developing an explicit analysis of its complete nonperturbative content, via the resurgence pathway, \textit{e.g.}, \cite{gs21}. Other closely related approaches to the construction of nonperturbative JT gravity include, \textit{e.g.}, \cite{gjk21} which focuses on understanding the strongly-coupled phase of the matrix model when eigenvalue contours extend into their complex plane; or \cite{j21} which focuses instead on maintaining the reality of these matrix-model eigenvalue integration-contours.

The string-theoretic (topological) genus expansion is expected to be \textit{asymptotic}, based on general grounds \cite{gp88, s90}; as well as \textit{resurgent}, based on a large body of both old and recent evidence within the realms of double-scaled/minimal and topological strings \cite{gz91, ez93, m06, msw07, m08, msw08, ps09, gikm10, kmr10, dmp11, asv11, sv13, as13, cesv13, gmz14, cesv14, gm21, gs21, bssv22, gm22}. On top, it has been shown to be \textit{resonant} \cite{gikm10, asv11, sv13, as13, gs21}, hence further including nonperturbative effects due to anti-eigenvalues or negative-tension D-branes \cite{mss22, sst23}. The same holds true for the analogous matrix-model free-energy perturbative-expansion \cite{e19} (and, as we shall later discuss, for multi-resolvent correlation functions). Now, it is important to stress that the aforementioned string-theoretic examples of resurgence were built upon extensive analysis of large-order data from their genus expansions. Generically, these data are hard to find. But, in practice, the above examples were also chosen so that such generic difficulties could be circumvented: the specific-heat of minimal string theory may be computed from its string equations \cite{biz80, gm90b, dss90, mss91, ss03, e16}, which are very amenable to resurgent analysis \cite{gikm10, asv11, sv13, as13, gs21, bssv22}; whereas the free energy of topological string theory may be computed via the holomorphic anomaly equations \cite{bcov93a, bcov93b, hkr08}, which have also been extensively dealt-with from a resurgence standpoint \cite{cesv13, cesv14, c15, csv16, gm22}. Unfortunately, for JT gravity both are lacking---in particular its complete string equation is yet unknown (but see the discussion in \cite{gs21}). One must then resort to some alternative means in order to move forward.

As already mentioned, one of the main results in \cite{sss19} was to compute $n$-point functions of JT-partition-function insertions via perturbative genus-expansions. In particular, their perturbative genus-$g$ contribution to some such $n$-point function is basically dictated by the corresponding Weil--Petersson volume, \textit{i.e.}, the volume of the moduli space of (hyperbolic) Riemann surfaces precisely with genus $g$ and $n$ geodesic boundaries (of prescribed lengths). It is a fascinating mathematical story how these volumes had previously been the focus of great attention in the work of Mirzakhani and collaborators \cite{m07a, m07b, z07, z08, m10, mz11, mp17}. One key point in this broader construct is that Weil--Petersson volumes have their higher-genus contributions recursively dictated out of their lower-genus contributions by the topological recursion \cite{eo07a, eo08, eo09}, as shown in \cite{eo07b}. This topological recursion iterates upon a specific spectral curve and, in this way, the relevant spectral curve for Weil--Petersson volumes \cite{eo07b} is precisely the one which was later used to construct the JT matrix model in \cite{sss19}. Now, the topological recursion in \cite{eo07a} originated from the perturbative expansion of matrix-model free-energies \cite{e04, ceo06}, hence it is inherently \textit{perturbative} by construction. This is to say, even though it can in principle yield large-genus results---hence immediately serving as an alternative means to the string equations or the holomorphic anomaly constructs---it cannot\footnote{One other reason why the topological recursion was not used in the aforementioned large-genus resurgent analyses of double-scaled/minimal and topological strings was also because it is algorithmically slower as compared to addressing either string equations or the holomorphic anomaly---but see as well \cite{gs21}.} immediately yield (multi) instanton results. But, at the matrix model level, such (multi) instanton results have been previously (generically) constructed in \cite{msw07, msw08}. As such, in the very same spirit, one may now extend \cite{msw07, msw08} into a systematized \textit{nonperturbative extension} of the topological recursion, which is herein both constructed and applied to JT gravity. This allows us to address instanton sectors of JT-gravity matrix-model free energy and multi-resolvent correlators, finally unveiling their resurgent properties. It should also be clear from above that there is an added bonus to this story: exploring and understanding the resurgent properties of JT gravity will immediately yield mathematical results on the large-genus resurgent properties of Weil--Petersson volumes---and this we shall also explore in the present work.

The contents of this paper are organized as follows. We begin by recalling the matrix model description of JT gravity \cite{sss19} in section~\ref{sec:setup}; starting off with the matrix model set-up in subsection~\ref{subsec:JTMM-setup}, its relation to Weil--Petersson volumes in subsection~\ref{subsec:WPvolumes}, and finally the topological recursion set-up including the JT gravity spectral-curve in subsection~\ref{subsec:classicalTR}. So far this is to a very large extent a purely perturbative construction, and we move-on to a first discussion of nonperturbative contributions to the matrix-model free-energy in section~\ref{sec:oldmmodels}. JT gravity has no (closed form) string equation, as do multicritical models or minimal strings (see, \textit{e.g.}, the recent discussion in \cite{gs21}), which implies that going nonperturbative must be achieved starting from spectral geometry alone. Obtaining such contributions as they follow from the matrix-model spectral-curve formulation is very well-known in the literature, corresponding to eigenvalue tunneling, or, equivalently, ZZ-brane effects---in particular, we shall be following \cite{msw07} rather closely throughout section~\ref{sec:oldmmodels}. In this way, the JT gravity nonperturbative ZZ contributions\footnote{A word on terminology. Albeit JT gravity is a two-dimensional \textit{spacetime} theory, whereas ZZ or FZZT contributions are associated to two-dimensional \textit{world-sheet} boundaries, we shall nonetheless be using this ZZ/FZZT standard matrix model/Liouville theory/minimal-string theoretic terminology throughout this paper.} are straightforward to write down, and match the large-central-charge limit of minimal string theory as addressed in \cite{gs21}. However, systematizing such a calculation for higher instanton sectors or higher loops around some fixed instanton sector is at this stage less clear. In section~\ref{sec:NPtoprec} we build upon \cite{eo07a, msw07} to present a generalization of the topological recursion towards computing nonperturbative contributions---essentially by building upon the standard topological recursion, but with an adequate use of the loop insertion operator which allows to control and compute instanton corrections. This construction holds not only for the free energy (systematizing results in the earlier section~\ref{sec:oldmmodels}), as described in subsection~\ref{subsec:NPTR-1inst-F}; but most importantly also extends to multi-resolvent correlation functions, as described in subsection~\ref{subsec:correlators}. The results of subsection~\ref{subsec:NPTR-1inst-F} build on a ``shifted'' matrix-model partition-function which implicitly played a key role in \cite{msw07}, and easily extend the results in section~\ref{sec:oldmmodels} to higher loops (herein with rather explicit seven-loop expressions). The results in subsection~\ref{subsec:correlators} then iteratively follow in the spirit of \cite{eo07a} by recursive use of the loop insertion operator, and are further immediately translatable to Weil--Petersson volumes again with many explicit higher-loop formulae. Having established a computational means towards ZZ nonperturbative effects, the next step is to validate the resurgent nature of JT gravity, at both free energy and correlation function levels. As it turns out, this further requires including FZZT nonperturbative effects on top of the aforementioned ZZ-instantons, as they both \textit{compete} for large-order \textit{dominance} already at \textit{leading} order in the asymptotics. This analysis is done in section~\ref{sec:resurgence-correlators}, where we further address the explicit resurgent large-genus asymptotics of these correlation functions. In particular, we address the one-point resolvent in subsection~\ref{subsec:onepointW}, and the two-point resolvent in subsection~\ref{subsec:twopointW}. Our Borel analysis and large-order tests provide for rather precise support of our formulae and a clear illustration of the different regimes of ZZ versus FZZT dominance. It would be interesting in future work to address instead the regions of ZZ versus FZZT dominance but at the level of \textit{Borel resummations} (eventually comparing to our present large-order results). Another feature of our Borel analysis is that all these nonperturbative effects display \textit{resonance} upon the Borel plane, with singularities always appearing in \textit{symmetric} pairs, indicating how negative-tension counterparts to ZZ and FZZT branes are an integral part of the nonperturbative structure of the theory. The resurgence requirement of negative-tension ZZ branes (corresponding to the anti-eigenvalue tunneling of \cite{mss22}) has already been made clear in \cite{sst23}; herein we now observe that on top of this there is also a resurgence requirement for negative-tension FZZT branes via the large-order behavior of multi-resolvent correlation functions. In this way, on top of the partition function results recently obtained in \cite{mss22, sst23}, it would be interesting in future work to address the complete resurgent transseries for all multi-resolvent correlators of generic matrix models. Due to the relation between JT gravity and Weil--Petersson volumes, large-order correlation-function formulae of section~\ref{sec:resurgence-correlators} translate to new and improved resurgence formulae describing the large-genus growth of Weil--Petersson volumes, which we address in section~\ref{sec:resurgence-volumes}. Large-order formulae are set-up in subsection~\ref{subsec:largegenusWP-formulae}, and then (successfully) tested against explicit large-genus behavior in subsection~\ref{subsec:largegenusWP-tests}. Note how this section further validates the resurgent nature of the generating functions of Weil--Petersson volumes. It is implicit that also at the level of Weil--Petersson volumes there will be regions of ZZ versus FZZT dominance in their large-genus behavior (which had already been somewhat partially seen in the literature, \textit{e.g.}, \cite{mp17, sss19, os19, k20, am20, k21}, but which we herein make clear and explicit). It goes without saying that the resurgent structure of multi-resolvent correlation functions, be it on what concerns large-order asymptotics, be it on what concerns their transseries transmonomial content, naturally translates to correlation functions of JT-partition-functions or eigenvalue spectral-densities---just like for the Weil--Petersson volumes addressed in section~\ref{sec:resurgence-correlators}. It would also be interesting in future work to make all such expressions fully explicit. Two appendices close our work. One appendix~\ref{app:shift}, includes further details on one of our main formulae concerning the shifted partition function for the nonperturbative topological recursion. One other appendix~\ref{app:corr_func}, addresses a different class of correlation functions (natural within the setting of two-dimensional quantum gravity), illustrating how herein the specific-heat two-point function immediately dictates the resurgent structure for all this class of observables. The present paper is partially complementary and companion to the earlier \cite{gs21}, and the reader might benefit from reading them both ensemble.

\section{Setting Up the JT Matrix Model Stage}
\label{sec:setup}

Let us begin by setting our stage on JT gravity and its matrix model description (see as well, \textit{e.g.}, the review \cite{mt22}); on Weil--Petersson volumes and how they serve as the building blocks of JT gravity (see as well, \textit{e.g.}, the review \cite{dw18}); and on the topological recursion for the JT matrix model and its relation to these Weil--Petersson volumes (see as well, \textit{e.g.}, the reviews \cite{eo08, eo09}).

\subsection{From JT Gravity to Its Double-Scaled Matrix Model}
\label{subsec:JTMM-setup}

The euclidean action describing JT gravity is given by \cite{t83, j84}
\be
\label{eq:JT-action}
\CS_{\text{JT}} \left[ g, \Phi \right] = - S_0\, \chi (\mathfrak{X}) - \frac{1}{2} \int_\mathfrak{X} \rmd^2 x\, \sqrt{g}\, \Phi \left( R+2 \right) - \int_{\partial \mathfrak{X}} \rmd s\ \sqrt{h}\, \Phi \left( K-1 \right),
\ee
\noindent
where $\mathfrak{X}$ is the bulk two dimensional spacetime, $\chi (\mathfrak{X})$ its Euler characteristic, $g$ the metric and $\Phi$ the dilaton which sets the curvature $R=-2$. This further implies the dynamics actually takes place on the one-dimensional boundary. With $\beta$ the (regularized) length of this asymptotic boundary, the disk partition function associated to \eqref{eq:JT-action} may be evaluated via localization (directly in the bulk) \textit{exactly} as \cite{msy16, sw17, sss19}
\be
\label{eq:Z-disk}
Z_{\text{disk}} (\beta) = \rme^{S_0}\, \frac{\rme^{\frac{\pi^2}{\beta}}}{\sqrt{16 \pi \beta^3}}.
\ee
\noindent
Note how this is only the \textit{leading} contribution to the one-point function $\ev{Z(\beta)}$, as spacetimes with different topologies will contribute to (generic) correlation functions---and in two dimensions these are classified by genus and number of boundaries, or else $\chi (\mathfrak{X})$ as in \eqref{eq:JT-action}.

The full spacetime genus expansion may be recursively obtained by turning instead to the formulation of JT gravity as a matrix model \cite{sss19}. To fix notation, we refer to an $N \times N$ hermitian one-matrix model with potential $V(M)$ and partition function given by the usual matrix integral
\be
\label{eq:ZN}
\CZ_N = \frac{1}{\text{vol} \left( \text{U}(N) \right)} \int \rmd M\, \rme^{- \frac{1}{g_{\text{s}}} \Tr\, V(M)},
\ee
\noindent
which we normalized with the volume factor of the gauge group. The string coupling $g_{\text{s}}$ matches $\mathrm{e}^{-S_0}$ in the dilaton gravity framework. In this set-up one may compute the (asymptotic) genus-expansion of the connected part of arbitrary partition-function ($\beta_i$ asymptotic boundaries) correlators \cite{sss19}
\be
\label{eq:PFcorr}
\ev{Z(\beta_1) \cdots Z(\beta_n)}_{(\text{c})} \simeq \sum_{g=0}^{+\infty} Z_{g,n} \left( \beta_1, \ldots, \beta_n\right)\, g_{\text{s}}^{2g+n-2},
\ee
\noindent
where the spacetime partition function is $Z(\beta) = \Tr\, \rme^{-\beta M}$ in the matrix model, and we used the subscript $\text{(c)}$ to denote the connected component of the correlator. As we shall see in the following, the above spacetime partition-function correlators are related in a simple way to the matrix-model correlation-functions of multi-resolvents,
\bea
\label{eq:Rn-multi}
W_{n} \left( x_1, \ldots, x_n \right) &=& \ev{\Tr\, \frac{1}{x_1-M} \cdots \Tr\, \frac{1}{x_n-M}}_{(\text{c})} = \\
\label{eq:Rn-multi2}
&\simeq& \sum_{g=0}^{+\infty} W_{g,n} \left( x_1, \ldots, x_n\right)\, g_{\text{s}}^{2g+n-2},
\eea
\noindent
which are the generating functions of multi-trace correlation functions
\be
\label{eq:Rn-multi-trace}
W_{n} \left( x_1, \ldots, x_n \right) = \sum_{\left\{ m_i \ge 1 \right\}} \frac{1}{x_1^{m_1+1} \cdots x_{n}^{m_n+1}} \ev{\Tr\, M^{m_1} \cdots \Tr\, M^{m_n}}_{(\text{c})}.
\ee

Making use of the matrix-model topological-recursion \cite{eo07a}, the computation of the $W_{g,n}$ is recursive---as very clearly illustrated in \cite{sss19}---starting off with the $E$-eigenvalue spectral density which is \textit{dictated from} the above disk partition function \eqref{eq:Z-disk}; \textit{i.e.}, via inverse Laplace transform:
\be
\label{eq:rho-from-Z-LAP}
\rho_0 (E) = \rme^{-S_0}\, \frac{1}{2\pi\rmi} \int_{0^+ -\rmi \infty}^{0^+ + \rmi \infty} \rmd\beta\, Z_{\text{disk}} (\beta)\, \rme^{\beta E} = \frac{1}{4\pi^2}\, \sinh \left( 2\pi \sqrt{E} \right).
\ee
\noindent
More in line with the whole topological-recursion set-up \cite{eo07a}, we shall mostly trade this spectral density by its corresponding spectral curve, via
\be
\label{eq:spec_dens}
\rho_0 (E) = \frac{1}{2\pi}\, \im\, y(E),
\ee
\noindent
which results in \cite{m07a, eo07b, sss19} (changing to the standard spectral curve $\left\{ x,y \right\}$ conventions)
\be
\label{eq:JT_spec}
y(x) = -\frac{1}{2\pi} \sin \left( 2\pi \sqrt{-x} \right).
\ee

Finally, for our purposes it is also convenient to introduce the holomorphic effective potential
\begin{equation}
\label{eq:hol-eff-pot}
V'_{\text{h;eff}}(x)=y(x),
\end{equation}
\noindent
whose real part is the potential acting on the eigenvalues of the matrix model,
\begin{equation}
\label{eq:eff-pot}
V_{\text{eff}}(x) = \re\, V_{\text{h;eff}}(x).
\end{equation} 

\paragraph{On the Double Scaling Limit:}

The JT spectral curve \eqref{eq:JT_spec} has an infinite branch-cut, running from $0$ to $\infty$, which is typical of double-scaled matrix models---in contrast to off-critical finite-cut matrix models which have spectral curves of the form
\be
\label{eq:finite-cut-SC}
y(x) = M(x)\, \sqrt{ \left(x-a\right) \left(x-b\right)},
\ee
\noindent
for some finite interval $(a,b)$ corresponding to the endpoints of the cut. At the level of the spectral geometry, the double-scaling limit can be seen as ``zooming-in'' into one of these two endpoints, sending the other one to infinity; see, \textit{e.g.}, \cite{dgz93}. This procedure affects the observables of the model in a precise way, which is straightforwardly taken into account by the topological recursion. Further, as a consequence of the spectral curve having a branch-cut, both in finite-cut and in double-scaled matrix models the multi-resolvent correlators \eqref{eq:Rn-multi} are multivalued functions of the $x_i$. This implies both the spectral curve and these correlation functions are best understood as single-valued functions on a double-cover of the complex plane.

\subsection{The Role of Weil--Petersson Volumes}
\label{subsec:WPvolumes}

In order to make the connection between the partition-function \eqref{eq:PFcorr} and the multi-resolvent \eqref{eq:Rn-multi} correlators explicit, it is convenient to introduce a few concepts from algebraic geometry in the following. Let $\CM_{g,n}$ be the moduli space\footnote{Note that $\CM_{g,n}$ is isomorphic to the moduli space of hyperbolic Riemann surfaces with $n$ cusps, which is of interest to us as recall from \eqref{eq:JT-action} that the dilaton $\Phi$ sets the curvature $R = -2$ to be constant negative. For this reason the same notation is typically used to denote the two moduli spaces.} of non-singular algebraic curves $(\Upsigma_{g,n}; p_1, \ldots, p_n)$, of genus $g$ and with $n$ distinct marked points $p_i$ on $\Upsigma_{g,n}$. Further, let $\overline{\CM}_{g,n}$ be a suitable compactification of $\CM_{g,n}$ called the Deligne--Mumford compactification (see, \textit{e.g.}, \cite{z12, dw18}). For each $1 \leq i \leq n$ it is possible to define a line-bundle $\mathcal{L}_i$ over $\CM_{g,n}$ whose fiber at each point is the cotangent space to $\Upsigma_{g,n}$ at $p_i$. This bundle extends over $\overline{\CM}_{g,n}$ and one can consider the first Chern classes
\be
\uppsi_i = c_1 ( {\mathcal L}_i )\, \in\, H^2 \left(\overline{\CM}_{g,n}\right), \qquad 1 \leq i \leq n.
\ee
\noindent
These $\uppsi$-classes in a way play the role of the building blocks of the cohomology of the moduli space. The forgetful morphism $\pi: \overline{\CM}_{g,n+1} \to \overline{\CM}_{g,n}$ drops the last point marked $n+1$ and leads to the introduction of the Miller--Morita--Mumford classes
\be
\upkappa_m = \pi_* \left( \uppsi_{n+1}^{m+1} \right)\, \in\, H^{2m} \left(\overline{\CM}_{g,n}\right), \qquad m \geq 0,
\ee
\noindent
which will finally allow us to construct the Weil--Petersson volumes, \textit{i.e.}, the volumes of the moduli space $\overline{\CM}_{g,n}$, $V_{g,n} = \text{vol}_{\text{WP}}\, \overline{\CM}_{g,n}$ (with specified integration measure). These volumes are obtained by integrating the Weil--Petersson symplectic form $\omega_{\text{WP}}$ over $\overline{\CM}_{g,n}$, which in turn is constructed in terms of the first Miller--Morita--Mumford class $\upkappa_1$ as
\begin{equation}
\omega_{\text{WP}} = 2\pi^{2}\upkappa_1.
\end{equation}
\noindent
The precise definition of Weil--Petersson volumes is then (see, \textit{e.g.}, \cite{dw18})
\begin{equation}
\label{eq:Vgn0WP}
V_{g,n} = \frac{1}{\left(3g-3+n\right)!} \int_{\overline{\CM}_{g,n}} \left( 2 \pi^2 \upkappa_1  \right)^{3g-3+n}.
\end{equation}
\noindent
In a completely similar fashion one defines Weil--Petersson volumes $V_{g,n} (\boldsymbol{b}) = \text{vol}_{\text{WP}}\, \CM_{g,n} (\boldsymbol{b})$ of the moduli space $\CM_{g,n} (b_1,\ldots,b_n)$ of hyperbolic Riemann surfaces of genus $g$ but now with $n$ labeled geodesic (in the hyperbolic metric on $\Upsigma$) boundary components of lengths $\boldsymbol{b} = \left( b_1,\ldots,b_n \right) \in \BR_+^n$,
\be
\label{eq:VgnWP}
V_{g,n} (b_1,\ldots,b_n) = \int_{\overline{\CM}_{g,n}} \exp \left( 2\pi^{2}\upkappa_1 \right) \exp \left( \frac{1}{2} \sum_{i=1}^n \uppsi_i b_i^2\right).
\ee
\noindent
These volumes $V_{g,n}(\boldsymbol{b})$ are \cite{m07a, m07b} polynomials\footnote{In the regime where $b_i \gg g \gg 1$, and with all $b_i$ of the same order, the Weil--Petersson volumes are approximated by their highest order terms, which correspond to the Witten--Kontsevich intersection numbers \cite{w91, k92}.} of degree $3g-3+n$ in $b_1^2, \ldots, b_n^2$ whose corresponding coefficients are rational multiples of specific powers of $\pi$. Further, their constant term is \eqref{eq:Vgn0WP}, \textit{i.e.},
\begin{equation}
\label{eq:zerobound}
V_{g,n} (\boldsymbol{0}) \equiv V_{g,n},
\end{equation}
\noindent
itself a rational multiple of $\pi^{6g-6+2n}$. In \cite{m07a, m07b}, Mirzakhani further showed that these Weil--Petersson volumes can be computed recursively.

As shown in \cite{sss19}, several JT gravity observables are directly related to the quantities which were introduced above. In particular, the JT-gravity euclidean path-integral dictated by the action \eqref{eq:JT-action} for a closed surface of genus $g \geq 2$ computes the Weil--Petersson volumes $V_{g,0}$. To make it explicit, the free energy of JT gravity can be written as a sum over topologies of the form
\begin{equation}
\label{eq:JTfree}
\mathcal{F}(g_{\text{s}}) \simeq \sum_{g=0}^{+\infty} V_{g,0}\, g_{\text{s}}^{2g-2}
\end{equation}
\noindent
(but where the genus zero and one contributions need to be defined separately). Since Weil--Petersson volumes grow factorially like $ \sim (2g)!$ at large genus \cite{z07, z08, m10, mz11}, we are dealing with an asymptotic perturbative expansion which needs to be completed through the inclusion of nonperturbative contributions, \textit{i.e.}, D-brane instantons \cite{s90}. The JT-gravity euclidean path-integral over a genus $g$ surface with $n$ Schwarzian boundaries, yielding the corresponding $Z_{g,n} \left( \beta_1, \ldots, \beta_n\right)$ contribution to the partition-function correlator introduced in \eqref{eq:PFcorr}, is obtained by glueing $n$ ``hyperbolic trumpets'', connecting the $b$-geodesic boundary to the $\beta$-asymptotic boundary, of the form
\be
Z_{\text{Sch}}^{\text{trump}} (\beta,b) = \frac{\mathrm{e}^{-\frac{b^{2}}{4\beta}}}{\sqrt{4\pi\beta}},
\ee
\noindent
to the corresponding $n$-boundary Weil--Petersson volume as \cite{sss19}
\be
\label{eq:PFtoWP}
Z_{g,n} \left( \beta_1, \ldots, \beta_n \right) = \int_{0}^{+\infty} b_1\, \rmd b_1\, Z_{\text{Sch}}^{\text{trump}} \left(\beta_1,b_1\right) \cdots \int_{0}^{+ \infty} b_n\, \rmd b_n\, Z_{\text{Sch}}^{\text{trump}}\left(\beta_n,b_n\right)\, V_{g,n} \left( b_1,\ldots,b_n \right).
\ee
\noindent
Again, the asymptotic $\sim (2g)!$ large-genus growth of the $V_{g,n} (\boldsymbol{b})$ implies that so will the genus expansion \eqref{eq:PFcorr} be asymptotic and need adequate nonperturbative completion. This expression makes explicit the relation between partition-function correlators \eqref{eq:PFcorr} and Weil--Petersson volumes \eqref{eq:VgnWP}---but we still would like to connect them to the multi-resolvent correlators \eqref{eq:Rn-multi}, which we will do in the following subsection.

In analogy with the definition of the JT free energy in \eqref{eq:JTfree}, it is convenient to introduce generating functions for the $n$-point Weil--Petersson volumes, which we define as the formal power series
\be
\label{eq:fullWP}
\CV_{n} \left(b_1,\dots,b_n\right) \simeq \sum_{g=0}^{+\infty} V_{g,n} \left( b_1,\dots,b_n \right) g_{\text{s}}^{2g+n-2}
\ee
\noindent
and
\be
\label{eq:fullWP2}
\CV_{n} \simeq \sum_{g=0}^{+\infty} V_{g,n}\, g_{\text{s}}^{2g+n-2}.
\ee
\noindent
By virtue of \eqref{eq:PFcorr} and \eqref{eq:PFtoWP}, the $\CV_n (\boldsymbol{b})$ turn out to be related to full partition-function correlation-functions through
\be
\ev{Z(\beta_1) \cdots Z(\beta_n)}_{(\text{c})} = \int_{0}^{+\infty} b_1\, \rmd b_1\, Z_{\text{Sch}}^{\text{trump}} \left(\beta_1,b_1\right) \cdots \int_{0}^{+ \infty} b_n\, \rmd b_n\, Z_{\text{Sch}}^{\text{trump}}\left(\beta_n,b_n\right)\, \CV_{n}\left(b_1,\ldots,b_n\right).
\ee
\noindent
In short, studying nonperturbative contributions associated to observables in JT gravity goes hand-in-hand with studying those associated to the $\CV_{n} (\boldsymbol{b})$. Moreover, as it will be shown in this work, it will allow us to obtain resurgent large-genus asymptotics for Weil--Petersson volumes.

\subsection{Topological Recursion and JT Gravity Spectral Curve}
\label{subsec:classicalTR}

The final ingredient we need is to set-up the topological recursion for the case of JT gravity. This formalism recursively and compactly computes the genus-$g$ multi-resolvent correlators $W_{g,n}$ in \eqref{eq:Rn-multi2} \cite{eo07a}; which, in turn, are related to the Weil--Petersson volumes in \eqref{eq:VgnWP} through a Laplace transform \cite{eo07b}. The method of topological recursion is built upon a slightly more precise notion of \textit{spectral curve}, denoted by $\mathscr{S}$, than used so far. The data encoding such a spectral curve $\mathscr{S}$ consists of a Riemann surface $\Sigma$ with coordinate $z$; a holomorphic projection $x : \Sigma \to  \mathbb{C}\mathbb{P}^1$ to the base $\mathbb{C}\mathbb{P}^1$ turning $\Sigma$ into a ramified cover of the sphere; a meromorphic one-form $y\, \dd x$ on $\Sigma$; and a fundamental differential of the second-kind, $B (z_1,z_2)$, the Bergman kernel, which is a symmetric $1\otimes 1$-form on $\Sigma \times \Sigma$ with a normalized double-pole on the diagonal (and no other pole), and behaving near the diagonal as:
\be
B(z_1,z_2) = \frac{\dd z_1\, \dd z_2}{\left(z_1-z_2\right)^2} + \text{holomorphic at } z_1=z_2.
\ee
\noindent
Henceforth by ``spectral curve'' we shall imply these data $\mathscr{S} \equiv \left\{ \Sigma, x, y\, \dd x, B (z_1,z_2) \right\}$. The topological recursion construction then associates to any such $\mathscr{S}$ a doubly-indexed family of meromorphic multi-differentials (the symplectic invariants of the spectral curve) $\omega_{g,n}$ on $\Sigma^{\times n}$. In the current case of a spectral curve of genus-$0$, \textit{i.e.}, $\Sigma = \mathbb{C}\mathbb{P}^1$, it is known that the Bergman kernel is unique and given by
\be
\label{eq:Bergman-genus0}
B(z_1,z_2) = \frac{\dd z_1\, \dd z_2}{\left(z_1-z_2\right)^2}.
\ee

\begin{figure}
\begin{center}
\includegraphics[width = 0.9\textwidth]{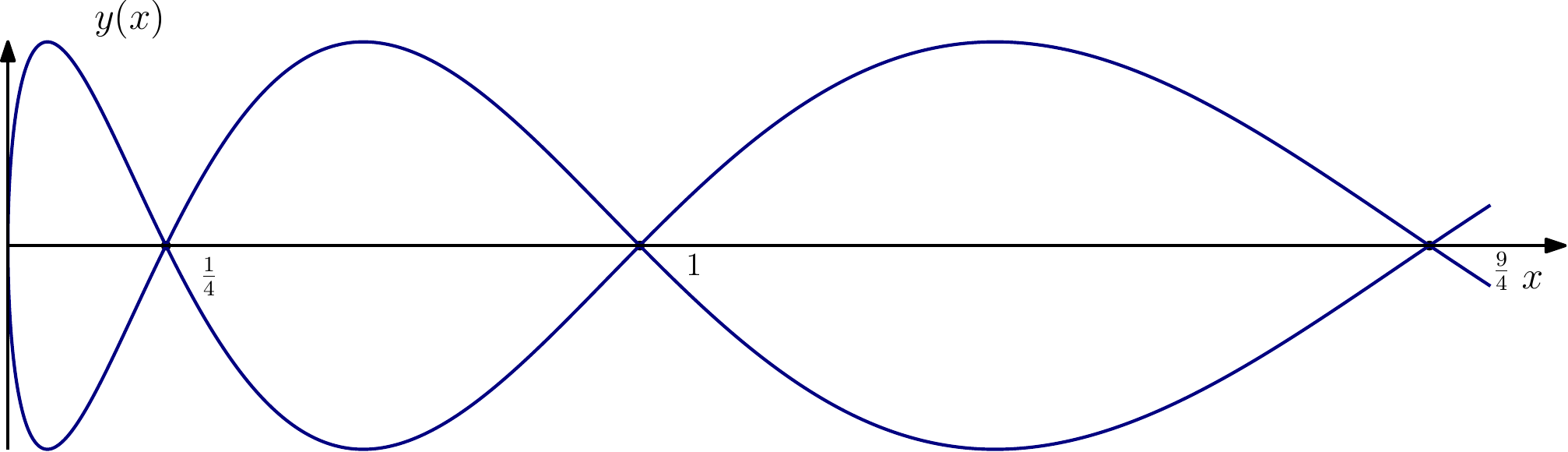}
\caption{The spectral curve of JT gravity in $(x,y)$ coordinates, for real values of $x$.}
\label{fig:spectralcurve}
\end{center}
\end{figure}

In order to align ourselves with the conventions in the mathematical literature concerning the topological recursion and Weil--Petersson volumes, we will not quite work with the spectral curve obtained in \eqref{eq:JT_spec}, but rather with a slight modification\footnote{This change has no effect on the $\widehat{W}_{g,n}$ computed via the topological recursion (see below).}, namely
\be
\label{eq:JT_spec1}
y(x) = \frac{1}{2\pi} \sin \left( 2\pi \sqrt{x} \right).
\ee
\noindent
Using the above spectral curve notation, this takes the form
\be
\label{eq:JT_spec2}
\mathscr{S} = \left\{ \Sigma = \mathbb{C}\mathbb{P}^1,\, x(z) = z^2,\, y(z) = \frac{1}{2\pi} \sin 2\pi z \right\}.
\ee
\noindent
The real part of $\mathscr{S}$, obtained for positive real values of $x$, is depicted in figure~\ref{fig:spectralcurve}. It has a single branchpoint (the zeroes of $\dd x$) at $0 \in \mathbb{C}\mathbb{P}^{1}$, and the involution $\iota : \mathbb{C}\mathbb{P}^{1} \to \mathbb{C}\mathbb{P}^{1}$ exchanging the two sheets around the branch-point---\textit{i.e.}, $x(\iota(z)) = x(z)$---is given by $\iota(z)=-z$. Further, on the $z$-plane we naturally define the $\mathcal{B}_{\ell}$-cycles via the segments from $-\ell/2 $ to $\ell/2$, for $\ell \in \mathbb{N}^+$. We refer to these segments as $\gamma_{-\ell/2 \rightarrow \ell/2}$. If we regard $\mathscr{S}$ as the graph given by $\left(x, y(x) \right) \in \mathbb{C} \times \mathbb{C}$, then the $\mathcal{B}_{\ell}$-cycles become the closed cycles obtained as the $x$-image of the $\gamma_{-\ell/2 \rightarrow \ell/2}$ segments. The only $\mathcal{A}$-cycle of the spectral curve is the single cycle winding around the cut of the matrix model. Cycles of the form $\gamma_{z_1 \rightarrow z_2} $ are called ``of the third kind'' (see \cite{e17} for more details).

In order to set-up the topological recursion, we first define its initial conditions. These are given by the spectral curve $y(x)$ and the Bergman kernel, as
\be
\label{eq:TR-init}
\omega_{0,1}(z) = -\frac{1}{2}\, y(z)\, \mathrm{d} x(z) = - \frac{z \sin{2\pi z}}{2\pi}\, \mathrm{d}z \qquad \text{ and } \qquad \omega_{0,2} (z_1,z_2) = B (z_1,z_2).
\ee
\noindent
The recursion then recursively computes the multi-differentials $\omega_{g,n}(\mathscr{S})$ via \cite{eo07a}
\be
\hspace{-5pt}
\label{eq:TRomega}
\omega_{g,n} \left( z_{1}, I \right) = \underset{z \to 0}{\operatorname{Res}}\, \frac{\frac{1}{2} \int_{-z}^{z} \omega_{0,2} \left(z_{1}, \bullet \right)}{\omega_{0,1}(z) - \omega_{0,1}(-z)} \left( \omega_{g-1,n+1} \left( z, -z ,I \right) + \sum'_{\substack{\mathclap{g_{1} + g_{2} = g}\\ \mathclap{I_{1} \sqcup I_{2} = I}}} \omega_{g_{1}, \abs{I_{1}} + 1} \left( z, I_{1} \right) \omega_{g_{2}, \abs{I_{2}} + 1} \left( -z, I_{2} \right) \right),
\ee
\noindent
where $I = \left\{ z_{2} ,\ldots, z_{n} \right\}$, and the prime in the summation indicates that we should not include the ``boundary'' contributions, $\left\{ I_1=I; g_1=g \right\}$ nor $\left\{ I_2=I; g_2=g \right\}$. One often denotes these multi-differentials as
\be
\label{eq:fromwtoW}
\omega_{g,n} \left( z_{1}, \ldots, z_{n} \right) = \widehat{W}_{g,n} \left( z_{1}, \ldots, z_{n} \right) \mathrm{d}z_{1} \cdots \mathrm{d}z_{n},
\ee
\noindent
where the $\widehat{W}_{g,n}$ functions---with the exception of $\widehat{W}_{0,1}$ and $\widehat{W}_{0,2}$, which are obtained from \eqref{eq:TR-init}---are finally related to our initial goal, the genus-$g$ multi-resolvent correlators $W_{g,n}$ in \eqref{eq:Rn-multi2}, through
\be
\label{eq:fromhattonohat}
\widehat{W}_{g,n} \left( z_1, \ldots, z_n \right) = \left(-2\right)^{n} z_1 \cdots z_n\, W_{g,n} \left( - z_1^2, \ldots, - z_n^2\right), \qquad \text{ for } \qquad (n,g)\neq (0,0).
\ee

Besides recursively computing all genus-$g$ multi-resolvent correlators, the topological recursion also recursively computes all genus-$g$ free energies $\CF_g$; precisely out of the $W_{g,n}$. One way to define these free energies $\mathcal{F}_{g} (\mathscr{S}) \equiv \omega_{g,0} (\mathscr{S})$, for $g>1$, is by
\begin{equation}
\label{eq:free-en-TR}
\mathcal{F}_{g} = \frac{1}{2-2g}\, \underset{z \to 0}{\operatorname{Res}}\, \Phi(z)\, \omega_{g,1}(z), \qquad \text{ where } \qquad \mathrm{d} \Phi(z) = \omega_{0,1}(z).
\end{equation}
\noindent
These can also be packaged into the partition function \eqref{eq:ZN} in the usual way,
\begin{equation}
\label{eq:Zg0}
\CZ (g_{\text{s}}^{-1}\mathscr{S}) = \exp\, \sum_{g=0}^{+\infty} \mathcal{F}_{g} (\mathscr{S})\, g_{\text{s}}^{2g-2}.
\end{equation}
\noindent
In particular, the notion of a partition function shifted by a third-kind cycle $\gamma = \gamma_{z_1 \rightarrow z_2}$ will also be useful in the following. It is defined as
\begin{equation}
\label{eq:Zgn}
\CZ (g_{\text{s}}^{-1}\mathscr{S} + \gamma) = \exp\, \sum_{g=0}^{+\infty} \sum_{n=0}^{+\infty} \frac{g_{\text{s}}^{2g-2+n}}{n!}\, \int_{z_1}^{z_2} \cdots \int_{z_1}^{z_2} \omega_{g,n}(\mathscr{S}).
\end{equation}

As we have already mentioned, all these data are related to the Weil--Petersson volumes $V_{g,n} \left( b_1, \ldots, b_n \right)$ via simple Laplace transform \cite{eo07b} (when focusing on the spectral curve \eqref{eq:JT_spec2}, of course). In particular, the free energies $\CF_g$ obtained by running the topological recursion with spectral curve \eqref{eq:JT_spec2} match the $V_{g,0}$ in \eqref{eq:Vgn0WP}, while the multi-resolvents $\widehat{W}_{g,n} (\boldsymbol{z})$ are related to the corresponding Weil--Petersson volumes $V_{g,n} (\boldsymbol{b})$ defined in \eqref{eq:VgnWP}; upon taking the (slightly unconventional) Laplace transform
\be
\NCL_{z} \left\{f\right\} = \int_{0}^{+\infty} b\, \mathrm{d}b\, f(b)\, \rme^{-z b}. 
\ee
\noindent
It acts on power series as the linear isomorphism
\bea
\NCL\, :\, \mathbb{C}[[b^2]] \quad &\longleftrightarrow& \quad z^{-2}\, \mathbb{C}[[z^{-2}]] \\
\label{eq:Laplace-b-vs-z}
\frac{b^{2k}}{\left(2k+1\right)!} \quad &\longleftrightarrow& \quad \frac{1}{z^{2k+2}}, \quad k \geq 0. 
\eea
\noindent
The relation between topological recursion data and Weil--Petersson volumes is then is simply
\be
\label{eq:fromWtoV}
\widehat{W}_{g,n} \left( z_1, \ldots, z_n\right) = \left( \prod_{i=1}^n \NCL_{z_i} \right) \cdot V_{g,n} \left( b_1, \ldots, b_n \right). 
\ee
\noindent
Conversely, we have
\begin{equation}
\label{eq:fromVtoW}
V_{g,n} \left( b_1, \ldots, b_n \right) = \left( \prod_{i=1}^n \NCL^{-1}_{b_i} \right) \cdot \widehat{W}_{g,n} \left( z_1, \ldots, z_n \right),
\end{equation}
\noindent
where
\be
\label{eq:antiLapl}
\NCL^{-1}_{b} \left\{f\right\} = \underset{z \to 0}{\operatorname{Res}}\, f(z)\, \frac{\mathrm{e}^{b z}}{b}.
\ee
\noindent
Because the JT partition-function correlators \eqref{eq:PFcorr} are dictated by the Weil--Petersson volumes \eqref{eq:VgnWP} according to \eqref{eq:PFtoWP}, then, via the above formulae, so will these JT correlators \eqref{eq:PFcorr} be dictated by the multi-resolvent correlation functions \eqref{eq:Rn-multi2}---and vice-versa. One is hence free to pick what is the convenient set of objects to compute, from which all else follows.

It was further shown in \cite{eo07b} that the Mirzakhani recursion-relation for the $V_{g,n} (\boldsymbol{b})$ in \cite{m07a, m07b} is equivalent to the topological recursion for the $\widehat{W}_{g,n} (\boldsymbol{z})$ in \eqref{eq:TRomega}, for the spectral curve \eqref{eq:JT_spec2} and upon application of the above Laplace transform. As an example, we list the first few $\widehat{W}_{g,n} (\boldsymbol{z})$ and their corresponding\footnote{In the same way that these JT correlation functions may be regarded as a deformation \cite{sss19} of the multi-resolvent correlators for the topological-gravity Airy spectral-curve (\textit{i.e.}, the simplest spectral-curve example of $y = \sqrt{x}$ \cite{w91, k92, dw18}, where recall that our JT spectral curve \eqref{eq:JT_spec2} is $y \approx \sqrt{x} - \frac{2}{3} \pi^2 x^{3/2} + \cdots$), also their corresponding Weil--Petersson volumes may be regarded as a deformation of the Kontsevich volumes; \textit{e.g.}, \cite{abcglw20}.} $V_{g,n} (\boldsymbol{b})$ (some of these already appeared in \cite{sss19})
\begin{align}
&\widehat{W}_{0,3} \left( z_{1}, z_{2}, z_{3} \right) =  \frac{1}{z_{1}^{2} z_{2}^{2} z_{3}^{2}} & & \longleftrightarrow & &V_{0,3} \left( b_1, b_2, b_3 \right) = 1, \\
&\widehat{W}_{1,1} \left(z\right) = \frac{1}{8 z^{4}} + \frac{\pi^{2}}{12 z^{2}} & & \longleftrightarrow & &V_{1,1} \left(b\right) = \frac{1}{48} b^2 + \frac{\pi^2}{12}, \\
&\widehat{W}_{0,4} \left( \boldsymbol{z} \right) = \frac{1}{z_{1}^2 z_{2}^2 z_{3}^2 z_{4}^2} \left( \sum_{i=1}^4\frac{3}{{z}_{i}^{2}} + 2\pi^{2} \right) & & \longleftrightarrow & &V_{0,4} \left( \boldsymbol{b} \right) = \sum_{i=1}^4 \frac{1}{2} b_i^2 + 2\pi^2, \\
&\widehat{W}_{1,2} \left( z_1, z_2 \right) = \frac{1}{z_{1}^2 z_{2}^2} \left( \frac{5}{8 z_{1}^{4}} + \frac{5}{8 z_{2}^{4}} + \right. & & \longleftrightarrow & &V_{1,2} \left( b_1, b_2 \right) = \frac{1}{196} \left( b_1^4 + b_2^4 \right) + \\
&\qquad\qquad\quad \left. + \frac{\pi^{2}}{2 z_{1}^{2}} + \frac{\pi^{2}}{2 z_{2}^{2}} + \frac{3}{8 z_{1}^{2} z_{2}^{2}} + \frac{\pi^{4}}{4} \right) & & \quad & &\qquad\qquad\quad + \frac{\pi^2}{12} \left( b_1^2 + b_2^2 \right) + \frac{1}{96} b_1^2 b_2^2 + \frac{\pi^2}{4}. \nonumber
\end{align}

\section{ZZ Instantons from Matrix Model Eigenvalue Tunneling}
\label{sec:oldmmodels}

Having set the basis of the JT-gravity matrix-model \cite{sss19}, we are ready to move towards computing its associated nonperturbative effects. Standard, instanton-type, nonperturbative effects in matrix models are associated to eigenvalue tunneling \cite{d91, d92}, which, in turn, from a minimal-string-theoretic viewpoint are associated to ZZ-branes; \textit{e.g.}, \cite{zz01, akk03, st04, sst23}. In particular, the computation of nonperturbative data associated to the one-instanton sector of a generic one-cut (hermitian) matrix model was done in \cite{msw07} (and extended to multi-instantons in \cite{msw08}). This was obtained by studying the $N$-eigenvalue configuration where $N-1$ eigenvalues sit at the perturbative minimum of the matrix-model effective-potential \eqref{eq:eff-pot}, while the one remaining eigenvalue is placed at a nonperturbative extremum of the potential. Such extrema $x_\star$, naturally located outside the cut, are defined via the matrix-model saddle-point requirement \cite{msw07}
\be
\label{eq:NP-saddle-def}
V'_{\text{h;eff}}(x_\star)=0 \qquad \Rightarrow \qquad M(x_\star) = 0.
\ee
\noindent
We used the holomorphic effective potential \eqref{eq:hol-eff-pot} and moment function $M(x)$ featured in the spectral curve \eqref{eq:finite-cut-SC}. In the present section, the results in \cite{msw07} will be briefly sketched and then adapted to the JT-gravity double-scaled infinite-cut case. In the following section~\ref{sec:NPtoprec} we will then systematize the procedure in \cite{msw07}, by exploiting the recursive nature of the topological recursion.

The complete transseries \cite{abs18} of JT gravity would have to take into account \textit{all} its (multi-instanton) nonperturbative contributions, which include \textit{resonant} (negative-brane) effects \cite{gs21, mss22, sst23}; but such a complete construction is out of the scope of the present work. Instead, we are herein focusing on first addressing \textit{multi-loop} corrections around the \textit{one}-instanton nonperturbative sector. In this case, out of the (already simplified) one-parameter transseries for the matrix-model free-energy
\be
\label{eq:F-1PTS}
\mathcal{F} \left( g_{\text{s}}, \sigma \right) = \CF^{(0)} + \sum_{\ell=1}^{+\infty} \sigma^\ell\,
\rme^{-\frac{\ell A}{g_{\text{s}}}}\, \sum_{h=0}^{+\infty} \CF^{(\ell)}_h\, g_{\text{s}}^{h+\beta^{(\ell)}} + \cdots,
\ee
\noindent
we focus solely on its \textit{one}-instanton sector (which we also rewrite in terms of the matrix-model partition-function as in \cite{hhikkmt04, msw07})
\be
\CF = \CF^{(0)} + \sigma\, \CF^{(1)} + \cdots = \log \CZ_N = \log \CZ_N^{(0)} + \sigma\,  \frac{\CZ_N^{(1)}}{\CZ_N^{(0)}} + \cdots. 
\ee
\noindent
In the above formulae, $\CF^{(0)}$ denotes the perturbative sector, $A$ is the instanton action, and each $\ell$-instanton sector has an associated characteristic exponent $\beta^{(\ell)}$ and a corresponding power of the \textit{transseries parameter} $\sigma$. In particular, the one-instanton sector is
\be
\label{eq:F(1)MM}
\CF^{(1)} = \frac{\CZ_N^{(1)}}{\CZ_N^{(0)}}.
\ee

Following \cite{msw07}, the one-instanton sector of the partition function is obtained by taking the matrix integral \eqref{eq:ZN} to diagonal ($\lambda$-eigenvalue) gauge, and then specifying eigenvalue integration-contours as 
\begin{equation}
\label{eq:fMSW}
\CZ^{(1)}_N = \frac{N}{N! \left(2\pi\right)^N} \int_{\mathcal{I}} \mathrm{d}x\, \mathrm{e}^{-\frac{V(x)}{g_s}}\, \int_{\mathcal{I}_0} \prod_{i=1}^{N-1} \mathrm{d}\lambda_i\, \Delta^2 \left(x, \lambda_1, \ldots, \lambda_{N-1}\right) \mathrm{e}^{- \frac{1}{g_{\text{s}}} \sum_{i=1}^{N-1} V(\lambda_i)}.
\end{equation}
\noindent
Here $\mathcal{I}$ is the saddle-point contour passing through the non-trivial nonperturbative saddle $x_{\star}$ defined in \eqref{eq:NP-saddle-def}, $\mathcal{I}_0$ is the standard perturbative saddle-point contour, and $\Delta$ denotes the usual Vandermonde determinant. This is rewritable as
\begin{equation}
\label{eq:f2MSW}
\CZ^{(1)}_N = \frac{1}{2\pi} \int_{\mathcal{I}} \mathrm{d}x\, f(x),
\end{equation}
\noindent
with
\be 
\label{eq:fff}
f(x) = \CZ^{(0)}_{N-1} \ev{\det \left( x \textbf{1} - M'\right)^2}_{N-1}^{(0)} \mathrm{e}^{- \frac{1}{g_{\text{s}}}V(x)}.
\ee

Double-scaled one-matrix models have flat effective potential along the semi-infinite interval $\CC = (-\infty,0]$, corresponding to the branch-cut of the double-scaled spectral-curve, and, in general, may have local minima or maxima outside of the cut. Each of the these local minima or maxima of the matrix-model effective-potential, located along $x>0$, corresponds to a (positive/negative) instanton configuration, whose action is given by \cite{msw07}
\be
\label{eq:instactMM}
A = V_{\text{h;eff}}(x_\star) - V_{\text{h;eff}}(0) = \int_{0}^{x_\star} \rmd x\, y(x),
\ee
\noindent
where the holomorphic effective potential $V_{\text{h;eff}}$ is defined in \eqref{eq:hol-eff-pot} and $x_\star$ is the position of the nonperturbative saddle we want to consider. Combining \eqref{eq:F(1)MM}-\eqref{eq:f2MSW}-\eqref{eq:fff}, we obtain
\be
\CF^{(1)} = \frac{1}{2\pi\CZ^{(0)}_N} \int_{\mathcal{I}} \mathrm{d}x\, f(x),
\ee
\noindent
whose general structure takes the form \cite{msw07}:
\be
\label{eq:MM-F(1)}
\CF^{(1)} \simeq  \sqrt{g_{\text{s}}}\, S_1\, \rme^{-\frac{A}{g_{\text{s}}}} \left\{ \CF^{(1)}_{1} + g_{\text{s}}\, \CF^{(1)}_{2} + \cdots \right\}.
\ee
\noindent
Herein $S_1$ is the Stokes coefficient, and the $\CF^{(1)}_{h}$ denote the $h$-loop contributions around the one-instanton configuration. Note that the $\sqrt{g_s}$ factor implies a characteristic exponent $\beta^{(1)}=\frac{1}{2}$. As it turns out \cite{msw07, msw08}, all nonperturbative data, \textit{i.e.}, all coefficients appearing in \eqref{eq:MM-F(1)}, are encoded via spectral geometry. In the present context of a spectral curve with an infinite cut from $-\infty$ to 0, it is convenient to first write it in terms of the moment function $M(x)$ in \eqref{eq:finite-cut-SC} as
\begin{equation}
y(x) = M(x)\, \sqrt{x}.
\end{equation}
\noindent
Then, adapting the results in \cite{msw07} to the present double-scaled infinite-cut case, we obtain the first loop around the one-instanton configuration as
\begin{equation}
\label{eq:1loopMM}
S_1 \cdot \CF_1^{(1)} = \frac{1}{4} \sqrt{\frac{1}{2\pi M'(x_\star)\, x_\star^{5/2}}}.
\end{equation}
\noindent
Analogously, one can proceed to obtain higher loops around the one-instanton configuration. For convenience, introduce the notation
\be
\label{eq:Ftilde-norm}
\widetilde{\CF}^{(1)}_h \equiv \frac{\CF^{(1)}_h}{\CF_1^{(1)}}.
\ee
\noindent
We then obtain \cite{msw07}, \textit{e.g.},
\bea
\widetilde{\CF}^{(1)}_2 &=& \frac{M' ( 0 )}{4 \sqrt{x_\star}\, M^2 ( 0 )} - \frac{17}{12 x_\star^{3/2}\, M ( 0 )} + \frac{5 \left( M'' ( x_\star ) \right)^2 - 3 M' ( x_\star )\, M^{(3)} ( x_\star )}{24 \sqrt{x_\star} \left( M' ( x_\star ) \right)^3} + \nonumber \\
&&
+ \frac{35 M'' ( x_\star )}{48 x_\star^{3/2} \left( M' ( x_\star ) \right)^2} + \frac{173}{96 x_\star^{5/2}\, M' ( x_\star )}.
\label{eq:2loopMM}
\eea

\begin{figure}
\begin{center}
\includegraphics[width = 0.7\textwidth]{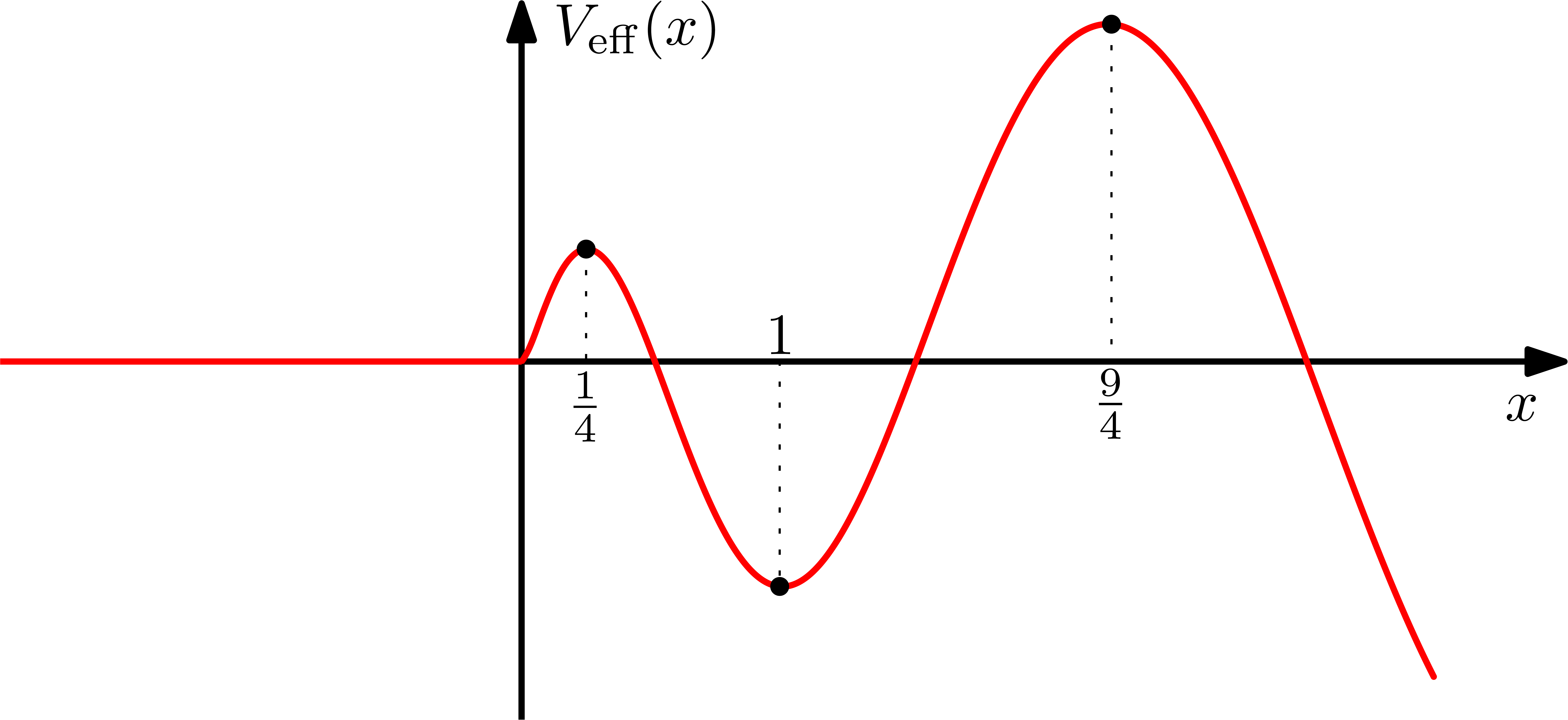}
\caption{The potential $V_{\text{eff}}(x)$ acting on a single eigenvalue of the JT matrix model.}
\label{fig:potential}
\end{center}
\end{figure}

It is straightforward to apply the above formulae to the spectral curve of the JT matrix model, \eqref{eq:JT_spec1}. The real part of the holomorphic effective potential \eqref{eq:eff-pot} yields the potential acting on a single eigenvalue \cite{sss19},
\begin{equation}
\label{eq:Veff}
V_{\text{eff}} (x) = \frac{1}{4\pi^3}\, \Big( \sin \left( 2\pi\sqrt{x} \right) - 2\pi \sqrt{x}\, \cos \left( 2\pi\sqrt{x} \right) \Big), \qquad x>0,
\end{equation}
\noindent
illustrated in figure~\ref{fig:potential} (this potential vanishes for $x<0$). The first stationary-point of the effective-potential \eqref{eq:Veff} (\textit{i.e.}, the smallest instanton action in absolute value) is located at $x_{\star,1}=1/4$; from where the corresponding instanton-action follows from \eqref{eq:instactMM} as \cite{sss19} 
\begin{equation}
\label{eq:A1}
A_1 = \int_{0}^{\frac{1}{4}} \mathrm{d}x\, y(x) = \frac{1}{4\pi^2}.
\end{equation}
\noindent
It is straightforward to extend this result to all the stationary points, corresponding to $x_{\star,\ell} = \ell^2/4$ for $\ell \in \BN^+$, obtaining the instanton actions
\begin{equation}
\label{eq:A_ll}
A_\ell = \int_{0}^{\frac{\ell^2}{4}} \mathrm{d}x\, y(x) = (-1)^{\ell+1} \frac{\ell}{4\pi^2}.
\end{equation}
\noindent
These had been previously computed in \cite{gs21} as the $k \to +\infty$ limit of the $(2,2k-1)$ minimal string result---and of course everything matches. Next, let us compute the coefficients $\CF_{h}^{(1)}$ appearing in \eqref{eq:MM-F(1)}, by using the formulae \eqref{eq:1loopMM} and \eqref{eq:2loopMM}. For the nonperturbative saddle $x_{\star,1}=1/4$ we obtain
\be
\label{eq:one-loop-msw}
S_1 \cdot \CF_1^{(1)} = \frac{\rmi}{\sqrt{2\pi}}
\ee
\noindent
and
\be
\label{eq:two-loops-msw}
\widetilde{\CF}^{(1)}_2 = -\frac{68}{3}-\frac{5}{6}\pi^{2}.
\ee
\noindent
These results are immediately generalizable to all other nonperturbative saddles labelled by the positive integer $\ell$. We obtain
\be
\label{eq:one-loop-msw-l}
S_1 \cdot \CF_{\ell,1}^{(1)} = \sqrt{\frac{(-1)^{\ell}}{2\pi\ell^{3}}}
\ee
\noindent
and
\be
\label{eq:two-loops-msw-l}
\widetilde{\CF}^{(1)}_{\ell,2}=\frac{1}{6\ell^3} \left( 68 \left(-1 + \left(-1\right)^\ell\right) + \pi^2 \ell^2 \left(-2 + 3 \left(-1\right)^\ell\right) \right).
\ee
\noindent
\eqref{eq:one-loop-msw-l} was previously computed in (Table 1 of) \cite{gs21} as the $k \to +\infty$ limit of the $(2,2k-1)$ minimal string results, with an exact match. In the next section, we show how the topological recursion can be used not only to obtain all these expressions, but, more importantly, to recursively generate in-principle \textit{all} loops around any given instanton configuration. Furthermore, it generalizes these results to arbitrary multi-resolvent correlation functions in a straightforward way.

One last discussion before closing this section pertains to the possible expected structure of the JT string equation. The analysis in \cite{gs21} (see as well \cite{os19}) raised the possibility that such string equation might be of \textit{finite-difference} rather than differential type. Such possibility now has additional support from the present results concerning the location of JT Borel singularities. Have in mind that the JT transseries is resonant; \textit{i.e.}, that all instanton actions, or corresponding Borel singularities, arise in symmetric pairs. This was shown in \cite{gs21} and further analyzed in \cite{sst23} where some resonant sectors of the JT transseries were explicitly computed (alongside a few non-trivial Stokes data). What this implies is that the complete set of JT Borel singularities includes \eqref{eq:A_ll} and all their symmetrics, \textit{i.e.}, one must extend $\ell \in \BZ^{\times}$. Moreover, the multi-instanton singularities will be integer multiples of this set, hence themselves on top of the latter. Such infinite ``towers'' of distinct Borel singularities running over the integers, with very similar ``local'' nonperturbative content as in \eqref{eq:one-loop-msw-l}-\eqref{eq:two-loops-msw-l}, are indeed a distinctive feature of the resurgent structure of finite-difference equations. Albeit this is not a proof, it is still compelling evidence to be explored in future work, possibly building on our present results alongside \cite{gs21, mss22, sst23}.

\section{The Nonperturbative Topological Recursion}
\label{sec:NPtoprec}

The topological recursion \cite{eo07a} as briefly reviewed in subsection~\ref{subsec:classicalTR} mainly concerns \textit{perturbative} computations; both within the matrix-model world and beyond it \cite{eo08, eo09}---concretely, this ``perturbative'' topological recursion computes the perturbative expansion associated to rather generic algebraic curves. However, such expansion cannot be the whole story. As discussed in the previous section, this is in fact already very clear at the matrix model level \cite{msw07, m08, msw08}. It was further realized in \cite{em08} that nonperturbative corrections were necessary in order to obtain holomorphic, background-independent partition-functions for curves describing either matrix models or topological strings. In the present section we show how to extend the ``perturbative'' topological recursion to a complete ``nonperturbative'' topological recursion, \textit{i.e.}, a recursion which further includes \textit{nonperturbative} contributions; and how in the process it provides a geometric interpretation for instanton corrections. Note that, just as in the previous section, what we are computing now are nonperturbative effects of ZZ-brane type. In particular, we shall obtain loop expansions around the ZZ one-instanton sector of both free energy and multi-resolvent correlators.

\subsection{The One-Instanton Sector of the Free Energy}
\label{subsec:NPTR-1inst-F}

Building upon the calculation in \cite{msw07} (see as well \cite{msw08, mss22})---which we applied to JT gravity in the preceding section---we start by computing the one-instanton contribution to the free energy. This will recover the results of section~\ref{sec:oldmmodels}, now in the language of the topological recursion, as well as show how to easily extend them to an arbitrary number of loops around a fixed one-instanton sector. Furthermore, our approach turns out to be particularly fruitful when considering nonperturbative contributions to correlation functions, as will be shown in our following subsection~\ref{subsec:correlators}.

The perturbative partition function $\CZ_N^{(0)}$ associated to the matrix integral \eqref{eq:ZN} encodes the perturbative free energies $\mathcal{F}^{(0)}_g = \omega_{g,0}$ in \eqref{eq:free-en-TR} associated to the spectral curve $\mathscr{S}$, as in equation \eqref{eq:Zg0}, \textit{i.e.}, as
\be
\CZ_N^{(0)} = \CZ (g_{\text{s}}^{-1} \mathscr{S}) \simeq \exp\, \sum_{g=0}^{+\infty} \mathcal{F}^{(0)}_{g} (\mathscr{S})\, g_{\text{s}}^{2g-2}.
\ee
\noindent
As we shall see in the following, the one-instanton sector of the free energy, $\CF^{(1)}$, may in fact be obtained starting from the above quantity. First, define 
\be
\label{eq:Fgn}
F_{g,n}(z) = \overbrace{\int_{-z}^z \cdots \int_{-z}^z}^{n} \omega_{g,n},
\ee
\noindent
for $(g,n)\neq (0,2)$. The $(g,n)= (0,2)$ case needs to be defined separately via regulatization \cite{eo07a}, as
\begin{equation}
F_{0,2}(z) = - 2 \log 4 z^2 = - 2 \log 4 x (z),
\end{equation}
\noindent
where in the last equality we solely use the adequate parametrization for a genus-$0$ spectral curve $x(z) = z^2$ as in \eqref{eq:JT_spec2}. Next, collect these $F_{g,n}$ according to their Euler characteristic $\chi$, by defining
\be
\label{eq:SSchi}
\mathbb{S}_{\chi} (z) = \sum_{\substack{2g-2+n=\chi-1\\ g\geq 0, n\geq 1}} \frac{F_{g,n}(z)}{n!}, \qquad \chi=0,1,\ldots,
\ee
\noindent
and
\be
\label{eq:SSchi-summed}
\BS \left(z;g_{\text{s}}\right) = \sum_{\chi=0}^{+\infty} \BS_{\chi} (z)\, g_{\text{s}}^{\chi-1}.
\ee
\noindent
Finally, introduce the \textit{wave function}
\be
\label{eq:psi-section4}
\psi \left( x(z); g_{\text{s}} \right) = \frac{\CZ \left( g_{\text{s}}^{-1} \mathscr{S} + \gamma_{-z \rightarrow z} \right)}{\CZ \left( g_{\text{s}}^{-1} \mathscr{S} \right)} = \exp\, \BS \left(z;g_{\text{s}}\right).
\ee
\noindent
We will often abuse notation and drop explicit $z$ or $g_{\text{s}}$ dependence in these expressions. 

\begin{figure}
\begin{center}
\includegraphics[width=\textwidth]{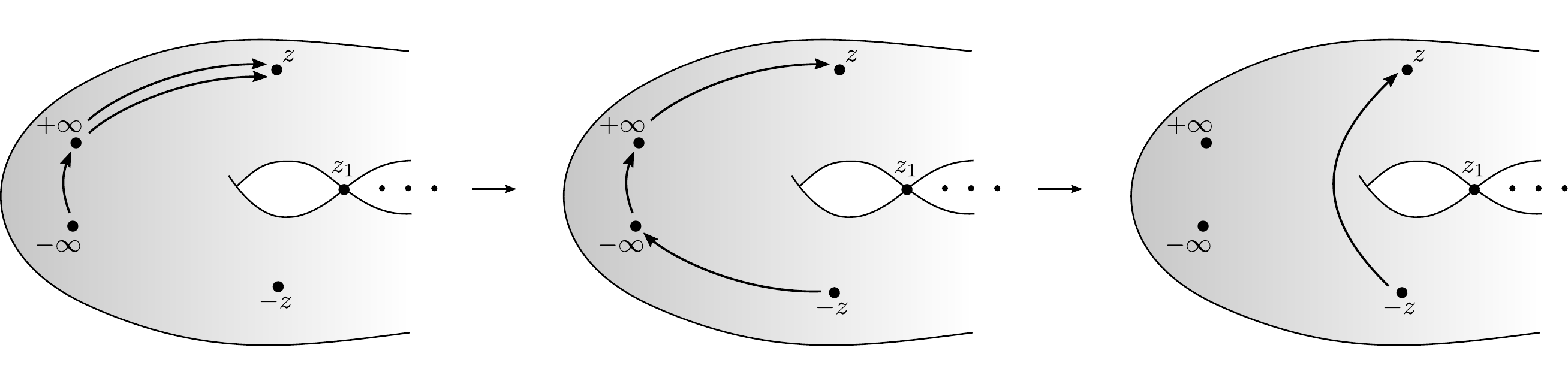}
\caption{A pictorial representation of the composition of cycles in \eqref{eq:cyclecomp}.}
\label{fig:sketch}
\end{center}
\end{figure}

\begin{figure}
\begin{center}
\includegraphics[width = 9cm]{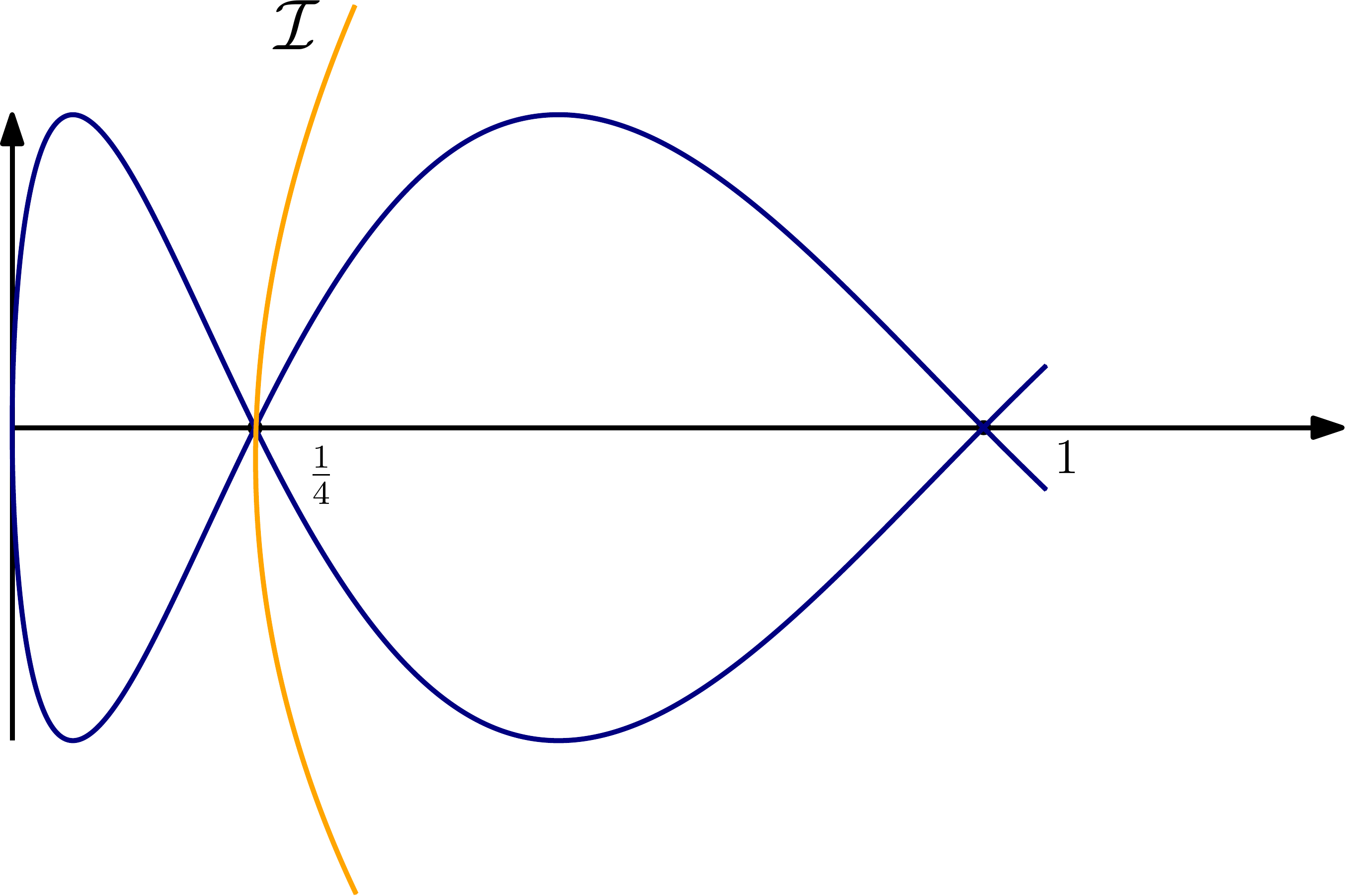}
\caption{Non-trivial saddle-point contour $\mathcal{I}$ through the maximum located at $x_{\star,1}=\frac{1}{4}$.}
\label{fig:path}
\end{center}
\end{figure}

Now, the point of these definitions is that one can use the above wave-function $\psi(x)$ to easily obtain the function $f(x)$ appearing in \eqref{eq:f2MSW}; which in turn allows for a  systematic computation of the free-energy one-instanton sector, $\CF^{(1)}$, using the tools of the topological recursion. Indeed we find:
\be
\label{eq:ftopsi}
f \left( x(z) \right) = \CZ \left( g_{\text{s}}^{-1} \mathscr{S} + \gamma_{-z \rightarrow z} \right) = \psi \left( x(z) \right) \CZ \left( g_{\text{s}}^{-1} \mathscr{S} \right).
\ee
\noindent
This equality is motivated as follows. One first realizes that the expectation-value of the squared-determinant (evaluated in a matrix model with $N-1$ eigenvalues) appearing in the definition of $f(x)$ in \eqref{eq:fff}, is equivalent to the partition function of a spectral curve which is obtained by \textit{shifting} the original one in a specific way (see, \textit{e.g.}, \cite{e17}). This is:
\be
\label{eq:shiftedZ}
\CZ_{N-1}^{(0)} \ev{\det \left( x \textbf{1} - M'\right)^2}_{N-1}^{(0)} \rme^{- \frac{1}{g_{\text{s}}}V(x)} = \CZ \left( g_{\text{s}}^{-1}\mathscr{S} + 2\gamma_{+ \infty \rightarrow z} + \gamma_{- \infty \rightarrow + \infty} \right).
\ee
\noindent
Herein, the curve needed to be shifted twice by the third-kind cycle $\gamma_{+\infty \rightarrow z}$ in order to account for the squared determinant, and once by the third-kind cycle $\gamma_{-\infty \rightarrow +\infty}$ in order to account for for the shift from $N$ to $N-1$ (more details on why these are the appropriate shifts are given in appendix~\ref{app:shift}). Further using the symmetry under the involution of the spectral curve and the composition of cycles, we obtain
\bea
\CZ \left( g_{\text{s}}^{-1} \mathscr{S} + 2 \gamma_{+\infty \rightarrow z} + \gamma_{-\infty \rightarrow +\infty} \right) &=& \CZ \left( g_{\text{s}}^{-1} \mathscr{S} + \gamma_{+\infty \rightarrow z} - \gamma_{-\infty \rightarrow -z} + \gamma_{-\infty \rightarrow +\infty} \right) = \nonumber \\
&=& \CZ \left( g_{\text{s}}^{-1} \mathscr{S} + \gamma_{-z \rightarrow z} \right),
\label{eq:cyclecomp}
\eea
\noindent
as sketched in figure~\ref{fig:sketch}. This establishes \eqref{eq:ftopsi}. Finally, combining \eqref{eq:F(1)MM}, \eqref{eq:f2MSW}, and \eqref{eq:ftopsi}, a rather compact expression for the all-orders one-instanton free-energy emerges as
\be
\label{eq:saddle-point}
\CF^{(1)} = \frac{1}{2\pi} \int_{\mathcal{I}} \rmd x\, \psi (x) = \frac{1}{2\pi} \int_{\mathcal{I}} \rmd x\, \exp \left( \frac{1}{g_{\text{s}}} \BS_0 (x) + \BS_1 (x) + \sum_{\chi>1} \BS_{\chi} (x)\, g_{\text{s}}^{\chi-1} \right),
\ee
\noindent
where $\mathcal{I}$ is the first non-trivial saddle-point contour (see figure~\ref{fig:path}), and the $\BS_{\chi} (x)$ are simply the $\BS_{\chi} (z)$ but using $x(z) = z^2$ from \eqref{eq:JT_spec2}. Recall from the previous section~\ref{sec:oldmmodels} that, for JT gravity, the first stationary-point of the matrix-model effective-potential sits at $x_{\star,1}=1/4$ (\textit{e.g.}, recall figure~\ref{fig:potential}). The path of steepest-descent through this saddle (a maximum along the real line) has the direction of $\rme^{\mathrm{i}\frac{\pi}{2}}=\mathrm{i}$. Therefore, we deform $x$ around $x_{\star,1}$ as $x=x_{\star,1}+\mathrm{i}\sqrt{g_{\text{s}}}\, \rho$ and the integral \eqref{eq:saddle-point} becomes
\bea
\label{eq:steep-des}
\int_{\mathcal{I}}\mathrm{d}x\, \psi(x) &=& \\
&&
\hspace{-60pt}
= \mathrm{i}\sqrt{g_{\text{s}}} \int_{-\infty}^{+\infty} \mathrm{d}\rho\, \exp \left( \frac{1}{g_{\text{s}}} \BS_0 \left(x_{\star,1}+\mathrm{i}\sqrt{g_{\text{s}}}\, \rho\right) + \BS_1 \left(x_{\star,1}+\mathrm{i}\sqrt{g_{\text{s}}}\, \rho\right) + \sum_{\chi>0} \BS_{\chi} \left(x_{\star,1}+\mathrm{i}\sqrt{g_{\text{s}}}\, \rho\right) g_{\text{s}}^{\chi-1} \right). \nonumber
\eea
\noindent
Expanding \eqref{eq:steep-des} in powers of $g_{\text{s}}$ we are left with a Gaussian integral\footnote{Note how all half-integer powers of $g_{\text{s}}$ combine with odd powers of $\rho$, hence do not contribute to the integral.}, yielding
\begin{equation} 
\label{eq:F(1)saddle}
\frac{1}{2\pi} \int_{\mathcal{I}} \mathrm{d}x\, \psi(x) \simeq \mathrm{i} \sqrt{\frac{g_{\text{s}}}{2\pi \BS_0'' (x_{\star,1})}}\, \rme^{\frac{1}{g_{\text{s}}} \BS_0 (x_{\star,1})}\, \rme^{\BS_{1} (x_{\star,1})} \left( 1 + \sum_{h=1}^{+\infty} \widetilde{\CF}^{(1)}_{h+1}\, g_{\text{s}}^h \right).
\end{equation}
\noindent
Herein, the $\widetilde{\CF}_h^{(1)} $ can be computed in terms of the functions $\BS_{\chi}(x)$ and their derivatives, evaluated at the saddle point $x_{\star,1}$. Note that \eqref{eq:F(1)saddle} immediately implies $\beta^{(1)}=1/2$. In fact, via \eqref{eq:saddle-point} this expression further makes the full \eqref{eq:MM-F(1)} precise, including expressions for the instanton action $A$ and the precise Stokes coefficient $S_1$ (which were both also made explicit in \cite{msw07}).

Let us see how these integrations over $\mathcal{B}$-cycles match the results obtained in section~\ref{sec:oldmmodels}. Begin addressing the instanton actions $A_{\ell}$, with the $\mathcal{B}_{\ell}$-cycles given by the cycles of the third-kind $\gamma_{-\ell/2\rightarrow \ell/2}$ (recall the discussion ensuing the JT spectral curve in \eqref{eq:JT_spec2} and figure~\ref{fig:spectralcurve}). We recover the instanton action \eqref{eq:instactMM} of JT gravity \eqref{eq:A1} as the first $\mathcal{B}$-period of the spectral curve, at the first stationary point $x_{\star,1} = 1/4$, by simply using the topological recursion initial condition \eqref{eq:TR-init},
\be
\BS_0 (x_{\star,1}) = \int_{\mathcal{B}_1} \omega_{0,1}(z) = - \frac{1}{2} \int_{-\frac{1}{2}}^{\frac{1}{2}} y(z)\, x'(z)\, \mathrm{d}z = - \int_{0}^{\frac{1}{4}} y(x)\, \mathrm{d} x = - \frac{1}{4\pi^2} = - A_1
\ee
\noindent
Similarly, we recover the instanton actions \eqref{eq:A_ll} at different stationary points $x_{\star,\ell} = \ell^2/4$ as the $\mathcal{B}_{\ell}$-periods of the spectral curve,
\begin{equation} 
A_{\ell} = - \int_{\mathcal{B}_{\ell}} \omega_{0,1}(z) = \frac{1}{2} \int_{-\frac{\ell}{2}}^{\frac{\ell}{2}} y(z)\, x'(z)\, \mathrm{d}z = \int_{0}^{\frac{\ell^2}{4}} y(x)\, \mathrm{d}x = (-1)^{\ell+1} \frac{\ell}{4\pi^2}.  
\end{equation}

Moving on to the computation of the coefficients in \eqref{eq:MM-F(1)} is similarly easy. The first few $\BS_{\chi}(x)$ in our case are:
\bea
\label{eq:SS0}
\BS_0 (x) &=& \frac{1}{4\pi^{3}}\, \Big( 2\pi \sqrt{x}\, \cos \left( 2\pi\sqrt{x} \right) - \sin \left( 2\pi\sqrt{x} \right) \Big), \\
\label{eq:SS1}
\BS_1(x) &=& -\frac{1}{2} \log \left(- 16 x^{2} \right), \\
\label{eq:SS2}
\BS_{2}(x) &=& - \frac{\pi^{2}}{6 \sqrt{x}} - \frac{17}{12 x^{\frac{3}{2}}}, \\ 
\label{eq:SS3}
\BS_{3}(x) &=& \frac{\pi^{4}}{2 x} + \frac{2\pi^{2}}{x^{2}} + \frac{13}{4 x^{3}}.
\eea
\noindent
Comparing \eqref{eq:F(1)saddle} with \eqref{eq:MM-F(1)}, we find Stokes data as
\be
S_1 \cdot \CF_1^{(1)} (x_{\star}) =  \mathrm{i} \frac{\mathrm{e}^{\BS_1 (x_{\star})}}{\sqrt{2\pi \BS_0'' (x_{\star})}}.
\ee
\noindent
The expressions for the following $\widetilde{\CF}_{h}^{(1)}$ as functions of the saddle-point position $x_{\star}$ quickly gain more terms as $h$ increases, in which case we only display the next two. We easily recover the expression for the second-loop around the one-instanton, which already appeared in \cite{msw07},
\be
\label{eq:F_2}
\widetilde{\CF}_2^{(1)} (x_{\star}) = \BS_{2} (x_{\star}) - \frac{\BS_{1}' (x_{\star})^{2}}{2 \BS''_{0} (x_{\star})} - \frac{\BS_{1}'' (x_{\star})}{2 \BS''_{0} (x_{\star})} + \frac{\BS_{0}''' (x_{\star})\, \BS_{1}' (x_{\star})}{2 \BS''_{0} (x_{\star})^{2}} - \frac{5 \BS_{0}''' (x_{\star})^{2}}{24 \BS''_{0} (x_{\star})^{3}} + \frac{\BS_{0}^{(4)} (x_{\star})}{8 \BS''_{0} (x_{\star})^{2}}.
\ee
\noindent
As for the expression at three-loops, it reads:
\begin{align}
\widetilde{\CF}_3^{(1)} (x_{\star}) &= \BS_{3} (x_{\star}) + \frac{1}{2} \BS_{2} (x_{\star}) + \frac{1}{1152} \bigg\{ \frac{1}{\BS''_{0} (x_{\star})} \bigg( - 576 \BS_{2} (x_{\star})\, \BS_{1}' (x_{\star})^{2} - \\
&
- 576 \BS_{2} (x_{\star})\, \BS_{1}'' (x_{\star}) - 1152 \BS_{1}' (x_{\star})\, \BS_{2}' (x_{\star}) - 576 \BS_{2}'' (x_{\star}) \bigg) + \nonumber \\
&
+ \frac{1}{\BS''_{0} (x_{\star})^{2}} \bigg( 576 \BS_{2} (x_{\star})\, \BS_{0}''' (x_{\star})\, \BS_{1}' (x_{\star}) + 144 \BS_{2} (x_{\star})\, \BS_{0}^{(4)} (x_{\star}) + 576 \BS_{0}''' (x_{\star})\, \BS_{2}' (x_{\star}) + \nonumber \\
&
+ 144 \BS_{1}' (x_{\star})^{4} + 864 \BS_{1}' (x_{\star})^{2}\, \BS_{1}'' (x_{\star}) + 576 \BS_{1}' (x_{\star})\, \BS_{1}''' (x_{\star}) + 432 \BS_{1}'' (x_{\star})^{2} +  144 \BS_{1}^{(4)} (x_{\star}) \bigg) + \nonumber \\
&
+ \frac{1}{\BS''_{0} (x_{\star})^{3}} \bigg( - 240 \BS_{2} (x_{\star})\, \BS_{0}''' (x_{\star})^{2} - 480 \BS_{0}''' (x_{\star})\, \BS_{1}' (x_{\star})^{3} - 1440 \BS_{0}''' (x_{\star})\, \BS_{1}' (x_{\star})\, \BS_{1}'' (x_{\star}) - \nonumber \\
&
- 480 \BS_{0}''' (x_{\star})\, \BS_{1}''' (x_{\star}) - 360 \BS_{0}^{(4)} (x_{\star})\, \BS_{1}' (x_{\star})^{2} - 360 \BS_{0}^{(4)} (x_{\star})\, \BS_{1}'' (x_{\star}) - 144 \BS_{0}^{(5)} (x_{\star})\, \BS_{1}' (x_{\star}) - \nonumber \\
&
- 24 \BS_{0}^{(6)} (x_{\star}) \bigg) + \frac{1}{\BS''_{0} (x_{\star})^{4}} \bigg( 840 \BS_{0}''' (x_{\star})^{2}\, \BS_{1}' (x_{\star})^{2} + 840 \BS_{0}''' (x_{\star})^{2}\, \BS_{1}'' (x_{\star}) + \nonumber \\
&
+ 840 \BS_{0}''' (x_{\star})\, \BS_{0}^{(4)} (x_{\star})\, \BS_{1}' (x_{\star}) + 168 \BS_{0}''' (x_{\star})\, \BS_{0}^{(5)} (x_{\star}) + 105 \BS_{0}^{(4)} (x_{\star})^{2} \bigg) + \nonumber \\
&
+ \frac{1}{\BS''_{0} (x_{\star})^{5}} \bigg( - 840 \BS_{0}''' (x_{\star})^{3}\, \BS_{1}' (x_{\star}) - 630 \BS_{0}''' (x_{\star})^{2}\, \BS_{0}^{(4)} (x_{\star}) \bigg) + \frac{1}{\BS''_{0} (x_{\star})^{6}} \bigg( 385 \BS_{0}''' (x_{\star})^{4} \bigg) \bigg\}. 
\nonumber
\end{align}
\noindent
In particular, evaluating the  $\widetilde{\CF}_h^{(1)}$ loop-coefficients at the first nonperturbative saddle, $x_{\star,1} = 1/4$, it follows up to seven loops
\bea
\label{eq:FEdata-1}
S_1 \cdot \CF^{(1)}_{1} &=& \frac{\mathrm{i}}{\sqrt{2\pi}}, \\
\label{eq:FEdata-2}
\widetilde{\CF}_2^{(1)} &=& -\frac{68}{3}-\frac{5\pi^{2}}{6}, \\
\label{eq:FEdata-3}
\widetilde{\CF}_3^{(1)} &=& \frac{12104}{9}+\frac{818\pi^{2}}{9}+\frac{241\pi^{4}}{72}, \\
\label{eq:FEdata-4}
\widetilde{\CF}_4^{(1)} &=& -\frac{10171120}{81}-\frac{311672\pi^{2}}{27}-\frac{175879\pi^{4}}{270}-\frac{163513\pi^{6}}{6480}-\frac{29 \pi^{8}}{48}, \\
\widetilde{\CF}_5^{(1)} &=& \frac{3859832480}{243}+\frac{442580824\pi^{2}}{243}+\frac{50891471\pi^{4}}{405}+\frac{33364187\pi^{6}}{4860}+\frac{9595009\pi^{8}}{31104}+ \nonumber \\
&&
\label{eq:FEdata-5}
+ \frac{19613\pi^{10}}{1440}, \\
\widetilde{\CF}_6^{(1)} &=& -\frac{1843270459520}{729}-\frac{252085765904\pi^{2}}{729}-\frac{100630430374\pi^{4}}{3645}-\frac{6418813856\pi^{6}}{3645}- \nonumber \\
&&
\label{eq:FEdata-6}
-\frac{80771001607\pi^{8}}{816480}-\frac{34287983171\pi^{10}}{6531840}-\frac{673456447\pi^{12}}{2332800}-\frac{9292841\pi^{14}}{1020600}, \\
\widetilde{\CF}_7^{(1)} &=& \frac{3178904639039360}{6561}+\frac{167991228323072\pi^{2}}{2187}+ \frac{15149825697628\pi^{4}}{2187}+ \nonumber \\
&&
+\frac{16095986544931\pi^{6}}{32805}+\frac{189853128357547\pi^{8}}{6123600}+\frac{45276568832183\pi^{10}}{24494400}+ \nonumber \\
&&
\label{eq:FEdata-7}
+\frac{131892469980659\pi^{12}}{1175731200}+\frac{137997129899\pi^{14}}{19595520}+\frac{68726074733\pi^{16}}{195955200}.
\eea
\noindent
Notice how the first two coefficients were already obtained in section~\ref{sec:oldmmodels}, as \eqref{eq:one-loop-msw} and \eqref{eq:two-loops-msw}, and how it is straightforward albeit increasingly lengthy to compute higher loops.

\subsection{The One-Instanton Sector of Multi-Resolvent Correlators} 
\label{subsec:correlators}

One clear advantage of our construction of nonperturbative-contributions to the partition-function via the use of topological-recursion tools, it that we can now---in complete analogy with what we did for the free energy in \eqref{eq:F-1PTS}---very naturally extend it to introduce transseries representations of multi-resolvent \textit{correlation functions}. Things are now slightly more intricate, as on top of the ZZ nonperturbative contributions which we have already seen for the free energy, multi-resolvent correlators also have FZZT nonperturbative contributions \cite{fzz00, t00, sss19} (see as well the discussion in \cite{sst23}, in both minimal string and JT contexts). Their complete transseries \cite{abs18} would then have to take into account all these nonperturbative contributions, ZZ and FZZT, alongside their resonant negative-brane counterparts \cite{gs21, mss22, sst23}---but which we shall not address in full generality in this work. We shall instead herein focus on first addressing \textit{multi-loop} corrections around the \textit{one} ZZ instanton sector, and then address FZZT effects in the following section~\ref{sec:resurgence-correlators}. In this restricted scenario, and in parallel with the one-parameter transseries in \eqref{eq:F-1PTS}, we write the one-parameter ZZ transseries for multi-resolvent (connected) correlators in \eqref{eq:Rn-multi} via their ``hatted'' counterparts in \eqref{eq:fromhattonohat} as
\be
\label{eq:Wn-1PTS}
\widehat{W}_{n} \left( z_{1}, \ldots, z_{n}; \sigma \right) = \widehat{W}_{n}^{(0)} \left( z_{1}, \ldots, z_{n} \right) + \sum_{\ell=1}^{+\infty} \sigma^\ell\, \widehat{W}_{n}^{(\ell)} \left( z_{1}, \ldots, z_{n} \right) + \cdots.
\ee
\noindent
Here, for the perturbative sector we recall the topological expansion \eqref{eq:Rn-multi2},
\be
\label{eq:Rn-multi2-revisited}
\widehat{W}^{(0)}_n \left( z_1, \ldots, z_n \right) \simeq \sum_{g= 0}^{+\infty} \widehat{W}_{g,n}^{(0)} \left( z_1, \ldots, z_n \right)\, g_{\text{s}}^{2g+n-2}.  
\ee
\noindent
The one-instanton sector which we shall focus upon takes the form (compare with the free energy case in \eqref{eq:MM-F(1)})
\be
\label{eq:top_exp_inst}
\widehat{W}_{n}^{(1)} \left( z_{1}, \ldots, z_{n} \right) \simeq \sqrt{g_{\text{s}}}\, S_1\, \mathrm{e}^{-\frac{A}{g_{\text{s}}}}\, \CF_1^{(1)} \sum_{h=0}^{+\infty} \widetilde{W}_{h+1,n}^{(1)} \left( z_{1}, \ldots, z_{n} \right) g_{\text{s}}^{h}, 
\ee
\noindent
where the instanton action and the Stokes coefficient are \textit{the same} as for the free energy in \eqref{eq:MM-F(1)} (due to \eqref{eq:F(1)saddle} and below), and where in parallel with \eqref{eq:Ftilde-norm} we have normalized the coefficients as
\be
\label{eq:tilde_rel}
\widetilde{W}_{h,n}^{(1)} \left( z_{1}, \ldots, z_{n} \right) \equiv \frac{1}{\CF_1^{(1)}}\, \widehat{W}_{h,n}^{(1)} \left( z_{1}, \ldots, z_{n} \right).
\ee

For our purposes it is now very convenient to define the \textit{loop-insertion operator}\footnote{A word on notation. This operator $\Updelta_z$ should not be confused with the alien derivative $\Delta_\omega$ in spite of very similar notation (this is just a minor warning for the resurgence reader; the alien derivative appears nowhere else).} $\Updelta_z$, which acts on the topological-recursion multi-differentials $\omega_{g,n} \left( z_{1}, \ldots, z_{n} \right)$ as \cite{eo07a}
\be
\label{eq:Delta}
\Updelta_{z} \omega_{g,n} \left(z_1, \ldots, z_n\right) = \omega_{g,n+1} \left( z_1, \ldots, z_n, z \right).
\ee
\noindent
It was first introduced in the context of matrix models, but it can be defined for a much wider class of spectral curves \cite{eo07a}. A crucial feature of the operator $\Updelta_z$ is that it acts as a derivation, satisfying Leibniz and chain rules, and it further commutes with the standard derivative. Moreover, since the $\omega_{g,n}$ are symmetric under permutations of their variables, loop insertion operators further commute amongst themselves, \textit{i.e.}, $\Updelta_{z} \Updelta_{\widetilde{z}} = \Updelta_{\widetilde{z}} \Updelta_{z}$. By extension, it further naturally acts on the quantities $F_{g,n} (z)$ and $\BS_{\chi} (z)$ defined \eqref{eq:Fgn} and \eqref{eq:SSchi} in the previous subsection as 
\begin{align}
\Updelta_{z_1} F_{g,n} (z) &= \overbrace{\int_{-z}^z \cdots \int_{-z}^z}^{n} \omega_{g,n+1} \left( \bullet, z_{1} \right), \\
\Updelta_{z_{1}} \BS_{\chi} (z) &= \sum_{\substack{2g-2+n=\chi-1\\ g\geq 0, n\geq 1}} \frac{\Updelta_{z_1} F_{g,n} \left(z\right)}{n!}.
\end{align}
\noindent
Now, in order to compute the ZZ one-instanton contribution to the $n$-point correlator \eqref{eq:top_exp_inst}, we simply need to apply the loop-insertion operator \eqref{eq:Delta} $n$ times to the one-instanton contribution to the free energy \eqref{eq:saddle-point}. In equations:
\bea
\widehat{W}_{n}^{(1)} \left( z_{1}, \ldots, z_{n} \right) &=& \left( \prod_{k=1}^{n} g_{\text{s}}\, \Updelta_{z_k} \right) \CF^{(1)} = \frac{1}{2\pi} \int_{\mathcal{I}} \dd x \left( \prod_{k=1}^{n} g_{\text{s}}\, \Updelta_{z_k} \right) \psi(x) \equiv \frac{g_{\text{s}}^n}{2\pi} \int_{\mathcal{I}} \mathrm{d}x\, \Updelta_{\llbracket n \rrbracket} \psi(x) = \nonumber \\
&=& \frac{g_{\text{s}}^n}{2\pi} \int_{\mathcal{I}} \dd x \left( \sum_{p=1}^{n}\, \sum_{\substack{J_1 \sqcup \cdots \sqcup J_p = \llbracket n \rrbracket\\ J_k \neq \emptyset}}\, \prod_{k=1}^{p} \Big\{ \Updelta_{J_k} \BS \left( x; g_{\text{s}} \right) \Big\} \right) \mathrm{e}^{\BS \left( x;g_{\text{s}} \right)}, 
\label{eq:Wparts}
\eea
\noindent
where the powers of $g_{\text{s}}$ are required to match the multi-resolvent perturbative-expansion \eqref{eq:Rn-multi2} out of the corresponding free-energy \eqref{eq:Zg0}; where we introduced the notation $\llbracket n \rrbracket$ to denote $\left\{ 1, \ldots, n \right\}$; and, for $J \subseteq \llbracket n \rrbracket$, we defined
\be
\Updelta_J \equiv \prod_{i\in J}\Updelta_{z_i}.
\ee
\noindent
Notice how the second line of \eqref{eq:Wparts} is just an immediate application of the Leibniz rule to multiple loop-insertion operators. For example, we have for $n=1,2$:
\bea
\label{eq:W1-int}
\widehat{W}_{1}^{(1)} \left(z_{1}\right) &=& \frac{g_{\text{s}}}{2\pi} \int_{\mathcal{I}} \dd x\, \Updelta_{z_1} \BS(x)\, \rme^{\BS(x)}, \\
\label{eq:W2-int}
\widehat{W}_{2}^{(1)} \left(z_{1},z_{2}\right) &=& \frac{g_{\text{s}}^2}{2\pi} \int_{\mathcal{I}} \dd x \left\{ \Updelta_{z_2} \Updelta_{z_1} \BS(x) + \Updelta_{z_1} \BS(x) \Updelta_{z_2} \BS(x) \right\} \rme^{\BS(x)} .
\eea
\noindent	
After having applyed the loop-insertion operator $n$ times, one can then use the saddle-point method as in \eqref{eq:steep-des}-\eqref{eq:F(1)saddle}, \textit{i.e.}, expanding around $x_{\star,1}$ in powers of $\sqrt{g_{\text{s}}}$ in the imaginary direction, which amounts to substituting $x=x_{\star,1}+\mathrm{i}\sqrt{g_{\text{s}}}\, \rho$ and integrating over $\rho \in \mathbb{R}$. This yields
\be
\widehat{W}_{n}^{(1)} \left( z_{1}, \ldots, z_{n} \right) \simeq \mathrm{i} \sqrt{\frac{g_{\text{s}}}{2\pi \BS_0'' (x_{\star,1})}}\, \mathrm{e}^{\frac{1}{g_{\text{s}}} \BS_0 (x_{\star,1})}\, \mathrm{e}^{\BS_1 (x_{\star,1})} \left( \prod_{j=1}^{n} \Delta_{z_j} \BS_0 (x_{\star,1}) + \sum_{h=1}^{+\infty} \widetilde{W}_{h+1}^{(1)}\, g_{\text{s}}^h \right),
\ee
\noindent
which now makes the full \eqref{eq:top_exp_inst} precise, including the appearance of free-energy instanton action and Stokes data (\textit{e.g.}, upon comparison with \eqref{eq:MM-F(1)}, \eqref{eq:saddle-point}, and \eqref{eq:F(1)saddle}). As an example, we write down formulae for the first two\footnote{The first loop \eqref{eq:W111} was also obtained in \cite{sss19}, albeit in a different fashion from herein.} loops around the one-instanton configuration of the one-point function, as functions of the saddle point $x_{\star}$,
\begin{align}
\label{eq:W111}
\left. \widetilde{W}_{1,1}^{(1)} \left(z_{1}\right) \right|_{x_{\star}} &= \Updelta_{{z}_{1}} \BS_{0} (x_{\star}) = - \frac{2\sqrt{x_{\star}}}{x_{\star}-z_1^{2}}, \\
\label{eq:W121}
\left. \widetilde{W}_{2,1}^{(1)} \left(z_{1}\right) \right|_{x_{\star}} &= \BS_{2} (x_{\star})\, \Updelta_{{z}_{1}} \BS_{0} (x_{\star}) + \Updelta_{{z}_{1}} \BS_{1} (x_{\star}) - \frac{\Updelta_{{z}_{1}} \BS_{0} (x_{\star})\, \BS_{1}' (x_{\star})^{2}}{2 \BS''_{0} (x_{\star})} - \frac{\Updelta_{{z}_{1}} \BS_{0} (x_{\star})\, \BS_{1}'' (x_{\star})}{2 \BS''_{0} (x_{\star})} - \nonumber \\
&
- \frac{\BS_{1}' (x_{\star})\, \Updelta_{{z}_{1}} \BS_{0}' (x_{\star})}{\BS''_{0} (x_{\star})} - \frac{\Updelta_{{z}_{1}} \BS_{0}'' (x_{\star})}{2 \BS''_{0} (x_{\star})} + \frac{\Updelta_{{z}_{1}} \BS_{0} (x_{\star})\, \BS_{0}''' (x_{\star})\, \BS_{1}' (x_{\star})}{2 \BS''_{0} (x_{\star})^{2}} + \nonumber \\
&
+ \frac{\Updelta_{{z}_{1}} \BS_{0} (x_{\star})\, \BS_{0}^{(4)} (x_{\star})}{8 \BS''_{0} (x_{\star})^{2}} + \frac{\BS_{0}''' (x_{\star})\, \Updelta_{{z}_{1}} \BS_{0}' (x_{\star})}{2 \BS''_{0} (x_{\star})^{2}} - \frac{5 \Updelta_{{z}_{1}} \BS_{0} (x_{\star})\, \BS_{0}''' (x_{\star})^{2}}{24 \BS''_{0} (x_{\star})^{3}}.
\end{align}
\noindent
The contribution we are interested in, namely the one associated to the first non-trivial saddle, is obtained by evaluating \eqref{eq:W111} and \eqref{eq:W121} at $x_{\star} = x_{\star,1} = 1/4$. We shall be more explicit in examples as we turn to the analysis of Weil--Petersson volumes below.

In fact, recalling subsections~\ref{subsec:WPvolumes} and~\ref{subsec:classicalTR}, it should be clear to the reader that if we understand the one-instanton sector of an $n$-point multi-resolvent correlator, as above, then we can immediately obtain the one-instanton contributions to the generating function of $n$-boundary Weil--Petersson volumes \eqref{eq:fullWP}---as both are related via the inverse Laplace transform \eqref{eq:antiLapl}. Such contributions must be included into the ``perturbative'' Weil--Petersson generating-function \eqref{eq:fullWP} akin to \eqref{eq:Wn-1PTS}, \textit{i.e.}, as a one-parameter transseries\footnote{One can of course equally write an analogous transseries for the $V_{g,n}$ generating function defined in \eqref{eq:fullWP2}. It is simply obtained by setting all the $b_i$ to zero in \eqref{eq:WPtransseries}.} of the sorts
\be
\label{eq:WPtransseries}
\mathcal{V}_n \left( b_1,\dots,b_n; \sigma \right) = \mathcal{V}^{(0)}_n \left(b_1,\dots,b_n\right) + \sum_{\ell=1}^{+\infty} \sigma^\ell\, \mathcal{V}^{(\ell)}_{n} \left(b_1,\dots,b_n\right) + \cdots.
\ee
\noindent
The perturbative sector $\mathcal{V}^{(0)}_n (\boldsymbol{b})$ is of course given by \eqref{eq:fullWP}. In parallel with the multi-resolvent correlators we focus on the one-instanton contribution which, via \eqref{eq:top_exp_inst}, is of the form 
\be
\label{eq:WPtransseries-1inst}
\mathcal{V}_{n}^{(1)} \left( b_{1}, \ldots, b_{n} \right) \simeq \sqrt{g_{\text{s}}}\, S_1\, \mathrm{e}^{-\frac{A}{g_{\text{s}}}}\, \CF_1^{(1)} \sum_{h=0}^{+\infty} \widetilde{{V}}_{h+1,n}^{(1)} \left( b_{1}, \ldots, b_{n} \right) g_{\text{s}}^{h}. 
\ee
\noindent
Herein, the coefficients are normalized as usual
\be
\widetilde{{V}}_{g,n}^{(1)} \left( b_{1}, \ldots, b_{n} \right) \equiv \frac{1}{\CF_1^{(1)}}\, {V}_{g,n}^{(1)} \left( b_{1}, \ldots, b_{n} \right).
\ee
\noindent
Further, as we move from multi-resolvent correlators to Weil--Petersson volumes \eqref{eq:fromVtoW} via inverse Laplace transform \eqref{eq:antiLapl}, it is important to note how the variable $z_i$ subject to the transform only appears within the loop-insertion operator; see, \textit{e.g.}, \eqref{eq:W111} or \eqref{eq:W121}. This implies that in practice there is only a handful of functions for which the inverse-Laplace transform needs to be computed. They are of the form $\Updelta_{z_1} \BS_{\chi}(x_{\star})$, and a few examples are:
\bea
\label{eq:antiLapl-ex-1}
\NCL^{-1}_{b_1} \left\{ \Updelta_{z_1} \BS_{0} (x_{\star}) \right\} &=& \frac{\sinh \left( \sqrt{x_{\star}}\, b_1 \right)}{b_1/2}, \\
\label{eq:antiLapl-ex-2}
\NCL^{-1}_{b_1} \left\{ \Updelta_{z_1} \BS_{1} (x_{\star}) \right\} &=& \frac{2}{x_{\star}}, \\
\label{eq:antiLapl-ex-3}
\NCL^{-1}_{b_1} \left\{ \Updelta_{z_1} \BS_{2} (x_{\star}) \right\} &=& - \frac{b_1^4}{96 \sqrt{x_{\star}}} + b_1^2 \left( - \frac{\pi^2}{6 \sqrt{x_{\star}}} - \frac{17}{24 x_{\star}^{3/2}} \right) - \frac{\pi^4}{2 \sqrt{x_{\star}}} - \frac{3\pi^2}{x_{\star}^{3/2}} - \frac{17}{4 x_{\star}^{5/2}}.
\eea
\noindent
For generic $n$-point correlators, terms of the form $\Updelta_{z_n} \cdots \Updelta_{z_1} \BS_{\chi}(x_{\star})$ will also appear. In this case, a few further examples of relevant inverse-Laplace transforms for $n=2$ are:
\bea
\label{eq:antiLapl-more-ex-1}
\NCL^{-1}_{b_1,b_2} \left\{ \Updelta_{z_2} \Updelta_{z_1} \BS_{0} (x_{\star}) \right\} &=& - \frac{2}{\sqrt{x_{\star}}}, \\
\label{eq:antiLapl-more-ex-2}
\NCL^{-1}_{b_1,b_2} \left\{ \Updelta_{z_2} \Updelta_{z_1} \BS_{1} (x_{\star}) \right\} &=& \frac{b_1^2}{x_{\star}} + \frac{b_2^2}{x_{\star}} + \frac{4\pi^2}{x_{\star}} + \frac{4}{x_{\star}^2}.
\eea
\noindent
With all these in hand, we are now ready to explicitly evaluate a few nonperturbative contributions to the Weil--Petersson generating function.

\paragraph{The $n=1$ Case:}

Let us first focus on the nonperturbative contributions to one-boundary Weil--Petersson volumes $V_{g,n} (b_1)$. For the ensuing analysis it will be convenient to introduce the notation
\begin{equation}
\CS(z) \equiv \frac{\sinh \left( z/2 \right)}{z/2}, \qquad \CC(z) \equiv \frac{\cosh \left( z/2 \right)}{2}.
\end{equation}
\noindent
The function $\CS(z)$ was introduced in \cite{op02}. It also appears in the context of tau-functions of the KP hierarchy, and in Fock space computations of charged fermions (see, \textit{e.g.}, \cite{op02}). In an analogous way, the function $\CC(z)$ arises from Fock space computations of uncharged fermions (see, \textit{e.g.}, \cite{gkl21}), which underlie tau-functions of the BKP hierarchy. With this notation, and focusing on the first nonperturbative saddle $x_{\star,1}$, \eqref{eq:fromVtoW} adapted to \eqref{eq:WPtransseries-1inst} immediately yields the first loop around the one-instanton contribution via \eqref{eq:W111} and \eqref{eq:antiLapl-ex-1} as
\begin{equation}
\label{eq:sinhL}
\widetilde{{V}}_{1,1}^{(1)} (b_{1}) = \CS (b_1) = 1 + \frac{b_1^2}{24} + \frac{b_1^4}{1920} + \cdots.
\end{equation}
\noindent
Using further results for the one-point correlator, say \eqref{eq:W121}, and the inverse-Laplace transforms in \eqref{eq:antiLapl-ex-1}-\eqref{eq:antiLapl-ex-2}-\eqref{eq:antiLapl-ex-3}, we can easily obtain more loops around this one-instanton configuration. For instance:
\bea
\label{eq:V121}
\widetilde{{V}}_{2,1}^{(1)} (b_1) &=& - \CS (b_1) \left( \frac{1}{2} b_1^2 + \frac{5 \pi^2}{6} + \frac{68}{3}\right) + 16\, \CC (b_1) + 8, \\
\label{eq:V131}
\widetilde{{V}}_{3,1}^{(1)} (b_{1}) &=& \CS (b_1) \left( \frac{1}{8} b_1^4 + \frac{100}{3} b_1^2 + \frac{17 \pi^2}{12} b_1^2 + \frac{241 \pi^4}{72} + \frac{818 \pi^2}{9} + \frac{12104}{9} \right) - \\
&&
- \CC (b_1) \left( \frac{32}{3} b_1^{2} + \frac{176 \pi^2}{3} + \frac{3248}{3} \right) - \frac{1}{48} b_1^4 - \frac{17}{3} b_1^2 - \frac{\pi^2}{3} b_1^2 - \pi^4 - \frac{92 \pi^2}{3} - \frac{1624}{3}.
\nonumber
\eea
\noindent
We could easily go beyond the third loop, but the size of the formulae grows rapidly. However, we can set $b_1 = 0$ to recover the one-instanton sector of the $V_{g,n}$ Weil--Petersson volumes. In this case we can display the first six loops in a reasonable amount of space,
\bea
\label{eq:Vnonpert1}
\widetilde{V}_{1,1}^{(1)} &=& 1, \\
\label{eq:Vnonpert2}
\widetilde{V}_{2,1}^{(1)} &=& - \frac{5 \pi^{2}}{6} - \frac{20}{3}, \\
\label{eq:Vnonpert3}
\widetilde{V}_{3,1}^{(1)} &=& \frac{169 \pi^4}{72} + \frac{278 \pi^2}{9} + \frac{2360}{9}, \\
\label{eq:Vnonpert4}
\widetilde{V}_{4,1}^{(1)} &=& - \frac{29 \pi^8}{48} - \frac{77473 \pi^6}{6480} - \frac{58759 \pi^4}{270} - \frac{65936 \pi^2}{27} - \frac{1502320}{81}, \\
\label{eq:Vnonpert5}
\widetilde{V}_{5,1}^{(1)} &=& \frac{10169 \pi^{10}}{1440} + \frac{3292273 \pi^8}{31104} + \frac{10011497 \pi^6}{4860} + \frac{2205541 \pi^4}{81} + \frac{68755240 \pi^2}{243} + \frac{459169760}{243}, \\
\widetilde{V}_{6,1}^{(1)} &=& - \frac{9292841 \pi^{14}}{1020600} - \frac{9114151 \pi^{12}}{93312} - \frac{8766258803 \pi^{10}}{6531840} - \frac{4166763887 \pi^8}{163296} - \frac{1356187733 \pi^6}{3645} - \nonumber \\
&&
\label{eq:Vnonpert6}
- \frac{3216498470 \pi^4}{729} - \frac{31032738800 \pi^2}{729}  - \frac{183168540800}{729}.
\eea

There are some straightforward observations from the above results. First, whereas Weil--Petersson volumes $V_{g,n} (b_1,\ldots,b_n)$ of surfaces with $n$ geodesic boundaries with lengths $\boldsymbol{b} = \left( b_1,\ldots,b_n \right) \in \BR_+^n$ belong to $\mathbb{Q} \left[\pi^2, b_1^2, \ldots, b_n^2 \right]$, with homogeneous degree $6g-6+2n$ in $\pi$ and $b_i$, their corresponding instanton contributions in, \textit{e.g.}, \eqref{eq:sinhL}, \eqref{eq:V121} or \eqref{eq:V131}, instead arise as \textit{power series} in the $b_i$---linear combinations of hyperbolic functions $\CS (z)$, $\CC (z)$ multiplying \textit{non-homogeneous} polynomials in $\pi$ and $b_i$ (albeit still even in each individual variable). At this moment we cannot provide a definite interpretation of the above quantities in terms of enumerative invariants (but see \cite{csv16, gm22} for discussions on the somewhat similar case of nonperturbative Gromov--Witten invariants). However, it will be shown in section~\ref{sec:resurgence-correlators} how these nonperturbative contributions receive a geometric meaning in terms of the large-genus asymptotics of Weil--Petersson volumes. Further, in \cite{abcdglw19} two similar situations were investigated perturbatively, for which it would also be interesting to perform a nonperturbative analysis. The first instance involves the Witten--Kontsevich case, \textit{i.e.}, the polynomials which are obtained from the $V_{g,n}$ by taking the large $b_i$ limit (equivalently, by removing the cohomological class $\rme^{2\pi^2\upkappa_1}$ from the integrand in \eqref{eq:VgnWP} in the moduli space of curves). The second instance involves the Masur--Veech volumes of the moduli space of quadratic differentials (see as well \cite{fm23}). Both cases (and in fact many other enumerative problems) are computed by the topological recursion and could therefore be tackled at nonperturbative level with the same methods which we use in the present paper.

\paragraph{The General $n$ Case:}

Having understood the case of one-boundary Weil--Petersson volumes, we may now proceed to tackle the general case. From the expression for the one-instanton sector of multi-resolvent correlators in terms of the one-instanton sector of the free energy \eqref{eq:Wparts}, we can always find a recursive expression in $n$, at \textit{fixed} $g$, for the $\widehat{W}_{g,n}^{(1)}$,
\begin{equation}
\widehat{W}_{g,n+1}^{(1)} \left( z_{1}, \ldots, z_{n}, z_{n+1} \right) = \Updelta_{z_{n+1}} \BS_{0} (x_{\star})\, \widehat{W}_{g,n}^{(1)} \left( z_{1}, \ldots, z_{n} \right) + \CR\widehat{W}_{g,n}^{(1)},
\end{equation}
\noindent
where $\CR\widehat{W}_{g,n}^{(1)}$ contains the contribution of all partitions where $z_{n+1}$ is not by itself (\textit{i.e}, where $J_k \neq \left\{z_{n+1}\right\}$), as well as contributions from higher order terms in the
expansion of $\Updelta_{z_{n+1}} \BS(x)$; \textit{i.e.}, where $x$ is again deformed in the imaginary direction as $x = x_{\star}+\rmi\sqrt{g_{\text{s}}}\, \rho$ so that
\begin{equation}
\frac{1}{g_{\text{s}}^2}\, \Updelta_z \BS(x) = \frac{1}{g_{\text{s}}}\, \Updelta_z \BS_{0} (x_{\star}) + \rmi\frac{\rho}{\sqrt{g_{\text{s}}}}\, \Updelta_z \BS'_{0} (x_{\star}) - \frac{\rho^2}{2}\, \Updelta_z \BS''_{0} (x_{\star}) + \Updelta_z \BS_{1} (x_{\star}) + \cdots.
\end{equation}
\noindent
Observe that the leading order in $g_{\text{s}}$ will always come from the first term on the right-hand side. Therefore, for $g=0$ we immediately get a straightforward generalization of \eqref{eq:W111} for the one-loop contribution around the one-instanton sector of a generic $n$-point function, 
\be
\label{eq:sinhLn-LAP}
\left. \widetilde{W}_{1,n}^{(1)} \left(z_{1}, \ldots, z_{n}\right) \right|_{x_{\star}} = \prod_{k=1}^{n} \left. \widetilde{W}_{1,1}^{(1)} \left(z_{k}\right) \right|_{x_{\star}};
\ee
\noindent
or, equivalently, via inverse Laplace transform \eqref{eq:fromVtoW}, a generalization of \eqref{eq:sinhL} for a generic $n$-boundary Weil--Petersson volume,
\begin{equation}
\label{eq:sinhLn}
\widetilde{{V}}_{1,n}^{(1)} (b_{1}, \ldots, b_n) = \prod_{k=1}^{n} \CS (b_k).
\end{equation}

Next, we address the two-loop contribution to the one-instanton sector of general $n$-point multi-resolvent correlation functions, eventually to be encoded in $\widetilde{{V}}_{2,n}^{(1)} (b_{1},\ldots,b_n)$. Let us show how we can exactly compute such a generic formula; in fact generalizing our previous formulae for the $n=0$ and $n=1$ cases, \eqref{eq:F_2} and \eqref{eq:W121}, respectively. For a generic saddle point $x_{\star}$, this two-loop contribution is given by
\begin{align}
\label{eq:2loopgeneraln}
\left. \widetilde{W}_{2,n}^{(1)} \left(z_{1}, \ldots, z_{n}\right) \right|_{x_{\star}} &= \widetilde{\CF}_{2}^{(1)} (x_{\star}) \prod_{j=1}^n \Updelta_{z_j} \BS_0 (x_{\star}) + \left( \frac{\BS_0''' (x_{\star})}{2 \BS_0'' (x_{\star})^{2}} - \frac{\BS_1' (x_{\star})}{\BS_0'' (x_{\star})} \right) \times \\
&
\hspace{-45pt}
\times \sum_{j=1}^n \Updelta_{z_j} \BS_0' (x_{\star}) \prod_{k\neq j}\Updelta_{z_k} \BS_0 (x_{\star}) + \sum_{j=1}^n \left( \Updelta_{z_j} \BS_1 (x_{\star}) - \frac{\Updelta_{z_j} \BS_0'' (x_{\star})}{2 \BS_0'' (x_{\star})} \right) \prod_{k\neq j} \Updelta_{z_k} \BS_0 (x_{\star}) - 
\nonumber \\
&
\hspace{-90pt}
- \frac{1}{\BS_0'' (x_{\star})^2} \sum_{1\leq i<j \leq n} \Updelta_{z_i} \BS_0' (x_{\star})\, \Updelta_{z_j} \BS_0' (x_{\star}) \prod_{k\notin\{i,j\}} \Updelta_{z_k} \BS_0 (x_{\star}) + \sum_{1\leq i<j \leq n} \Updelta_{z_i} \Updelta_{z_j} \BS_0 (x_{\star}) \prod_{k\notin\{i,j\}} \Updelta_{z_k} \BS_0 (x_{\star}),
\nonumber
\end{align}
\noindent
where $\widetilde{\CF}_{2}^{(1)}(x_1)$ was given in~\eqref{eq:F_2}. The expression above assumes that empty sums equal zero and empty products equal one---for instance the second term of the above right-hand side is linear in the factors $\Updelta_z \BS$ for $n=1$ and quadratic for $n = 2$, whilst the fourth term is quadratic for $n=2$ and cubic for $n = 3$. Let us discuss in some detail how \eqref{eq:2loopgeneraln} was obtained.

One starts by returning to the formula for $\widehat{W}^{(1)}_n\left(z_1,\dots,z_n\right)$ given in \eqref{eq:Wparts}, and expands the integral around some chosen saddle-point $x = x_{\star}+\rmi\sqrt{g_{\text{s}}}\, \rho$ as usual (one may compare to \eqref{eq:steep-des} or \eqref{eq:F(1)saddle}),
\begin{align}
\label{eq:integral_correlators}
\frac{1}{2\pi} \int_{\mathcal{I}} \rmd x\, \Updelta_{\llbracket n \rrbracket} \psi(x) &= \frac{\mathrm{i}\sqrt{g_{\text{s}}}}{2\pi} \int_{\mathcal{I}} \dd\rho\, \Updelta_{\llbracket n \rrbracket} \psi (x_{\star}+\rmi \sqrt{g_{\text{s}}}\, \rho) = \\
&
\hspace{-20pt}
= \frac{\mathrm{i}\sqrt{g_{\text{s}}}}{2\pi} \int_{-\infty}^{+\infty} \rmd\rho\, \Updelta_{\llbracket n \rrbracket} \left( \rme^{\frac{1}{g_{\text{s}}} \BS_0 (x_{\star}) + \frac{\rmi}{\sqrt{g_{\text{s}}}} \BS_0' (x_{\star})\, \rho - \frac{1}{2} \BS_0'' (x_{\star})\, \rho^2 + \BS_1 (x_{\star})} \sum_{r \in \frac{1}{2} \mathbb{Z}_{\geq 0}} \BP_r (\rho)\, g_{\text{s}}^r \right), \nonumber
\end{align}
\noindent
where we introduced a family of polynomials in $\rho$ simply serving the purpose of keeping track of the powers of $g_{\text{s}}$ appearing in the above integrand,
\be
\label{eq:p-polynomials}
\BP_r (\rho) \equiv \sum_{\ell=1}^{2r} \frac{1}{\ell!}\, \sum_{\substack{\chi_j,k_j \in \mathbb{Z}_{\geq 0}\, : \\ \sum_{j=1}^{\ell} \left( \chi_j + \frac{1}{2} k_{j} \right) = \ell + r}}' \prod_{j=1}^{\ell} \frac{\BS_{\chi_j}^{(k_j)} (x_{\star})}{k_j!} \left( \rmi \rho \right)^{\sum_{j=1}^{\ell} k_j}.
\ee
\noindent
The primed sum above means that we are excluding the cases $(\chi_j,k_j) = (0,0), (0,1), (0,2), (1,0)$. In what follows it is also convenient to denote the exponent in the exponential pre-factor of the integrand in \eqref{eq:integral_correlators} as the single combination
\be
\widetilde{\BS}_{\star} (\rho,g_{\text{s}}) \equiv \frac{1}{g_{\text{s}}} \BS_0 (x_{\star}) + \frac{\rmi}{\sqrt{g_{\text{s}}}} \BS_0' (x_{\star})\, \rho - \frac{1}{2} \BS_0'' (x_{\star})\, \rho^2 + \BS_1 (x_{\star}).
\ee
\noindent
Now, in order to extract the two-loop contribution that we are looking for \eqref{eq:2loopgeneraln}, we need to isolate the order $g_{\text{s}}^{1-n}$ term appearing in the integrand of \eqref{eq:integral_correlators} which we denote $N_{\star} (\rho)$, \textit{i.e.},
\be
\Updelta_{\llbracket n \rrbracket} \left( \rme^{\widetilde{\BS}_{\star} (\rho,g_{\text{s}})} \sum_{r \in \frac{1}{2} \mathbb{Z}_{\geq 0}} \BP_r (\rho)\, g_{\text{s}}^r \right) = \rme^{\widetilde{\BS}_{\star} (\rho,g_{\text{s}})} \left( g_{\text{s}}^{-n}\, \prod_{j=1}^n \Delta_{z_j} \BS_0 (x_{\star}) + g_{\text{s}}^{1-n}\, N_{\star} (\rho) + \cdots \right),
\ee
\noindent
where the one-loop contribution of order $g_{\text{s}}^{-n}$ has already been addressed. Explicit computation yields
\begin{align}
N_{\star} (\rho) &= \BP_1 (\rho)\, \prod_{j=1}^n \Updelta_{z_j} \BS_0 (x_{\star}) + \mathrm{i} \rho\, \BP_{1/2} (\rho)\, \sum_{j=1}^n \Updelta_{z_j} \BS_0' (x_{\star}) \prod_{k\neq j} \Updelta_{z_k} \BS_0 (x_{\star}) + \nonumber \\
&
+ \BP_0 (\rho)\, \sum_{j=1}^n \left( \Updelta_{z_j} \BS_1 (x_{\star}) - \frac{1}{2} \rho^2\, \Updelta_{z_j} \BS_0'' (x_{\star}) \right) \prod_{k\neq j} \Updelta_{z_k} \BS_0 (x_{\star}) - \nonumber \\
&
- \rho^2\, \BP_0 (\rho) \sum_{1 \leq i < j \leq n} \Updelta_{z_i} \BS_0' (x_{\star})\, \Updelta_{z_j} \BS_0' (x_{\star}) \prod_{k \notin \{i,j\}} \Updelta_{z_k} \BS_0 (x_{\star}) + \nonumber \\
&
+ \BP_0 (\rho) \sum_{1 \leq j < k \leq n} \Updelta_{z_j} \Updelta_{z_k} \BS_0 (x) \prod_{i \notin \{j,k\}} \Updelta_{z_i} \BS_0 (x_{\star}).
\label{eq:intermediate-step}
\end{align}
\noindent
Herein, $\BP_0 (\rho) = 1$, and:
\begin{align}
\BP_{1/2} (\rho) &= \BS_1' (x_{\star})\, \mathrm{i} \rho + \frac{1}{3!} \BS_0''' (x_{\star}) \left( \mathrm{i}\rho \right)^3, \\
\BP_1 (\rho) &= \BS_2 (x_{\star}) + \left( \frac{1}{2!} \BS_1'' (x_{\star}) + \frac{1}{2!} \left( \BS_1' (x_{\star}) \right)^2 \right) \left( \mathrm{i}\rho \right)^2 + \left( \frac{1}{4!} \BS_0^{(4)} (x_{\star}) + \frac{1}{3!} \BS_1' (x_{\star})\, \BS_0''' (x_{\star}) \right) \rho^4 + \nonumber \\
&
+ \frac{1}{2} \left( \frac{\BS_0'''(x_{\star})}{3!} \right)^2 \left( \mathrm{i}\rho \right)^6.
\end{align}
\noindent
Putting all the above ensemble into \eqref{eq:Wparts}---to be evaluated via \eqref{eq:integral_correlators}---and focusing on its two-loop coefficient, one obtains
\bea
\label{eq:W12n-almost-1}
\left. \widetilde{W}_{2,n}^{(1)} \left(z_{1}, \ldots, z_{n}\right) \right|_{x_{\star}} &=& \sqrt{\frac{\BS''_0 (x_{\star})}{2\pi}}\, \mathrm{e}^{\frac{A_{\star}}{g_{\text{s}}} - \BS_1 (x_{\star})} \int_{-\infty}^{+\infty} \dd\rho\, \rme^{\widetilde{\BS}_{\star} (\rho,g_{\text{s}})}\, N_{\star} (\rho) = \\
&=& \sqrt{\frac{\BS''_0 (x_{\star})}{2\pi}}\, \int_{-\infty}^{+\infty} \dd\rho\, \rme^{- \frac{1}{2} \BS_0'' (x_{\star})\, \rho^2}\, N_{\star} (\rho).
\label{eq:W12n-almost-2}
\eea
\noindent
Herein we have included a pre-factor in \eqref{eq:W12n-almost-1} to remove the terms $\mathrm{e}^{-A/g_{\text{s}}}$ and $S_1 \cdot \CF^{(1)}_1$, so that we focus uniquely on the two-loop coefficient we are interested in. Further recall that $A_{\star} = - \BS_0 (x_{\star})$ and that $\BS_0' (x_{\star}) = 0$. Evaluating the integrals we have\footnote{Simply use the well-known formula for Gaussian moment integrals,
\be
\int_{-\infty}^{+\infty} \rmd\rho\, \rho^{m}\, \rme^{- \frac{1}{2} \BS_0'' (x_{\star})\, \rho^2} = \begin{cases} 0, & \text{ if } m \text{ is odd,} \\
\sqrt{\frac{2\pi}{\BS_0'' (x_{\star})}}\, \frac{\left(m-1\right)!!}{\left( \BS_0'' (x_{\star}) \right)^{m/2}}, & \text{ if } m \text{ is even.}
\end{cases}
\ee}
\begin{align}
\int_{-\infty}^{+\infty} \dd\rho\, \mathrm{e}^{- \frac{1}{2} \BS_0'' (x_{\star})\, \rho^2}\, \BP_1 (\rho) &= \sqrt{\frac{2\pi}{\BS_0'' (x_{\star})}}\, \widetilde{\CF}_2^{(1)} (x_{\star}), \\ 
\int_{-\infty}^{+\infty} \dd\rho\, \rmi\rho\, \mathrm{e}^{- \frac{1}{2} \BS_0'' (x_{\star})\,  \rho^2}\, \BP_{1/2} (\rho) &= \sqrt{\frac{2\pi}{\BS_0'' (x_{\star})}} \left( \frac{\BS_0''' (x_{\star})}{2 \BS_0'' (x_{\star})^{2}} - \frac{\BS_1' (x_{\star})}{\BS_0'' (x_{\star})} \right).
\end{align}
\noindent
These, combined with \eqref{eq:W12n-almost-2} and \eqref{eq:intermediate-step}, finally yields the desired result \eqref{eq:2loopgeneraln}. Observe that it is important to \textit{first} apply the loop insertion operator and \textit{then} perform the saddle-point integration, because $\BS_0' (x_{\star}) = 0$ but $\Updelta_{z} \BS_0' (x_{\star}) \neq 0$---hence the order of these two steps matters.

It is easy to check that for $n=0,1$ equation \eqref{eq:2loopgeneraln} reduces to the results for the free energy \eqref{eq:F_2} and for the one-point correlators \eqref{eq:W121} which were previously obtained. For example, for $n=2$ \eqref{eq:2loopgeneraln} becomes
\begin{align}
\label{eq:W221xstar}
\left. \widetilde{W}_{2,2}^{(1)} \left(z_{1}, z_{2}\right) \right|_{x_{\star}} &= \widetilde{\CF}_{2}^{(1)} (x_{\star})\, \Updelta_{z_1} \BS_0 (x_{\star})\, \Updelta_{z_2} \BS_0 (x_{\star}) + \left( \frac{\BS_0''' (x_{\star})}{2 \BS_0'' (x_{\star})^{2}} - \frac{\BS_1' (x_{\star})}{\BS_0'' (x_{\star})} \right) \times \\
&
\hspace{-60pt}
\times \Big( \Updelta_{z_1} \BS_0' (x_{\star})\, \Updelta_{z_2} \BS_0 (x_{\star}) + \Updelta_{z_2} \BS_0' (x_{\star}) \Updelta_{z_1} \BS_0 (x_{\star}) \Big) + \left( \Updelta_{z_1} \BS_1 (x_{\star}) - \frac{\Updelta_{z_1} \BS_0'' (x_{\star})}{2 \BS_0'' (x_{\star})} \right) \Updelta_{z_2} \BS_0 (x_{\star}) + \nonumber \\
&
\hspace{-60pt}
+
\left( \Updelta_{z_2} \BS_1 (x_{\star}) - \frac{\Updelta_{z_2} \BS_0'' (x_{\star})}{2 \BS_0'' (x_{\star})} \right) \Updelta_{z_1} \BS_0 (x_{\star})
- \frac{1}{\BS_0'' (x_{\star})^2}\, \Updelta_{z_1} \BS_0' (x_{\star})\, \Updelta_{z_2} \BS_0' (x_{\star})
+ \Updelta_{z_1} \Updelta_{z_2} \BS_0 (x_{\star}). \nonumber
\end{align}
\noindent
For the first non-trivial saddle-point, $x_{\star,1} = 1/4$, this expression further simplifies,
\begin{align}
\left. \widetilde{W}_{2,2}^{(1)} \left(z_{1}, z_{2}\right) \right|_{1/4} &= \widetilde{\CF}_{2}^{(1)} (1/4)\, \Updelta_{z_1} \BS_0 (1/4)\, \Updelta_{z_2} \BS_0 (1/4) + \\
&+
\left( 3 \Updelta_{z_1} \BS_0' (1/4) + \Updelta_{z_1} \BS_1 (1/4) - \frac{1}{2} \Updelta_{z_1} \BS_0'' (1/4) \right) \Updelta_{z_2} \BS_0 (1/4) + \nonumber \\
&+
\left( 3 \Updelta_{z_2} \BS_0' (1/4) + \Updelta_{z_2} \BS_1 (1/4) - \frac{1}{2} \Updelta_{z_2} \BS_0'' (1/4) \right) \Updelta_{z_1} \BS_0 (1/4) - \nonumber \\
&-
\Updelta_{z_1} \BS_0' (1/4)\, \Updelta_{z_2} \BS_0' (1/4) + \Updelta_{z_1} \Updelta_{z_2} \BS_0 (1/4), \nonumber
\end{align}
\noindent
where we used explicit values, \textit{e.g.}, $\BS_0 (1/4) = - \left( 4\pi^2 \right)^{-1}$, $\BS'_0 (1/4) = 0$, $\BS_0'' (1/4) = 1$, $\BS_0''' (1/4) = -2$, $\BS_1 (1/4) = \log \rmi$, $\BS_1' (1/4) = -4$, and where one can further use explicit expressions such as $\widetilde{\CF}_{2}^{(1)}(1/4) = - \frac{68}{3} - \frac{5\pi^{2}}{6}$ or $\Updelta_{z} \BS_{0} (1/4) = \frac{4}{4z^{2}-1}$.

Moving towards the more user-friendly realm of Weil--Petersson volumes, one considers the inverse-Laplace transform of \eqref{eq:2loopgeneraln} via \eqref{eq:fromVtoW} to obtain them as
\be
\widetilde{V}_{2,n}^{(1)} \left( b_1, \ldots, b_n \right) = \left( \prod_{i=1}^{n} \NCL^{-1}_{b_i} \right) \cdot \widetilde{W}_{2,n}^{(1)} \left( z_1, \ldots, z_n \right),
\ee
\noindent
and then setting $x_{\star,1}=1/4$. In this way, we immediately find the two-loop contribution to the one-instanton sector of the Weil--Petersson-volumes generating-function, as the rather neat formula (compare with \eqref{eq:V121} for the $n=1$ case)
\begin{align}
\label{eq:WP2loop}
\widetilde{V}_{2,n}^{(1)} \left( b_{1}, \ldots, b_{n} \right) &= - \left( \frac{5\pi^{2}}{6} + \frac{68}{3} \right) \prod_{j=1}^n \CS (b_i) - 4 \sum_{1 \leq i < j \leq n} \Big( 4\, \CC (b_i)\, \CC (b_j) + 1 \Big) \prod_{k \notin \{i,j\}} \CS (b_k) + \nonumber \\
&
+ \sum_{i=1}^n \left( - \frac{1}{2} b_i^2\, \CS (b_i) + 16\, \CC (b_i) + 8 \right) \prod_{k \neq i} \CS (b_k).
\end{align}
\noindent
Finally, we can evaluate \eqref{eq:WP2loop} at $\boldsymbol{b} = \boldsymbol{0}$ to obtain nonperturbative corrections to the Weil--Petersson volumes with marked points, herein $\widetilde{V}_{2,n}^{(1)}$ which is (compare with \eqref{eq:Vnonpert2} for the $n=1$ case):
\begin{equation}
\label{eq:V12n}
\widetilde{V}_{2,n}^{(1)} = - \frac{5 \pi^2}{6} - 4 n^2 + 20 n - \frac{68}{3}.
\end{equation}

Our method allows for the recursive computation of an arbitrarily high number of loops around the one-instanton configuration, but given the size of such formulae we refrain from writing expressions for $\widetilde{W}^{(1)}_{g,n}$ or $\widetilde{V}^{(1)}_{g,n}$ with $g>2$. Instead, as usual, we can obtain more manageable formulae by setting the $b_i$ to zero, obtaining for example the third-loop contribution around the one-instanton sector of the $n$-point Weil--Petersson volumes as (compare with \eqref{eq:Vnonpert3} for the $n=1$ case) 
\begin{align}
\label{eq:V13n}
\widetilde{V}_{3,n}^{(1)} &= \frac{241\pi^{4}}{72} - n \pi^{4} + \frac{818 \pi^2}{9} + \frac{2}{9} \left( - 6 n^3 + 87 n^2 - 351 n \right) \pi^2 + \frac{12104}{9} + \\
&+ \frac{8}{9} \left( 9 n^4 - 144 n^3 + 813 n^2 - 1896 n \right).
\nonumber
\end{align}
\noindent
All these nonperturbative results can be tested numerically through large-genus resurgent asymptotic relations, as it will be shown in some examples in the next sections.

\section{Resurgent Asymptotics of Multi-Resolvent Correlation Functions}
\label{sec:resurgence-correlators}

Up to this stage we focused on nonperturbative effects arising from eigenvalue tunneling \cite{d91, d92, msw07, msw08}; equivalently, ZZ-brane contributions \cite{zz01, akk03, st04, sst23}. Albeit there are none others when it comes to describing the complete nonperturbative content of the matrix-model partition-function (at least if all such effects are considered, \textit{i.e.}, if we further include anti-eigenvalues \cite{mss22} or negative-tension ZZ-branes \cite{sst23}) the same does not hold true\footnote{Albeit there are still large classes of operators in generic two-dimensional quantum gravity whose correlation functions \textit{only} receive ZZ-brane nonperturbative corrections. These are discussed in appendix~\ref{app:corr_func}.} when it comes to multi-resolvent correlation functions. In fact, for these there will now be FZZT nonperturbative contributions \cite{fzz00, t00, sss19} on top of the aforementioned ZZ contributions. Studying the complete, general transseries of these correlation functions, with both ZZ and FZZT transmonomials, alongside their resonant negative-brane counterparts, is out of scope in this work. However, and as we shall see, already at \textit{leading} order in the seemingly ZZ-dominated asymptotics it is not possible to ignore FZZT effects---which implies at the very least we must address the leading FZZT contributions to generic multi-resolvent correlation functions. This is what we shall do in the present section. Naturally, and as already discussed in the ZZ case, all these contributions will have analogues at the level of Weil--Petersson volumes, which we discuss in the next section~\ref{sec:resurgence-volumes}.

Having studied the ZZ world in some detail, let us enter the FZZT realm. Whereas minimal-string ZZ-branes consist of a discrete family labeled by two integers as $\ket{n,m}_{\text{ZZ}}$ and correspond to eigenvalue tunneling in the matrix model description, FZZT-branes, on the other hand, have minimal-string target-space\footnote{Hence identifying the minimal-string target-space with the FZZT D-brane moduli space \cite{mmss04}.} dependence via the uniformization variable $z \in \BC$ as $\ket{z}_{\text{FZZT}}$ and---at disk level---correspond to the holomorphic effective potential in the matrix model description (one recent discussion in the present context may be found in \cite{gs21}). Rather importantly, the ZZ-branes are known to be given by adequate differences of FZZT-branes \cite{zz01, m03},
\be
\label{eq:ZZfromFZZT}
\ket{m,n}_{\text{ZZ}} = \ket{z(m,-n)}_{\text{FZZT}} - \ket{z(m,n)}_{\text{FZZT}}
\ee
\noindent
(where the $m$ and $n$ label the nonperturbative saddle-points $x^{\star}_{m,n}$ in more generic cases than those of JT gravity in figure~\ref{fig:potential}, and $z(m,\pm n)$ their corresponding locations on both sheets of the spectral geometry). A more direct matrix-model way to characterize the (double-scaled) FZZT-brane is via a determinant\footnote{This is somewhat in the spirit of our shifted partition function, discussed in \eqref{eq:fff} and \eqref{eq:shiftedZ}, and in appendix~\ref{app:shift}. However, it should also be noted that while $\Psi_E$ can be constructed via the topological recursion in a similar way to the wave function $\psi (x)$ we earlier defined in \eqref{eq:psi-section4}, the two objects should not be confused. In particular, $\psi(x)$ \textit{does not} compute the observable \eqref{eq:FZZT-obs} associated to the insertion of an FZZT brane. Introducing orthogonal polynomials with respect to the measure induced by the matrix integral \eqref{eq:ZN}, $\left\{ p_{n} (E) \right\}$, then off-criticality the ``pure'' determinant correlator precisely corresponds to $p_{N} (E) = \ev{\det \left( E-M \right)}_{N}$.} insertion \cite{m93, mmss04, dw18, os21},
\be
\label{eq:FZZT-obs}
\Psi_E = \ev{\det \left( E-M \right) \rme^{- \frac{1}{2g_{\text{s}}} V(E)}}, \qquad \overline{\Psi}_E = \ev{\frac{1}{\det \left( E-M \right)}\, \rme^{\frac{1}{2g_{\text{s}}} V(E)}}.
\ee
\noindent
We switched back to the $E$-eigenvalue notation of \eqref{eq:spec_dens} for clarity, $\Psi$ is the FZZT wave-function (sometimes denoted Baker--Akhiezer function), and $\overline{\Psi}$ its FZZT negative-brane counterpart.  

There are two standard ways to explicitly compute these FZZT contributions \eqref{eq:FZZT-obs}. One is to consider the double-scaled version of the matrix-model orthogonal-polynomial defining recursion-relation; see, \textit{e.g.}, \cite{m90}. One then finds that the FZZT wave-functions satisfy the Schr\"odinger equation
\be
\label{eq:FZZT-schro-eq}
\left( - \frac{\rmd^2}{\rmd \kappa^2} + u(\kappa) \right) \Psi_E (\kappa) = E\, \Psi_E (\kappa),
\ee
\noindent
where $\kappa$ is the double-scaled variable incorporating both 't~Hooft and string couplings (which we omitted in \eqref{eq:FZZT-obs}) and the potential $u(\kappa)$ is the solution to the JT string equation. Of course not knowing $u(\kappa)$ is an obvious obstacle, but actually not a very serious one at the leading order we are interested in. Indeed \cite{os19}, consider the resolvent\footnote{We follow in parallel with the discussion in section~3.1 of \cite{gs21}, to where we refer the reader for further details.} associated to the above Hamiltonian in \eqref{eq:FZZT-schro-eq}, and let $R_{E} (\kappa)$ denote its diagonal integral kernel. Then making use of the Gel'fand--Dikii equation for the $R_{E} (\kappa)$ \cite{gd75} it can be shown \cite{os19} that
\be
\Psi_{\pm} (\kappa) = \sqrt{R_{E} (\kappa)}\, \exp \left( \pm \frac{1}{2} \int^\kappa \frac{\rmd \kappa'}{R_{E} (\kappa')} \right)
\ee
\noindent
are WKB-like solutions to \eqref{eq:FZZT-schro-eq}. The argument of the exponent may be computed in the semiclassical limit (making use of the genus-$0$ string equation---see \cite{gs21}) to obtain\footnote{Herein $u_0$ is the genus-$0$ variable of \cite{biz80, iz92}, and the $t_k$ are the KdV times (see \cite{gs21} for further details).}
\be
\frac{1}{2} \int^\kappa \frac{\rmd \kappa'}{R_{E} (\kappa')} = \int^{u_0} \rmd u\, \sum_{k=1}^{+\infty} k\, t_k\, u^{k-1} \sqrt{E+u} + \cdots \simeq \frac{1}{2} V_{\text{h;eff}} (E) + \cdots.
\ee

Another approach to explicitly computing FZZT contributions is via the loop equations for the determinant correlators in \eqref{eq:FZZT-obs}. These were computed in \cite{dw18} directly in the off-critical matrix model where, to leading order and at large $N$, one finds
\be
\left( - g_{\text{s}}^2\, \frac{\partial^2}{\partial E^2} + \frac{1}{4} y^2 (E) + \cdots \right) \Psi_E = 0
\ee
\noindent
again resulting in the WKB-like expression
\be
\Psi_E \simeq \exp \left( \pm \frac{1}{2g_{\text{s}}} V_{\text{h;eff}} (E) \right) + \cdots.
\ee
\noindent
Both these approaches show how, at semiclassical level, the FZZT disk amplitude corresponds to the (non-single-valued) matrix-model holomorphic effective potential
\be
\label{eq:PhiVeff}
\text{Disk} (E) = \frac{1}{2} \int^E \rmd x'\, y(x') = \frac{1}{2} V_{\text{h;eff}} (E),
\ee
\noindent
earlier introduced in \eqref{eq:hol-eff-pot}. At next-to-leading order in the WKB expansion, the FZZT wave-function is \cite{mmss04, sss19, os19, os21}
\be
\label{eq:FZZT-WKB-sol}
\Psi_E \simeq \frac{1}{\sqrt{\partial_z x(z)}}\, \rme^{\pm \frac{1}{2 g_{\text{s}}} \int^{x(z)} \rmd x'\, y (x')} + \cdots,
\ee
\noindent
where we have reverted back to the uniformization variable $z$ to be used in the following.

\subsection{Large Genus Asymptotics of the One-Point Function: ZZ and FZZT Branes}
\label{subsec:onepointW}

Let us readdress the nonperturbative content of multi-resolvent correlators, whose (partial) ZZ-transseries already figured in \eqref{eq:Wn-1PTS} and with one-instanton sector depicted in \eqref{eq:top_exp_inst}. Focusing on the one-point function $\widehat{W}_1(z)$ for starters, we would correspondingly write
\be
\label{eq:W1-1inst-1PTS}
\widehat{W}_{1} \left( z; \sigma \right) = \widehat{W}_{1}^{(0)} ( z ) + \sigma\, \sqrt{g_{\text{s}}}\, S_1\, \mathrm{e}^{-\frac{A}{g_{\text{s}}}}\, \CF_1^{(1)} \sum_{h=0}^{+\infty} \widetilde{W}_{h+1,1}^{(1)} ( z )\, g_{\text{s}}^{h} + \cdots.
\ee
\noindent
What our analysis will highlight in a clear way is that, already at leading order in the asymptotics, the above alone is incomplete. Instead, the asymptotics distinctively sees the presence of two kinds of well-known nonperturbative effects: ZZ and FZZT branes. These two competing effects appear to dominate one over the other, depending on the value of $z$. As such, and even though our main focus throughout the paper so far was in the computation of nonperturbative data associated to ZZ branes, the complete transseries describing the one-point function must further include FZZT brane contributions from scratch. This implies that schematically, and up to one-instanton contributions, the transseries we must consider is of the form
\be
\label{eq:W1=ZZ+FZZT}
\widehat{W}_1 \left(z; \textcolor{blue}{\sigma_{\text{ZZ}}}, \textcolor{red}{\sigma_{\text{FZZT}}} \right) = \widehat{W}_1^{(0)} ( z ) + \textcolor{blue}{\sigma_{\text{ZZ}}\, \widehat{W}_1^{(\text{ZZ})} ( z )} + \textcolor{red}{\sigma_{\text{FZZT}}\, \widehat{W}_1^{(\text{FZZT})} ( z )} + \cdots.
\ee
\noindent
The ZZ contribution corresponds to the one-instanton sector of the one-point function already written in \eqref{eq:W1-1inst-1PTS} above. As for the FZZT contribution---which was studied in \cite{sss19} in the context of the spectral density---to the one-point function $\widehat{W}_1(z)$, it takes the form\footnote{Rather schematically, this FZZT contribution is $\widehat{W}_1^{\text{FZZT}} \sim \Psi_E \Psi_E$, quadratic in the FZZT wave-function \eqref{eq:FZZT-WKB-sol} and its derivatives \cite{sss19} (as one observes by comparing leading orders of \eqref{eq:FZZT-WKB-sol} and \eqref{eq:W1-fzzt}). It was computed up to two-loops in \cite{os19}, and we used their very same method to extend such calculation to the present three-loop result.} \cite{sss19, os19}
\be
\label{eq:W1-fzzt}
\widehat{W}_1^{\text{FZZT}} (z) \simeq - \frac{1}{2z}\, \mathrm{e}^{-\frac{V_{\text{eff}}(z^2)}{g_{\text{s}}}} \left\{ 1 - \frac{17 + 2 \pi^2 z^2}{12 z^3}\, g_{\text{s}} + \frac{1225 + 644 \pi^2 z^2 + 148 \pi^4 z^4}{288 z^6}\, g_{\text{s}}^2 + \cdots \right\}.
\ee
\noindent
Via \eqref{eq:PhiVeff}, the ``instanton'' action associated to the FZZT contribution is the effective potential \eqref{eq:Veff}. As such, it depends on $z$, in contrast with the constant, ``pure'' instanton action \eqref{eq:instactMM} of the ZZ-brane one-instanton sector. In this way, while the Borel singularity associated to the ZZ brane remains fixed at $A_{\text{ZZ}}=1/4\pi^2$, the FZZT Borel-singularity moves around as $z$ varies. Let us understand what this implies at the level of large-order asymptotics.

Start with the perturbative, asymptotic genus expansion of our one-point function $\widehat{W}_{1}^{(0)} \left( z \right)$ (albeit what follows is completely generic),
\be
\label{eq:W1-pert-exp}
\widehat{W}^{(0)}_1 ( z ) \simeq \sum_{g= 0}^{+\infty} \widehat{W}_{g,1}^{(0)} ( z )\, g_{\text{s}}^{2g+\beta^{(0)}}
\ee
\noindent
as follows from \eqref{eq:Rn-multi2} or \eqref{eq:Rn-multi2-revisited}, where the perturbative coefficients grow factorially fast $\widehat{W}_{g,1}^{(0)} \sim \left( 2g \right)!$ and where we left the characteristic exponent $\beta^{(0)} = -1$ unspecified for the sake of (upcoming) generality. Its transseries completion of course follows from \eqref{eq:Wn-1PTS}-\eqref{eq:top_exp_inst} as
\be
\widehat{W}_{1} \left( z; \sigma \right) = \widehat{W}_{1}^{(0)} ( z ) + \sum_{\ell=1}^{+\infty} \sigma^\ell\, \mathrm{e}^{-\frac{\ell A}{g_{\text{s}}}} \sum_{h=0}^{+\infty} \widehat{W}_{h+1,1}^{(\ell)} ( z )\, g_{\text{s}}^{h + \beta^{(\ell)}} + \cdots,
\ee
\noindent
but where we are now ambiguous whether the transmonomials have ZZ or FZZT origin (hence this more generic rewrite as compared to \eqref{eq:W1-1inst-1PTS}). In fact if there are \textit{both} such contributions, we must then be in the presence of a \textit{two}-parameter\footnote{Two parameters at the very \textit{least}; \textit{i.e.}, in string theoretic problems transseries are \textit{resonant}, doubling the number of ``na\"\i ve'' parameters \cite{abs18, mss22, sst23} (more below).} transseries which could be of the form (see, \textit{e.g.}, \cite{abs18})
\be
\label{eq:W1-2PT-detailed}
\widehat{W}_{1} \left( z; \textcolor{blue}{\sigma_{\text{ZZ}}}, \textcolor{red}{\sigma_{\text{FZZT}}} \right) = \widehat{W}_{1}^{(0)} ( z ) + \sum_{\textcolor{blue}{n}=1}^{+\infty} \sum_{\textcolor{red}{m}=1}^{+\infty} \textcolor{blue}{\sigma_{\text{ZZ}}^n}\,  \textcolor{red}{\sigma_{\text{FZZT}}^m}\, \mathrm{e}^{-\frac{\textcolor{blue}{n A_{\text{ZZ}}} + \textcolor{red}{m A_{\text{FZZT}}}}{g_{\text{s}}}} \sum_{h=0}^{+\infty} \widehat{W}_{h+1,1}^{(\textcolor{blue}{n},\textcolor{red}{m})} ( z )\, g_{\text{s}}^{h + \beta^{(\textcolor{blue}{n},\textcolor{red}{m})}} + \cdots.
\ee
\noindent
This makes \eqref{eq:W1=ZZ+FZZT} more precise, but in fact the actual case of JT gravity is even more complicated due to the presence of an infinite number of distinct ZZ-brane contributions associated to the infinite number of nonperturbative saddle-points of the JT effective potential in \eqref{eq:Veff} and depicted in figure~\ref{fig:potential}, hence requiring an infinite number of transseries parameters (see more on these transseries in \cite{gs21, sst23}). On top, as already mentioned, resonance is ubiquitous in two-dimensional gravitational models belonging to the KdV hierarchy \cite{gs21} and effectively doubles the number of transseries parameters by inclusion of all negative-brane counterparts in the above expression. Focusing solely on the one-instanton contributions\footnote{Recall how resurgent large-order asymptotics are dominated by the closest singularity to the origin on the Borel plane, whilst all other singularities lead to exponentially-suppressed large-order contributions \cite{abs18}. As such, the leading large-genus asymptotics will be determined by the one-instanton sectors of the transseries which is the regime we herein focus upon. For an in-depth mathematical treatment of the relationship between Borel singularities, instanton actions and large-genus asymptotics, see also \cite{egggl23}.}, $\textcolor{blue}{n}=1$ and $\textcolor{red}{m}=1$, as we do in the current work, all these considerations simplify; singularities on the Borel plane of the perturbative sector \eqref{eq:W1-pert-exp} are essentially described by the ZZ and FZZT ``instanton'' actions $\textcolor{blue}{A_{\text{ZZ}}}$ and $\textcolor{red}{A_{\text{FZZT}}}$; and the large-genus asymptotic behavior of its perturbative coefficients is dictated by (resonant) resurgence as usual \cite{abs18}
\bea
\widehat{W}_{g,1}^{(0)} &\simeq&
\frac{\textcolor{blue}{S_{\text{ZZ}}}}{\mathrm{i}\pi}\, \frac{\Gamma \left( 2g-\textcolor{blue}{\beta_{\text{ZZ}}} \right)}{\textcolor{blue}{A_{\text{ZZ}}^{2g-\beta_{\text{ZZ}}}}} \left( \widehat{W}_{1,1}^{(\textcolor{blue}{1},\textcolor{red}{0})} + \frac{\textcolor{blue}{A_{\text{ZZ}}}\, \widehat{W}_{2,1}^{(\textcolor{blue}{1},\textcolor{red}{0})}}{2g-\textcolor{blue}{\beta_{\text{ZZ}}}-1} + \frac{\textcolor{blue}{A_{\text{ZZ}}^2}\, \widehat{W}_{3,1}^{(\textcolor{blue}{1},\textcolor{red}{0})}}{\left(2g-\textcolor{blue}{\beta_{\text{ZZ}}}-1\right) \left(2g-\textcolor{blue}{\beta_{\text{ZZ}}}-2\right)} + \cdots \right) + \nonumber \\
&&
+ \frac{\textcolor{red}{S_{\text{FZZT}}}}{\mathrm{i}\pi}\, \frac{\Gamma \left( 2g-\textcolor{red}{\beta_{\text{FZZT}}} \right)}{\textcolor{red}{A_{\text{FZZT}}^{2g-\beta_{\text{FZZT}}}}} \left( \widehat{W}_{1,1}^{(\textcolor{blue}{0},\textcolor{red}{1})} + \frac{\textcolor{red}{A_{\text{FZZT}}}\, \widehat{W}_{2,1}^{(\textcolor{blue}{0},\textcolor{red}{1})}}{2g-\textcolor{red}{\beta_{\text{FZZT}}}-1} + \cdots \right) + \cdots,
\label{eq:largeorder}
\eea
\noindent
where $\beta_{\textcolor{blue}{\text{ZZ}}} \equiv \beta^{(\textcolor{blue}{1},\textcolor{red}{0})}-\beta^{(0)}$ and $\beta_{\textcolor{red}{\text{FZZT}}} \equiv \beta^{(\textcolor{blue}{0},\textcolor{red}{1})}-\beta^{(0)}$. Transseries nonperturbative data clearly resurges in the large-order behavior of the perturbative sector, at least up to Stokes data.

The resurgence framework hence results in the \textit{explicit}, leading, large-genus asymptotics for the one-point function $\widehat{W}^{(0)}_{g,1} (z)$, with \textit{both} ZZ and FZZT nonperturbative contributions as
\bea
\label{eq:W1asymp}
\widehat{W}^{(0)}_{g,1}(z) &\simeq& \textcolor{blue}{\frac{1}{\sqrt{2}\, \pi^\frac{3}{2}} \left( 4\pi^2 \right)^{2g-\frac{3}{2}}\, \Gamma \left(2g-\frac{3}{2}\right) \left\{ \frac{4}{4 z^{2}-1} + \cdots \right\}} + \cdots \\
&&
+ \textcolor{red}{\frac{1}{\pi} \left( V_{\text{eff}} \left(z^{2}\right) \right)^{-2g+1}\, \Gamma \left(2g-1\right) \frac{1}{2z} \left\{ 1 - \frac{V_{\text{eff}} \left(z^2\right)}{2g-2}\,  \frac{17+2\pi^2 z^2}{12 z^3} + \cdots \right\}} + \cdots. \nonumber
\eea
\noindent
The (blue) ZZ-brane contribution follows from our earlier discussion, and we omit higher corrections computed in subsection~\ref{subsec:correlators}. The (red) FZZT-brane contribution follows from feeding FZZT data as in \eqref{eq:W1-fzzt} into \eqref{eq:largeorder}; up to its Stokes coefficient $\frac{1}{\pi}$ which may be fixed by requiring that in the small $z$ limit the (exactly solvable) Airy result\footnote{Recall the JT spectral curve is $y \approx \sqrt{x} - \frac{2}{3} \pi^2 x^{3/2} + \cdots$, with $y = \sqrt{x}$ the Airy curve \cite{w91, k92, dw18, sss19}.} is recovered \cite{egggl23}. Notice how whether it is the (blue) ZZ brane contribution or the (red) FZZT one to dominate the (leading) large-genus asymptotics explicitly depends on the value of $z$. We will next corroborate our claim for the large-genus asymptotics of the $\widehat{W}^{(0)}_{g,1}(z)$ with a study of the Borel singularities of our one-point correlator. Although we do not have control over its full ZZ/FZZT nonperturbative content, its Borel-plane singularities can nonetheless be obtained in approximate ways.

\paragraph{On Approximate Borel Transforms:}

Standard Borel resummation of an asymptotic series, \textit{e.g.}, our perturbative one-point function $\widehat{W}^{(0)}_1 ( z )$ in \eqref{eq:W1-pert-exp}, starts with evaluation of the Borel transform out of its coefficients; which is then analytically continued across the complex Borel plane into a (multi-branched) function $\CB [ W_1^{(0)} ] (s)$. Laplace transform along an adequate $\theta$-ray then yields the corresponding Borel resummation as (see, \textit{e.g.}, \cite{abs18}),
\be
\label{eq:Borel-resum}
\CS_{\theta} \widehat{W}^{(0)}_1 ( z ) = \int_{0}^{\rme^{\rmi \theta} \infty} \rmd s\, \CB [ W_1^{(0)} ] (s)\, \rme^{- \frac{s}{g_{\text{s}}}}.
\ee
\noindent
One alternative, albeit non-constructive, approach to $\CB [ W_1^{(0)} ] (s)$ is to try and directly find an integral representation for some given observable as a Laplace-type integral---hence allowing for an identification of its Borel transform directly from this integral representation by comparison with the above expression (see \cite{abs18} as well). This is what we shall do in the following.

Consider our one-point function $\widehat{W}_1 \left(z; g_{\text{s}} \right)$ in \eqref{eq:W1-1inst-1PTS}, where we emphasize its dependence on $g_{\text{s}}$ since it is the variable on which the Borel transform acts. Recall from \eqref{eq:W1-int} that its ZZ one-instanton sector may be written as an exponential integral after action of the loop insertion operator as in \eqref{eq:Wparts}; \textit{i.e.},
\be
\label{eq:int-rep-W1}
\widehat{W}_{1}^{(1)} \left(z\right) = \frac{g_{\text{s}}}{2\pi} \int_{\mathcal{I}} \dd x\, \Delta_{z} \BS(x)\, \rme^{\BS(x)}.
\ee
\noindent
If we just write out this expression formally, without any intent to a saddle-point evaluation as in subsection~\ref{subsec:NPTR-1inst-F}, then the argument of the exponential is, via \eqref{eq:SSchi-summed} with \eqref{eq:SS0} and \eqref{eq:SS1},
\be
\BS \left(x;g_{\text{s}}\right) \simeq \frac{1}{g_{\text{s}}} \BS_{0} (x) + \BS_{1} (x) + \cdots = - \frac{1}{g_{\text{s}}} V_{\text{eff}} (x) - \frac{1}{2} \log \left(- 16 x^{2} \right) + \cdots,
\ee
\noindent
and the loop insertion operator yields, for instance via \eqref{eq:W111},
\be
\Updelta_{z} \BS(x) = \Updelta_{z} \BS_{0} (x) + \cdots = - \frac{2\sqrt{x}}{x-z^{2}} + \cdots,
\ee
\noindent
leading us to:
\be
\label{eq:W11-approx-Borel-resum-first}
\widehat{W}^{(1)}_1 \left(z; g_{\text{s}} \right) \approx \frac{\rmi}{4\pi} \int_{\mathcal{I}} \dd x\, \frac{1}{x-z^{2}}\, \frac{1}{\sqrt{x}}\, \rme^{- \frac{V_{\text{eff}}(x)}{g_s}}.
\ee
\noindent
It should be now clear how to obtain an approximate Borel transform for $\widehat{W}^{(1)}_1 \left(z; g_{\text{s}} \right)$ given its above integral representation. The obvious trick (see \cite{abs18}; or else in the present JT-gravity context a slightly different version of which was already used in \cite{sss19}) is to perform the change of variable $s = V_{\text{eff}}(x)$ and upon comparison with \eqref{eq:Borel-resum} interpret the resulting integral,
\be
\label{eq:W11-approx-Borel-resum}
\widehat{W}^{(1)}_1 \left(z; g_{\text{s}} \right) \approx \frac{\rmi}{4\pi}  \int_{\widetilde{\mathcal{I}}} \dd s\, \textcolor{blue}{\frac{1}{V_{\text{eff}}' \left( x (s) \right)}}\, \textcolor{red}{\frac{1}{x (s)-z^{2}}}\, \frac{1}{\sqrt{x (s)}}\, \rme^{-\frac{s}{g_{\text{s}}}},
\ee
\noindent
as a Borel resummation. We will interpret \eqref{eq:W11-approx-Borel-resum-first} as an integral representation of the full one-point function $\widehat{W}_1(z)$, with different choices of the yet-unspecified integration-contour yielding different explicit sectors of its transseries. For instance, on top of the ZZ contours discussed in section~\ref{sec:NPtoprec}, there is also a natural FZZT contour picking the contribution of the simple pole at $x=z^2$ in \eqref{eq:W11-approx-Borel-resum-first}. Note how given the relation between ZZ and FZZT branes in \eqref{eq:ZZfromFZZT} it should not come as a big surprise that an initial ``ZZ expression'' is now giving rise to FZZT information. As such, if in \eqref{eq:W11-approx-Borel-resum-first} we pick the perturbative contour $\CI_0$, we claim that the integrand of \eqref{eq:W11-approx-Borel-resum} is the (truncated) Borel transform of the perturbative sector $\widehat{W}_1^{(0)} \left(z; g_{\text{s}} \right)$. While this is just a crude approximation to the actual Borel transform, given that it only uses the leading contribution to the integrand in \eqref{eq:int-rep-W1} and requires a heuristic contour argument, it nevertheless remarkably captures part of the actual Borel singularities. In fact, the piece we highlighted in blue is singular when $s = A_{\text{ZZ}}$, namely it captures the singularity at the ZZ-brane instanton action. The red piece instead diverges when $s = V_{\text{eff}} (z^2)$, namely at the FZZT brane instanton action. It is important however to point out that, since $V_{\text{eff}} (x)$ is invertible only in the $[0,1/4]$ interval, the singularity coming from the red piece belongs to the principal sheet of the Borel plane only in that interval. Finally, the square-root contribution does not produce any poles.

\begin{figure}
\centering
     \begin{subfigure}[h]{0.75\textwidth}
         \centering
         \includegraphics[width=\textwidth]{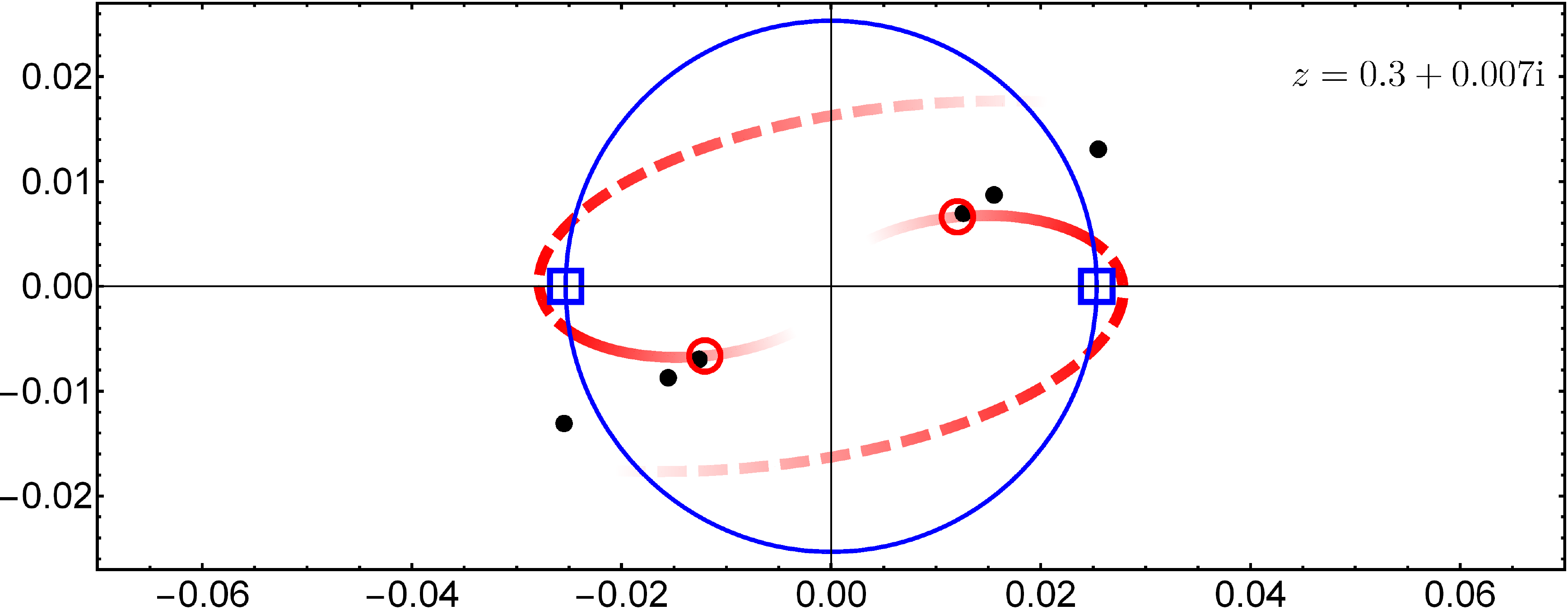}
         \vspace{1mm}
     \end{subfigure}
     \begin{subfigure}[h]{0.75\textwidth}
         \centering
         \includegraphics[width=\textwidth]{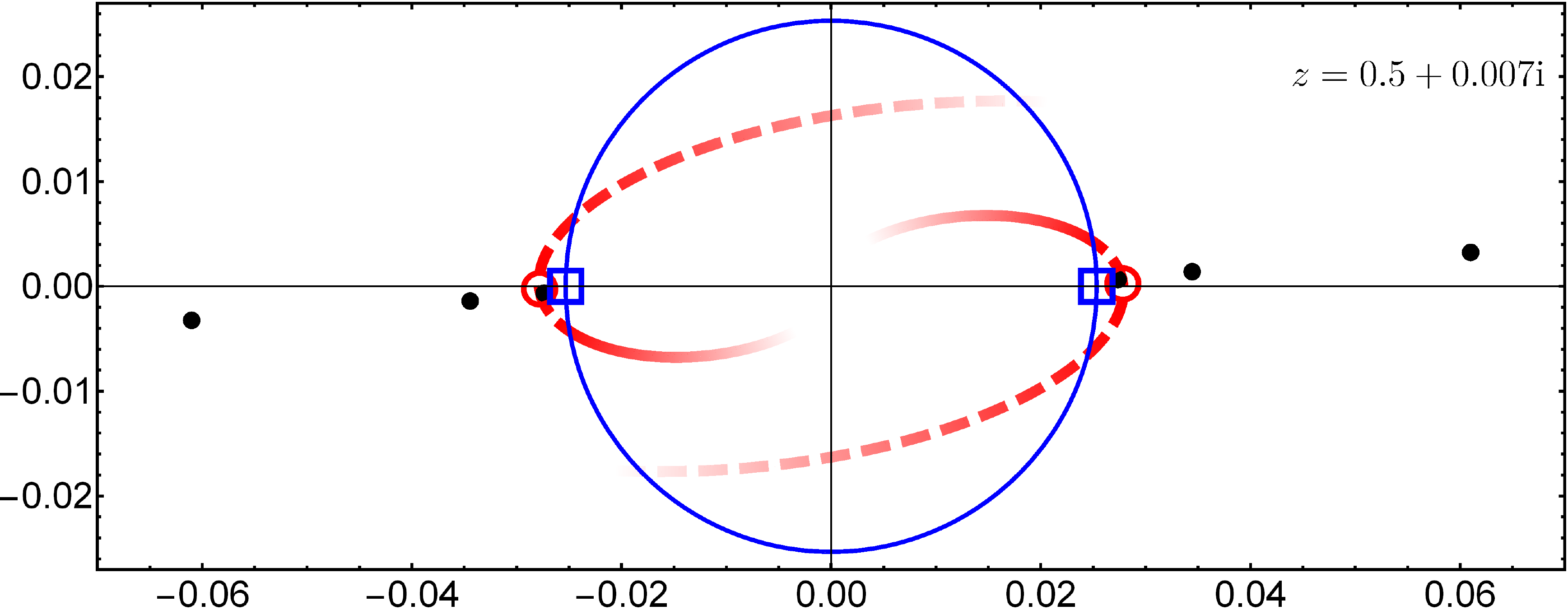}
         \vspace{1mm}
     \end{subfigure}
     \begin{subfigure}[h]{0.75\textwidth}
         \centering
         \includegraphics[width=\textwidth]{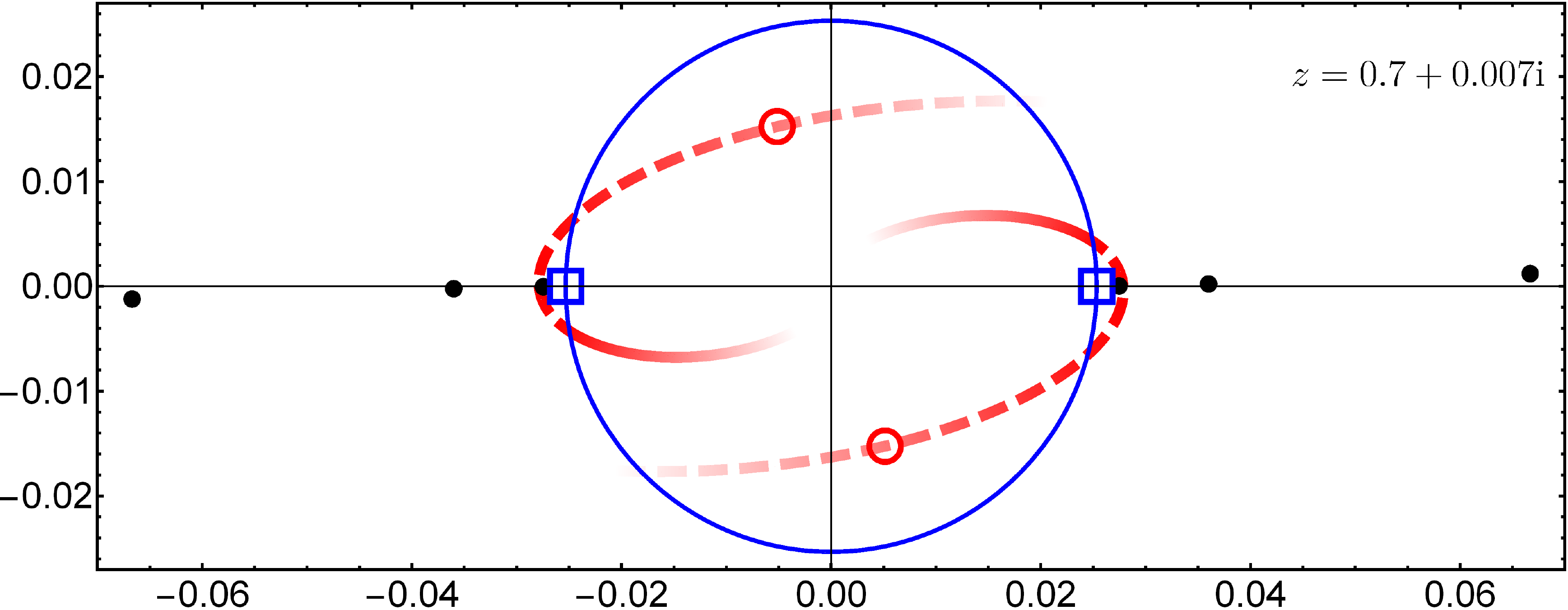}
     \end{subfigure}
     \caption{Borel--Pad\'e analysis of the Borel singularity structure of the perturbative one-point function $\widehat{W}^{(0)}_1 ( z )$ in \eqref{eq:W1-pert-exp}, for three different values of its argument, $z$. The blue squares represent the ZZ instanton actions, ``fixed'' at $\pm 1/4\pi^2$; while the red circles represent the FZZT instanton actions ``moving'' as $\pm V_{\text{eff}}(z^{2})$. This symmetric feature of Borel singularities is a hallmark of resonance \cite{sst23}. The red curves follow the trajectories of the FZZT actions, as the real part of $z$ varies, and they become dashed as soon as they cross the (blue) circle determined by the absolute value of the ZZ actions---hence losing dominance and leaving the principal-sheet of the Borel plane. In the first plot, the FZZT actions are well inside the blue circle, dominating the large-order behavior, and consequently the Pad\'e poles (the black dots) accumulate close to them. In the second plot, the FZZT actions leave the principal-sheet of the Borel plane. In the third plot, even though the absolute value of the FZZT actions is again smaller than the ZZ one, they have now left the principal-sheet and the Pad\'e poles remained attached to the ZZ singularities.}
\label{fig:Pade-1}
\end{figure}

Having an analytic approximation to the Borel transform of the one-point function asymptotic perturbative expansion \eqref{eq:W1-pert-exp}, one may ask if we can also find a numerical approximation to the very same function---eventually validating our assumptions above. Such numerical approximations to Borel transforms are commonly achieved via the use of Borel--Pad\'e approximants (see, \textit{e.g.}, \cite{abs18}). Let us see how to implement them in our current example. Go back to the perturbative series \eqref{eq:W1-pert-exp}, remove the genus-zero term and truncate the sum at some finite order $N$ (in our case we computed the $\widehat{W}_{g,1}^{(0)}(z)$ coefficients up to $g=7$ using the standard topological recursion). This correspondingly leads to the truncated Borel transform
\be
\label{eq:trunc-borel}
\CB_N [ \widehat{W}_1^{(0)} (z) ] (s) \approx \sum_{m=0}^{N-1} \frac{\widehat{W}_{m+1,1}^{(0)} (z)}{\Gamma \left(2m+1\right)}\, s^{2m},
\ee
\noindent
which we now approximate by an order-$N$ (diagonal) Pad\'e approximant, the Borel--Pad\'e approximant, which is a rational function of the form
\be
\label{eq:borel-pade}
\text{BP}_N [ \widehat{W}_1^{(0)} (z) ] (s) = \frac{\sum_{k=0}^{N} a_k\, s^{k}}{\sum_{k=0}^{N} b_k\, s^{k}}.
\ee
\noindent
Herein, the coefficients $\{a_k\}$, $\{b_k\}$ are fixed such that the Taylor expansion of \eqref{eq:borel-pade} around $s=0$ matches that of \eqref{eq:trunc-borel}. Using a Borel--Pad\'e approximant has the advantage that we are now dealing with a rational function whose poles indicate rather well the actual location of the Borel singularities of the original series. In fact, these poles tend to ``condensate'' into the branch cuts emerging from the Borel singularities. In our present case, even though with only a very limited set of perturbative data, the results of the Borel--Pad\'e analysis are surprisingly clear---confirming the Borel singularity structure predicted by the above approximate Borel transform \eqref{eq:W11-approx-Borel-resum} obtained from the integral representation. In particular, this Borel--Pad\'e analysis highlights how the FZZT brane singularity ends-up leaving the principal-sheet of the Borel plane, as shown in figure~\ref{fig:Pade-1} illustrating the ``fixed'' ZZ singularities versus the ``moving'' FZZT singularities for different values of $z \in \BC$. Further, this figure clearly illustrates how all ZZ or FZZT singularities arise in \textit{symmetric pairs}, which is to say it provides for strong supporting evidence on the \textit{resonance} features of both ZZ and FZZT nonperturbative corrections in the context of JT gravity \cite{gs21, sst23}.

\begin{figure}
\centering
     \begin{subfigure}[h]{0.49\textwidth}
         \includegraphics[width=\textwidth]{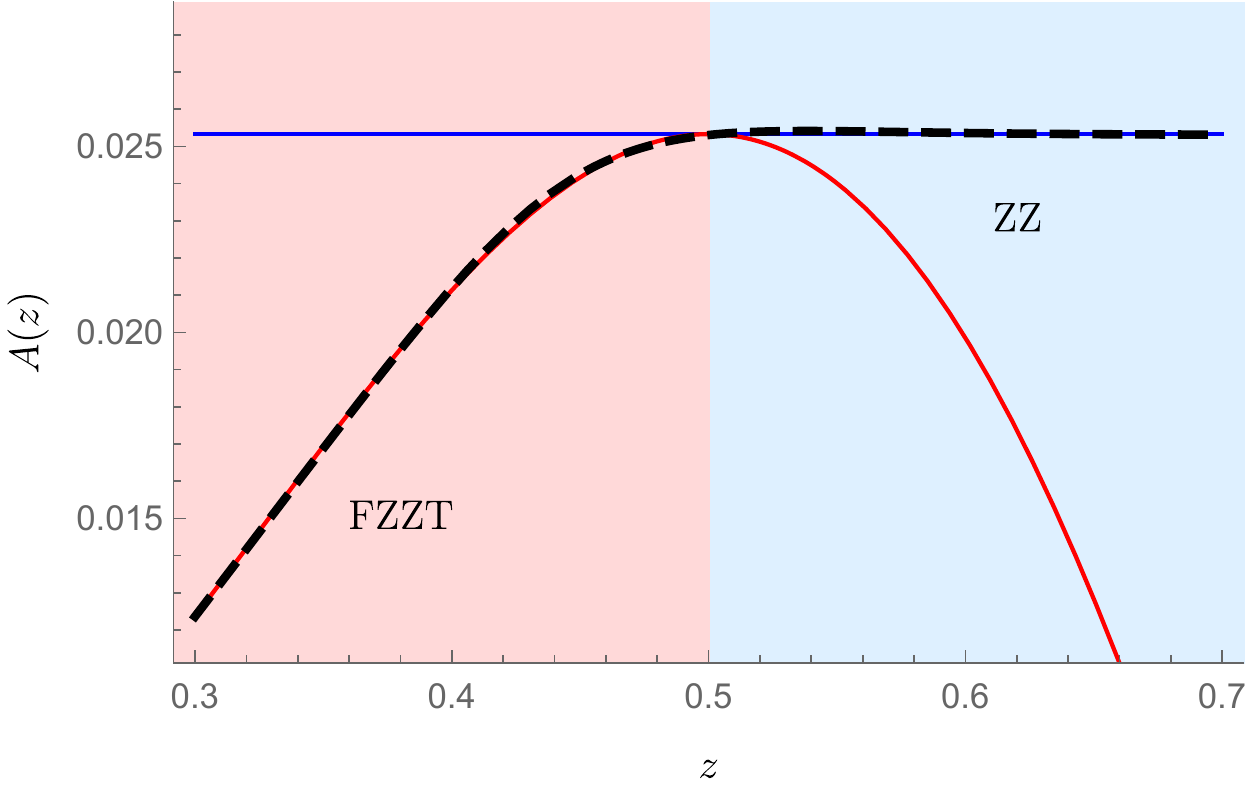}
     \end{subfigure}
     \begin{subfigure}[h]{0.49\textwidth}
         \includegraphics[width=\textwidth]{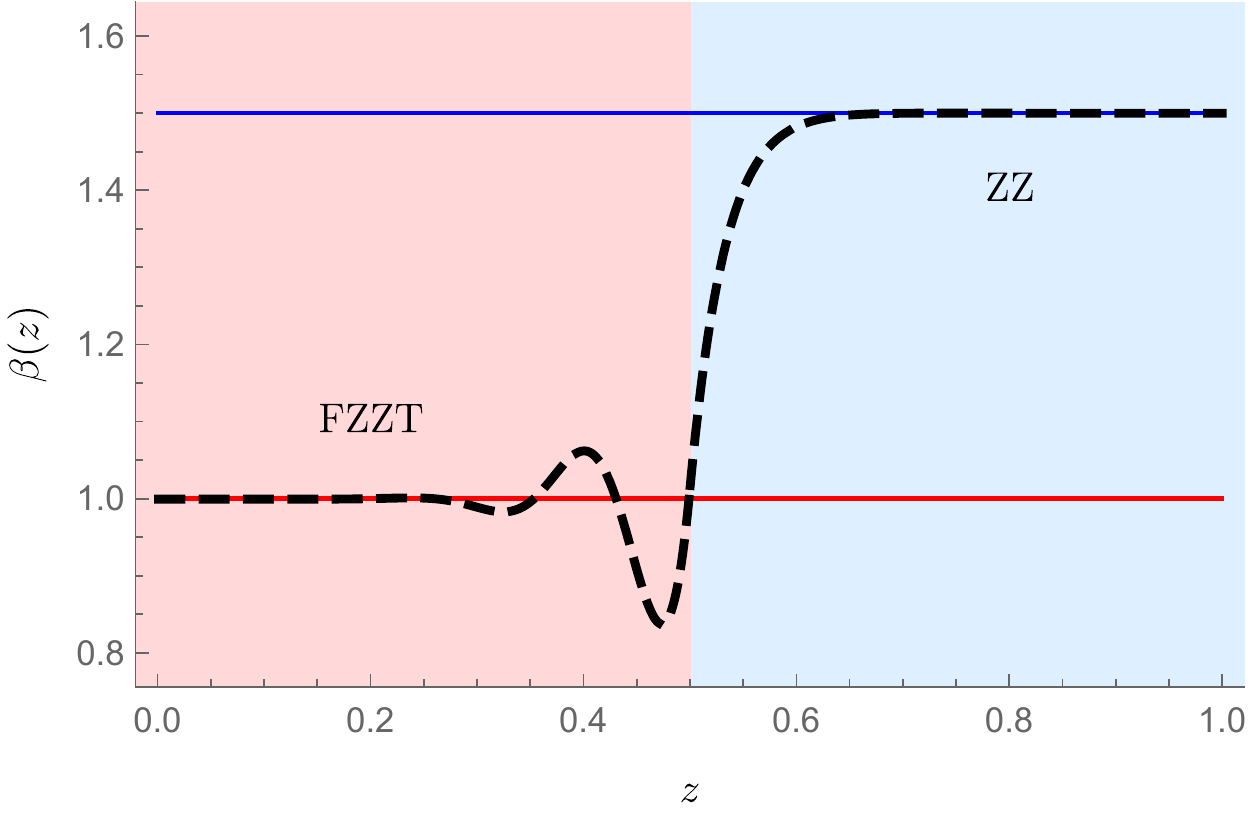}
     \end{subfigure}
\caption{The black-dashed lines are the fifth Richardson transforms of the sequences that determine the instanton action via the inverse square-root of \eqref{eq:Qg} (left), and the characteristic exponent $\beta$ \eqref{eq:Bg} (right) of $\widehat{W}_1^{(0)} (z)$, as function of real $z$ (in a range where the crossover is clear). In red and blue, the theoretical ``predictions'' associated to FZZT and ZZ branes, respectively.}
\label{fig:1point-act}
\end{figure}

\begin{figure}
\centering
     \begin{subfigure}[h]{0.5\textwidth}
         \centering
         \includegraphics[width=\textwidth]{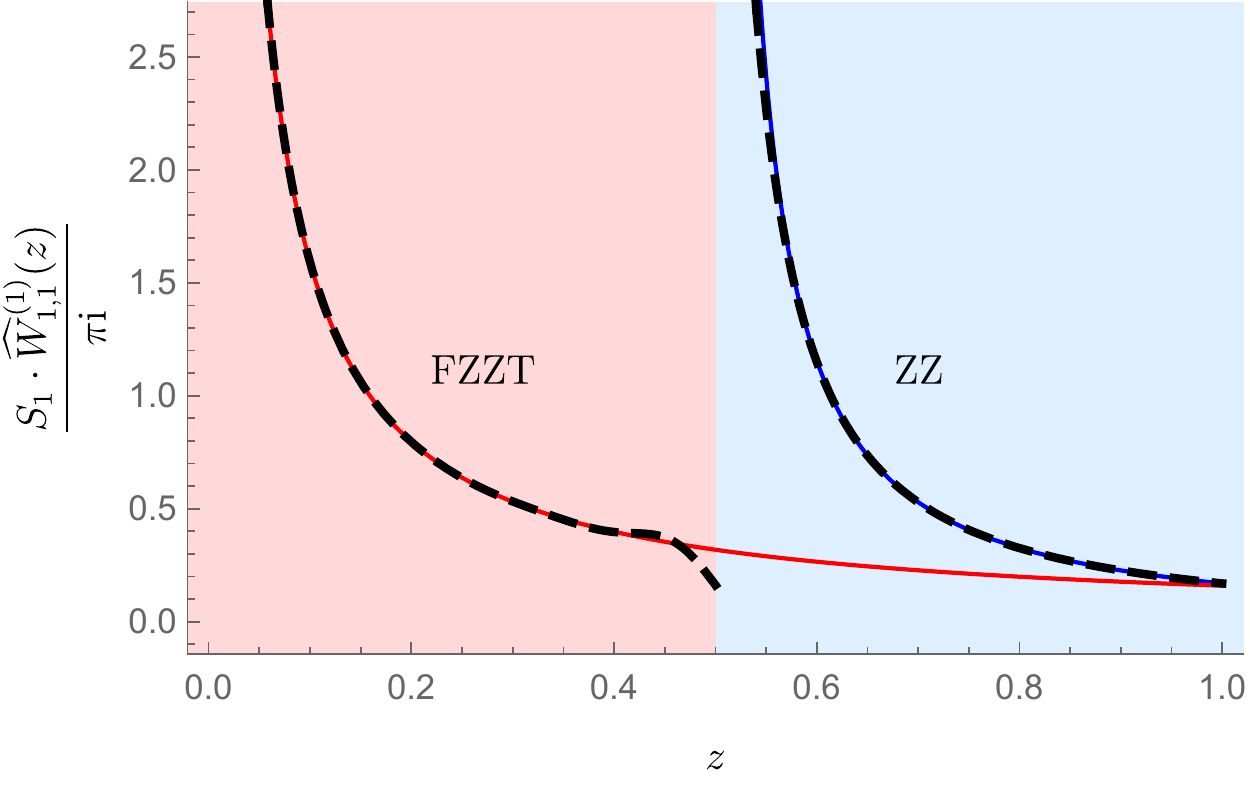}
     \end{subfigure}
     \caption{The black-dashed line is the sixth Richardson transform of the sequence that determines the one-loop around the one-instanton contribution \eqref{eq:M1g} to $\widehat{W}_1^{(0)} (z)$, as a function of $z$. Again, in red and blue, the ``predictions'' associated to FZZT and ZZ branes, respectively.}
\label{fig:1point-one}
\end{figure}

\begin{figure}
\centering
     \begin{subfigure}[h]{0.49\textwidth}
         \includegraphics[width=\textwidth]{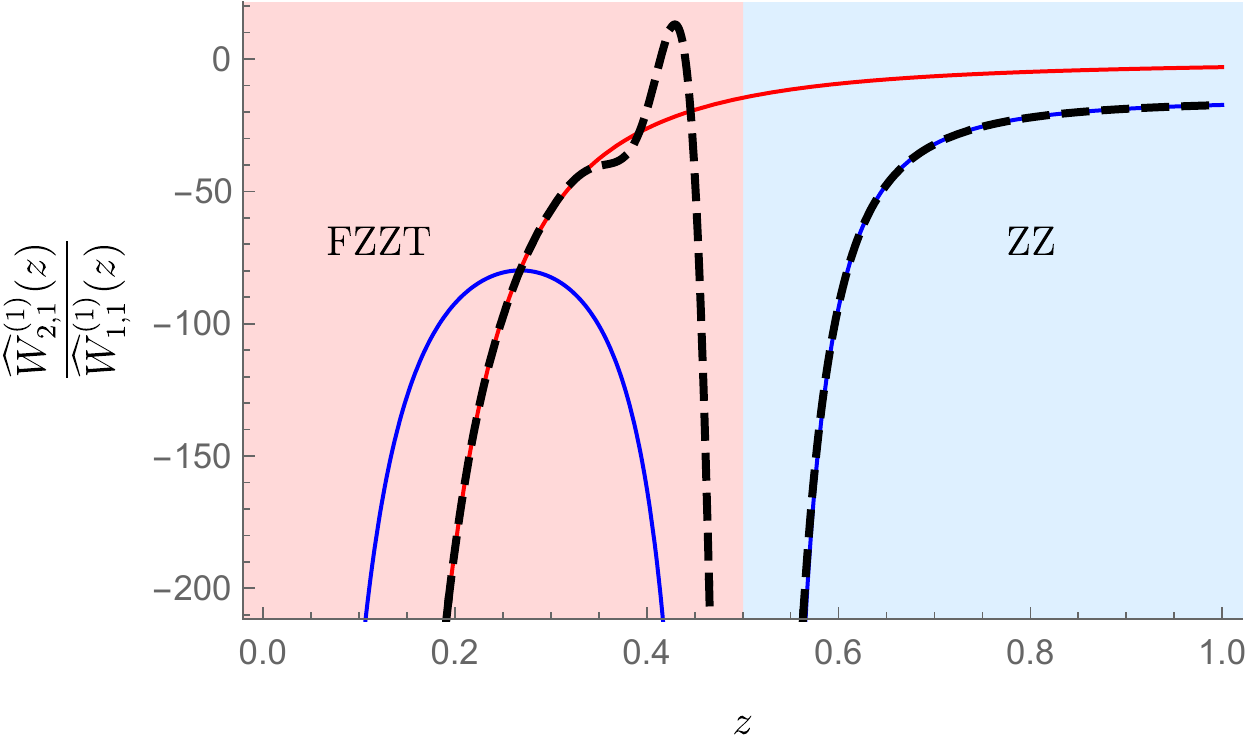}
     \end{subfigure}
     \begin{subfigure}[h]{0.49\textwidth}
         \includegraphics[width=\textwidth]{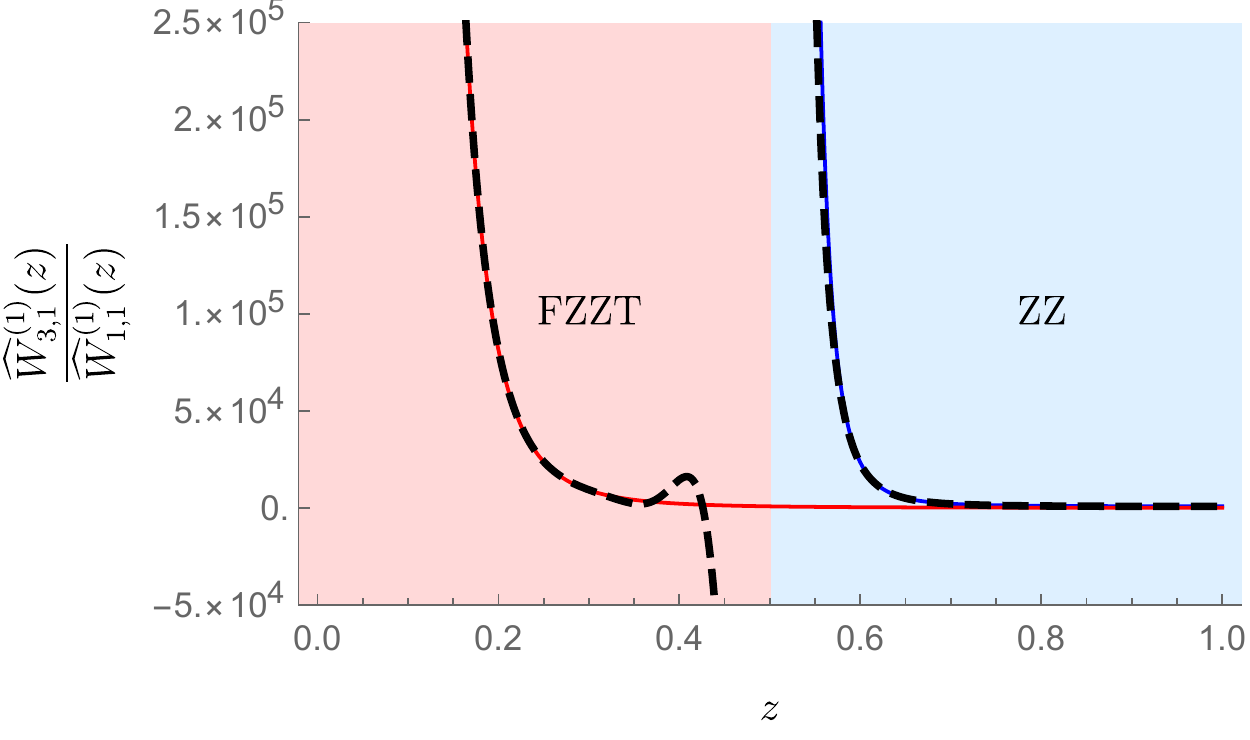}
     \end{subfigure}
\caption{The black-dashed lines are the sixth Richardson transforms of the sequences that determine the two- (left) and three- (right) loop contributions around the one-instanton sector, as functions of $z$ (given by \eqref{eq:M2g} and its straightforward three-loop extension). Red/blue are the usual FZZT/ZZ ``predictions''.}
\label{fig:1point-two-and-three}
\end{figure}

\paragraph{Numerical Large-Order Analysis:}

The features observed in the Borel--Pad\'e analysis which was illustrated in figure~\ref{fig:Pade-1} also show up in the analyses of large-order asymptotics displayed in figures~\ref{fig:1point-act}, \ref{fig:1point-one}, and~\ref{fig:1point-two-and-three}. Let us first recall these large-order tests (see, \textit{e.g.}, \cite{msw07, abs18}). In fact, large-order relations of the form \eqref{eq:largeorder} are not only a clean illustration of the interplay between perturbative and nonperturbative sectors in resurgent transseries, but also a rather useful tool for numerically extracting nonperturbative data out from the coefficients of a perturbative expansion. For simplicity, focus for a moment on the first line of equation \eqref{eq:largeorder}, addressing ZZ content (but let us drop the ZZ subscripts for greater clarity). It is immediate to introduce the new sequence
\begin{equation}
\label{eq:Qg}
Q_{g} \equiv \frac{\widehat{W}^{(0)}_{g+1,1}}{4g^{2}\, \widehat{W}^{(0)}_{g,1}} \simeq \frac{1}{A^{2}} \left( 1 + \frac{1-2\beta}{2g} + \cdots \right),
\end{equation}
\noindent
which, given access to perturbative data at sufficiently large orders, may be used to extract the value of the instanton action $A$ numerically. Moreover, we can further use it to extract the value of the characteristic exponent $\beta$, via the new sequence
\begin{equation}
\label{eq:Bg}
B_g \equiv \frac{1}{2} - g \left( A^{2}\, Q_g - 1 \right) \simeq \beta + \cdots.
\end{equation}
\noindent
These sequences naturally have infinite $1/g$ corrections, hence are amenable to the use of Richardson extrapolation in order to improve their convergence as $g$ goes to infinity. This numerical method allows us to obtain reliable numerical checks even in cases when the number of available perturbative coefficients is not large (see, \textit{e.g.}, \cite{abs18}). Similar sequences may be set-up extracting further nonperturbative data out of the perturbative series; for example the one and two-loop coefficients around the one-instanton sector follow as
\begin{equation}
\label{eq:M1g}
M_{1,g} \equiv \frac{A^{2g-\beta}}{\Gamma \left(2g-\beta\right)}\, \widehat{W}^{(0)}_{g,1} \simeq \frac{S_1 \widehat{W}_{1,1}^{(1)}}{\mathrm{i}\pi} \left( 1 + \frac{A}{2g}\, \frac{\widehat{W}_{2,1}^{(1)}}{\widehat{W}_{1,1}^{(1)}} + \cdots \right)
\end{equation}
\noindent
and
\begin{equation}
\label{eq:M2g}
M_{2,g} \equiv \frac{2g}{A} \left( \frac{\mathrm{i}\pi}{S_1 \widehat{W}_{1,1}^{(1)}}\, \frac{A^{2g-\beta}}{\Gamma \left( 2g-\beta \right)}\, \widehat{W}^{(0)}_{g,1} - 1 \right) \simeq \frac{\widehat{W}_{2,1}^{(1)}}{\widehat{W}_{1,1}^{(1)}} + \cdots.
\end{equation}
\noindent
Following this strategy, one can easily construct the general sequence $M_{\ell,g}$ which can be used to test the $\ell$-loop contribution around the one-instanton configuration.

Let us then consider the asymptotic analyses in figures~\ref{fig:1point-act}, \ref{fig:1point-one}, and~\ref{fig:1point-two-and-three}, addressing the large-order behavior of the perturbative sequence \eqref{eq:largeorder}-\eqref{eq:W1asymp}, with both ZZ and FZZT contributions, and for varying argument $z$. Even though these checks were performed using only the first eight perturbative coefficients, the numerics show surprisingly well how for $z<1/2$ the FZZT contribution, having the smaller instanton action, is the one that dominates the large-genus asymptotics of the perturbative coefficients $\widehat{W}^{(0)}_{g,1}$. As soon as $z$ hits $1/2$, namely the value at which the FZZT action coincides with the ZZ action, large-order dominance is interchanged, as the FZZT singularity leaves the principal-sheet of the Borel plane. This change of regime is supported by numerical checks on the instanton action, on the characteristic exponent $\beta$, and on the one-, two- and three-loop contributions to the one-instanton sector. The check on the instanton action neatly shows how it changes from the $z$-dependent FZZT action $V_{\text{eff}} (z^{2})$ to the constant $1/4\pi^2$ ZZ action. The $\beta$ numerics instead show the expected jump of the characteristic exponent from $1$ to $3/2$, while at one-loop order we clearly see two distinct $z$-dependent regimes in figure~\ref{fig:1point-one}. The first one is dominated by the FZZT contribution proportional to $z^{-1}$ as shown in \eqref{eq:W1-fzzt}. The second instead is given by the one-loop term around the one-instanton ZZ sector computed in \eqref{eq:W111},
\be
S_1 \cdot \widehat{W}_{1}^{(1)}(z) = \frac{\rmi}{\sqrt{2\pi}}\, \frac{4}{4z^{2}-1}.
\ee
\noindent
As we shall see in the next section, it is this ZZ term which, upon Laplace transform, results in the characteristic $\sinh (b)/b$ behavior of the Weil--Petersson volumes. Analogous changes in the $z$-dependent regimes of FZZT versus ZZ dominance can be observed in figure~\ref{fig:1point-two-and-three} for the two- and three-loop contributions.

\subsection{Large Genus Asymptotics of the Two-Point Function: ZZ and FZZT Branes}
\label{subsec:twopointW}

Having understood the one-point function, the extension to higher-point multi-resolvent correlation functions should now be straightforward---simply follow the strategy in the previous subsection, now with the addition of an extra number of FZZT contributions, one for each new possible variable in the multi-resolvent correlators. We will illustrate such generic strategy in the present subsection within the setting of the two-point function, which already includes all these new FZZT features as compared to the one-point function.

Consider the two-point function $\widehat{W}_2(z_1,z_2)$. Because it depends on \textit{two} distinct points, $z_1$ and $z_2$, there will correspondingly be \textit{two} distinct FZZT contributions. Comparing with \eqref{eq:W1=ZZ+FZZT} it is then immediate to write down its transseries, up to one-instanton contributions, as being of the form
\bea
\label{eq:W2-zz-fzzt}
\widehat{W}_2 \left( z_1,z_2; \textcolor{blue}{\sigma_{\text{ZZ}}}, \textcolor{red}{\sigma_{\text{FZZT}_1}}, \textcolor{violet}{\sigma_{\text{FZZT}_2}} \right) &=& \widehat{W}_2^{(0)} (z_1,z_2) + \textcolor{blue}{\sigma_{\text{ZZ}}\, \widehat{W}_2^{(\text{ZZ})} (z_1,z_2)} + \\
&&
+ \textcolor{red}{\sigma_{\text{FZZT}_1}\, \widehat{W}_2^{(\text{FZZT}_1)} (z_1,z_2)} + \textcolor{violet}{\sigma_{\text{FZZT}_2}\, \widehat{W}_2^{(\text{FZZT}_2)} (z_1,z_2)} + \cdots. \nonumber
\eea
\noindent
Being more precise on this (at least) \textit{three}-parameter transseries (compare with \eqref{eq:W1-2PT-detailed}), it should be clear that the ZZ-brane contribution takes the usual form $\mathrm{e}^{-A_{\text{ZZ}}/g_{\text{s}}}\, \widehat{W}^{(1)}_2\left(z_1,z_2\right)$ which was explored in detail in the previous section and which at leading order is, via \eqref{eq:sinhLn-LAP},
\be
\widetilde{W}_{1,2}^{(1)} \left(z_{1}, z_{2}\right) = \frac{4 x_{\star}}{\left( z_1^{2} - x_{\star} \right) \left( z_2^{2} - x_{\star} \right)}.
\ee
\noindent
As for the FZZT contributions, they take the form\footnote{In the exact same way as we did in \eqref{eq:W1-fzzt}, we are herein using the FZZT wave-function \eqref{eq:FZZT-WKB-sol} as computed in \cite{os19} to go beyond the one-loop result of \cite{sss19} for the two-point function, extending it to the present two-loop result.} \cite{sss19, os19}
\bea
\label{eq:W2-fzzt-a}
\widehat{W}_2^{(\text{FZZT}_1)} (z_1,z_2) &\simeq& - \frac{1}{z_1^2-z_2^2}\, \rme^{-\frac{V_{\text{eff}} (z_1^2)}{g_{\text{s}}}} \left\{ 1 - \frac{12 z_1^2 + 5 z_2^2 + 2 \pi^2 z_1^2 z_2^2}{12 z_1^3 z_2^2}\, g_{\text{s}} + \cdots \right\}, \\
\label{eq:W2-fzzt-b}
\widehat{W}_2^{(\text{FZZT}_2)} (z_1,z_2) &=& \widehat{W}_2^{(\text{FZZT}_1)} (z_2,z_1).
\eea
\noindent
This of course will result in a more involved singularity structure upon the Borel plane, with more possibilities for dominance of the large-genus asymptotics, as now it is three instanton actions which are competing for the leading contribution as $z_1$ and $z_2$ vary. The large-genus asymptotics for the two-point function, with both ZZ and FZZT contributions, is hence an immediate generalization of \eqref{eq:W1asymp} where we now find
\bea
\label{eq:W2asymp}
\widehat{W}^{(0)}_{g,2} (z_1,z_2) &\simeq& \textcolor{blue}{\frac{1}{\sqrt{2}\, \pi^\frac{3}{2}} \left( 4\pi^2 \right)^{2g-\frac{1}{2}}\, \Gamma \left(2g-\frac{1}{2}\right) \left\{ \frac{4}{4 z_1^{2}-1}\, \frac{4}{4 z_2^{2}-1} + \cdots \right\}} + \cdots \\
&&
\hspace{-25pt}
+ \textcolor{red}{\frac{1}{\pi} \left( V_{\text{eff}} \left(z_1^{2}\right) \right)^{-2g}\, \Gamma \left(2g\right) \frac{1}{z_2^2-z_1^2} \left\{ 1 - \frac{V_{\text{eff}} \left(z_1^2\right)}{2g-1}\, \frac{12 z_1^2 + 5 z_2^2 + 2\pi^2 z_1^2 z_2^2}{12 z_1^3 z_2^2} + \cdots \right\}} + \cdots \nonumber \\
&&
\hspace{-25pt}
+ \textcolor{violet}{\frac{1}{\pi} \left( V_{\text{eff}} \left(z_2^{2}\right) \right)^{-2g}\, \Gamma \left(2g\right) \frac{1}{z_1^2-z_2^2} \left\{ 1 - \frac{V_{\text{eff}} \left(z_2^2\right)}{2g-1}\, \frac{12 z_2^2 + 5 z_1^2 + 2\pi^2 z_1^2 z_2^2}{12 z_2^3 z_1^2} + \cdots \right\}} + \cdots. \nonumber
\eea
\noindent
Recall from \eqref{eq:Rn-multi2-revisited} and, say, \eqref{eq:W1-pert-exp}, that the characteristic exponent for the perturbative $n$-point correlator is $\beta^{(0)}=n-2$, and that the one featured at large order is given by the difference between one-instanton and perturbative $\beta \equiv \beta^{(1)} - \beta^{(0)}$ (explicitly in \eqref{eq:largeorder}). Moreover, ZZ Stokes data follows from our earlier considerations and FZZT Stokes data is again fixed by the small-$z$ Airy limit. This very same structure generalizes in an obvious way to higher $n$-point correlators, with $n>2$, and will translate straightforwardly into the large-genus asymptotics for Weil--Petersson volumes upon inverse Laplace transform, as we shall see in the next section.

\begin{figure}
\centering
     \begin{subfigure}[h]{0.75\textwidth}
         \centering
         \includegraphics[width=\textwidth]{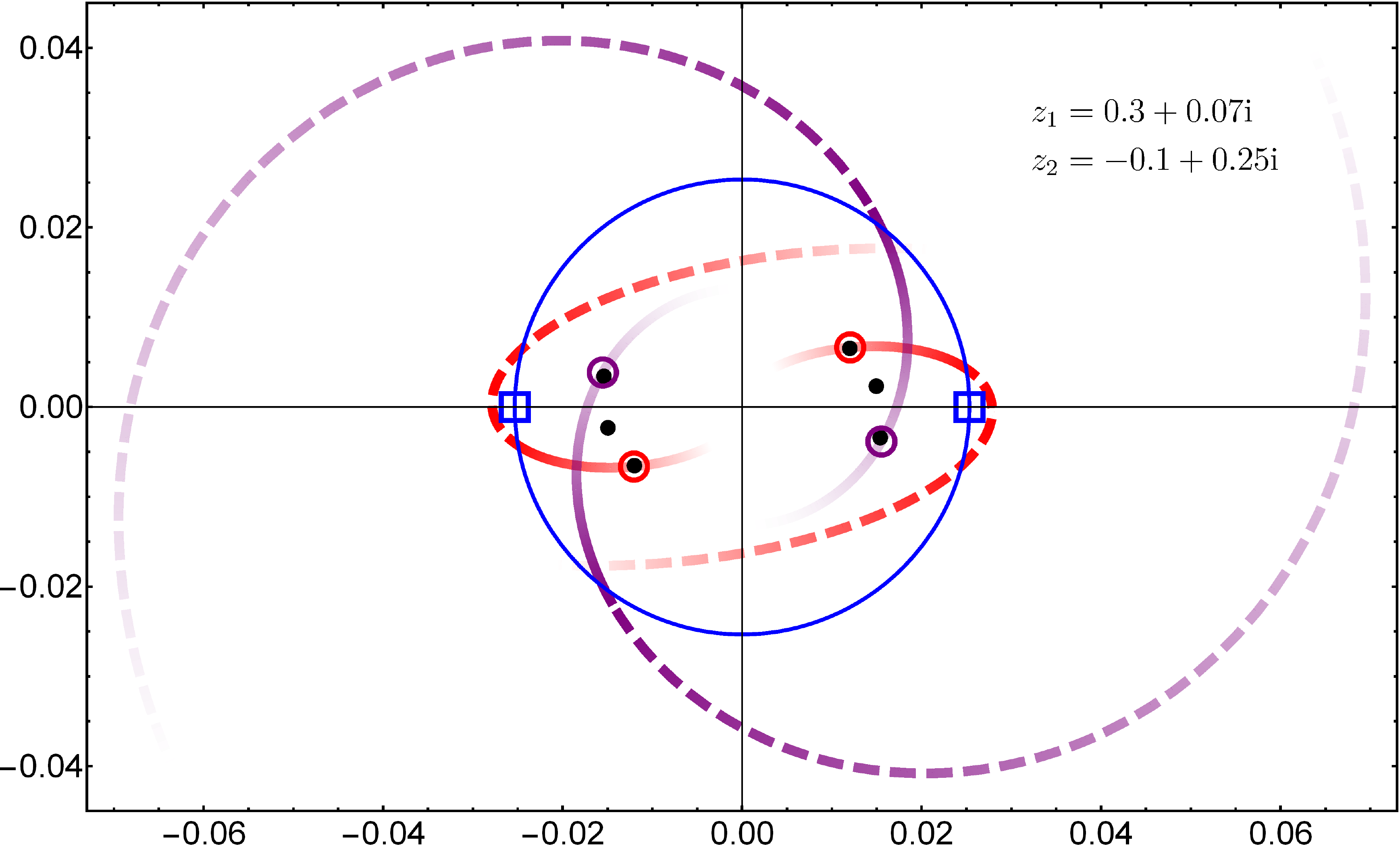}
         \vspace{1mm}
     \end{subfigure}
     \begin{subfigure}[h]{0.75\textwidth}
         \centering
         \includegraphics[width=\textwidth]{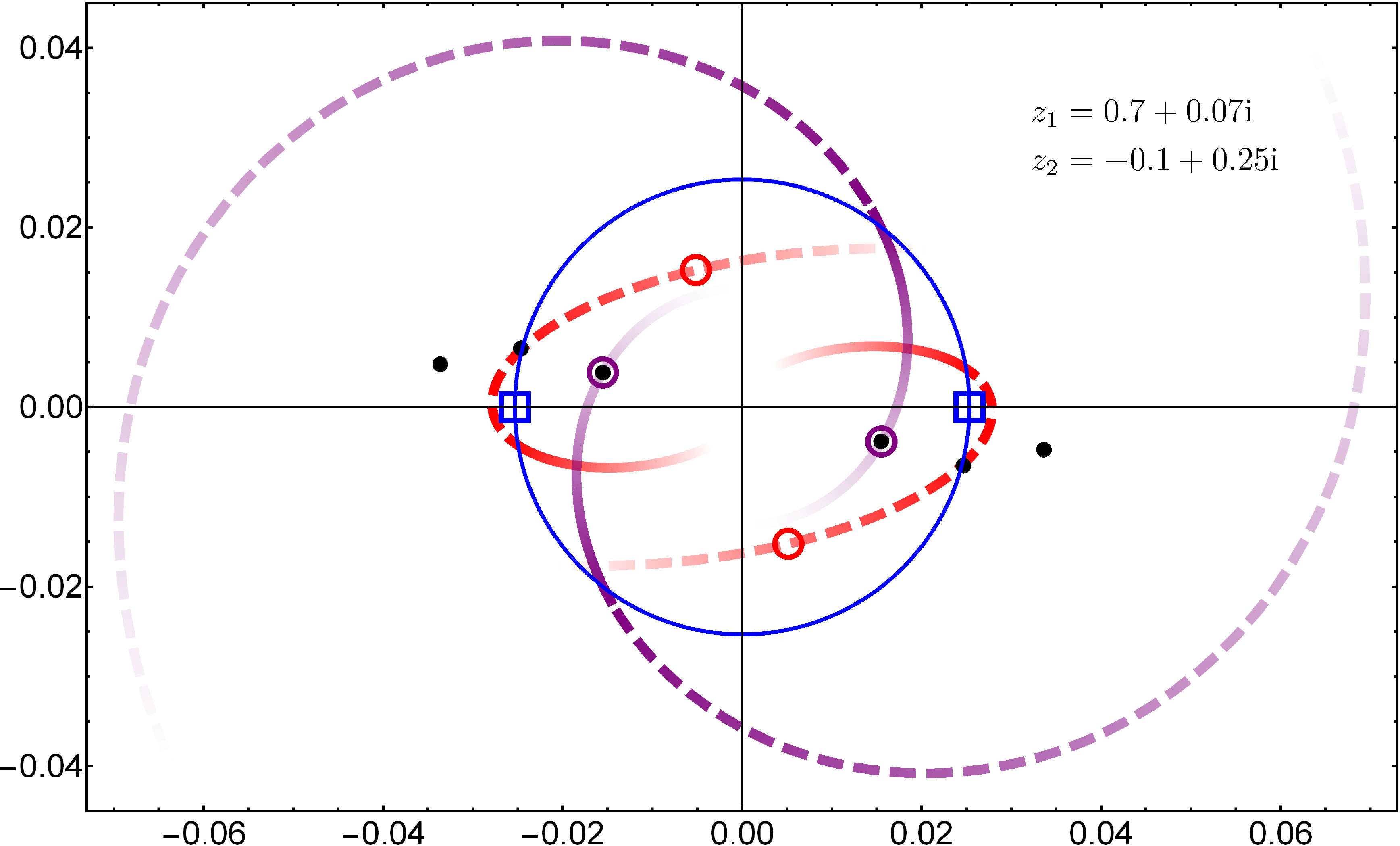}
     \end{subfigure}
          \caption{Borel--Pad\'e analysis of the Borel singularity structure of the perturbative two-point function $\widehat{W}_2^{(0)} (z_1,z_2)$, for two different values of its arguments, $\left(z_1,z_2\right)$. The blue squares represent the ZZ instanton actions, ``fixed'' at $\pm 1/4\pi^2$; while the red circles represent the first (resonant) pair of FZZT instanton actions ``moving'' as $\pm V_{\text{eff}} (z_1^{2})$; and the purple circles represent the second (resonant) pair of FZZT instanton actions now ``moving'' as $\pm V_{\text{eff}} (z_2^{2})$. The red and purple curves follow the trajectories of the two pairs of FZZT actions, as the real parts of $z_1$ and $z_2$ vary, and they become dashed as soon as they cross the (blue) circle determined by absolute value of the ZZ actions---potentially losing dominance and leaving the principal-sheet of the Borel plane. In the first plot, both pairs of FZZT actions are inside the blue circle, dominating the large-order behavior, and consequently the Pad\'e poles (the black dots) accumulate close to them. In the second plot, the first resonant pair of FZZT actions has left the principal sheet of the Borel plane but the second resonant pair remains inside it, hence the Pad\'e poles stay attached to the latter.}
\label{fig:Pade-2}
\end{figure}

\begin{figure}
\centering
      \begin{subfigure}[h]{0.75\textwidth}
         \centering
         \includegraphics[width=\textwidth]{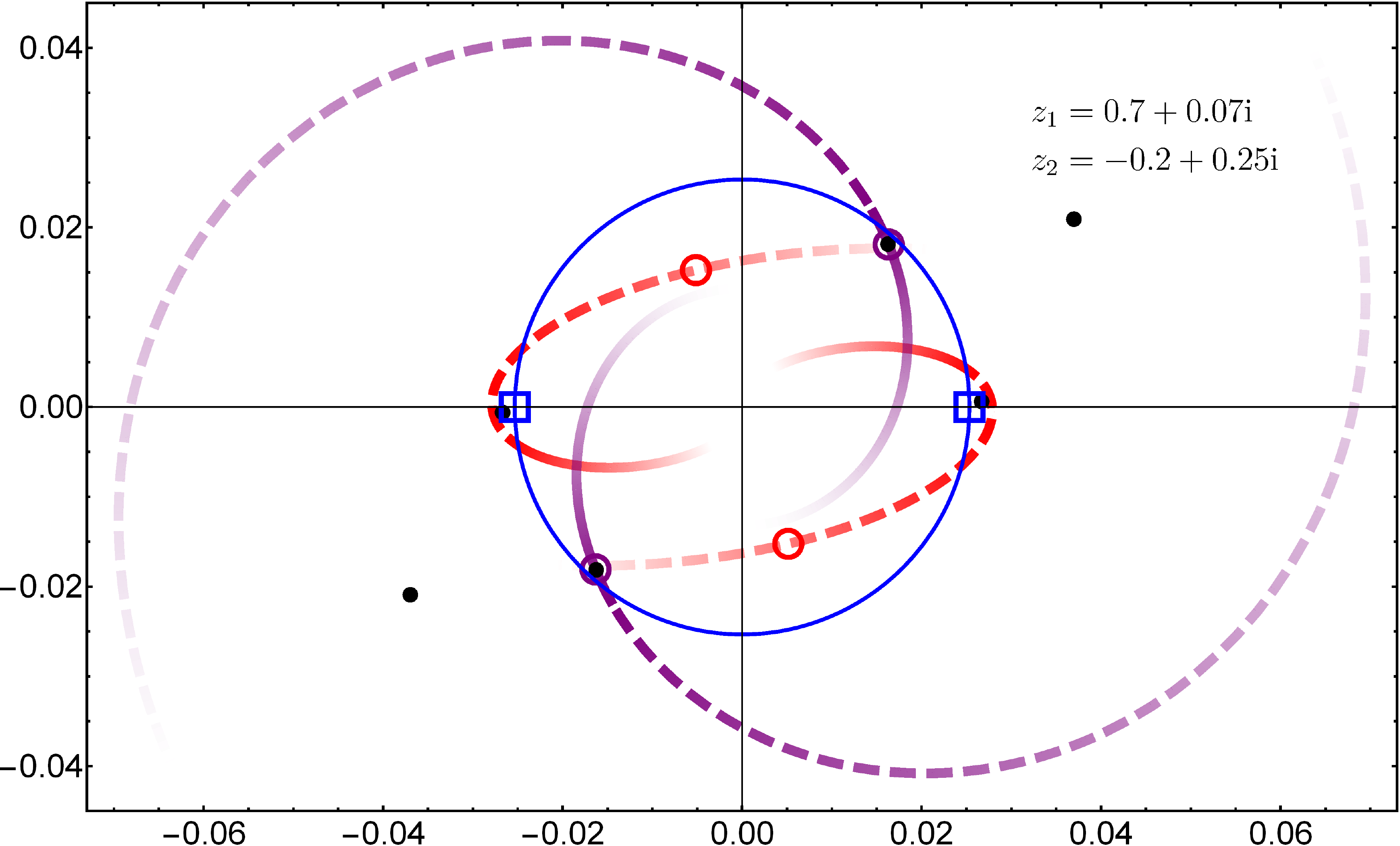}
     \vspace{1mm}
     \end{subfigure}
     \begin{subfigure}[h]{0.75\textwidth}
         \centering
         \includegraphics[width=\textwidth]{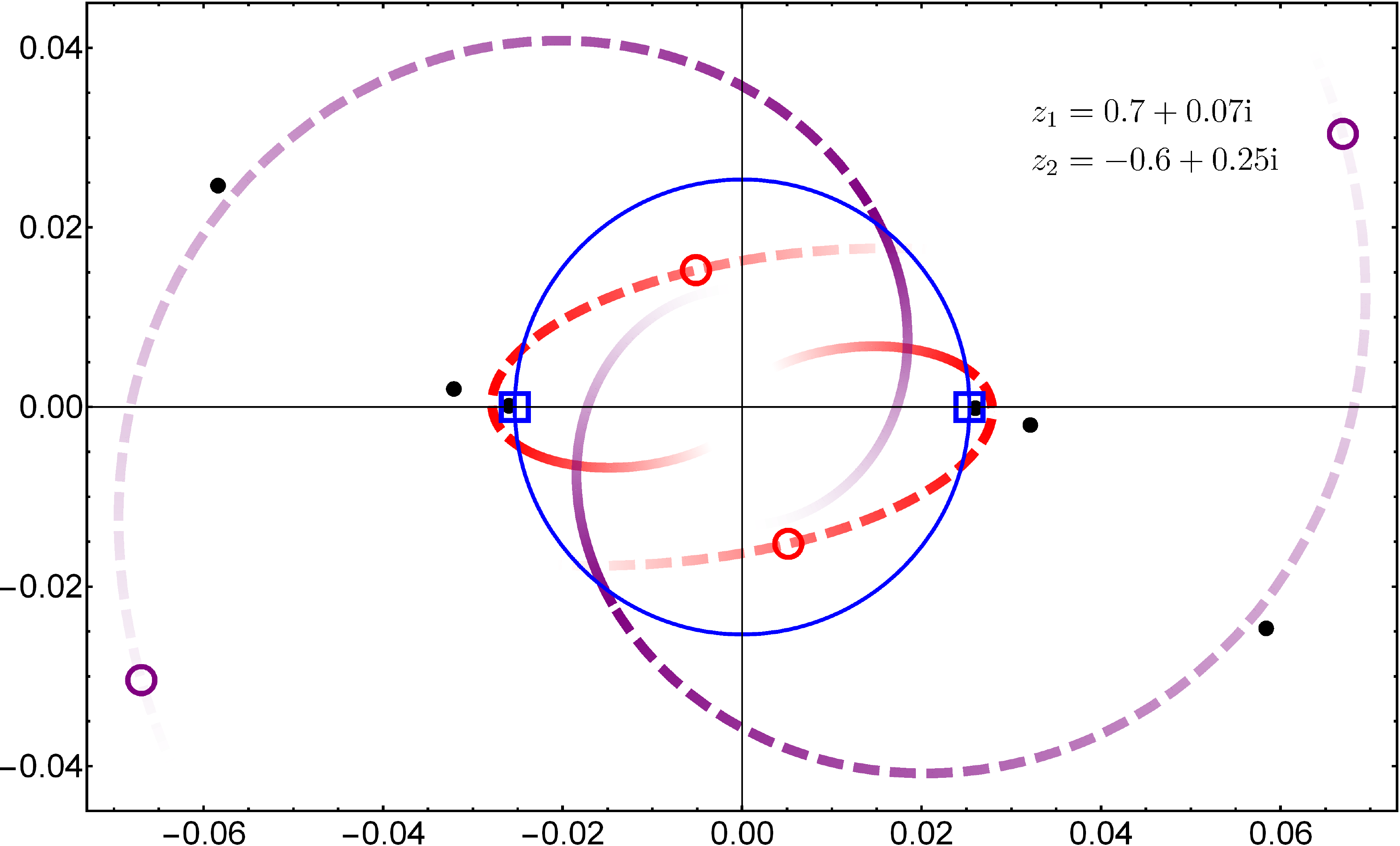}
     \end{subfigure}
     \caption{Borel--Pad\'e analysis of the Borel singularity structure of the perturbative two-point function $\widehat{W}_2^{(0)} (z_1,z_2)$, for two different values of its arguments, $\left(z_1,z_2\right)$ (continued from figure~\ref{fig:Pade-2}). In the first plot, the first resonant pair of FZZT actions has left the principal-sheet of the Borel plane. The second resonant pair remains inside the principal sheet, but now with an absolute value which is close to that of the ZZ actions---hence the Pad\'e poles (the black dots) attach themselves both to the second pair of FZZT actions and to the ZZ actions (hence both roughly equally contributing to the large-order behavior). In the second plot, all FZZT actions have left the principal sheet. Consequently, the Pad\'e poles now accumulate solely close to the ZZ actions.}
\label{fig:Pade-3}
\end{figure}

As we move to the discussion on approximate Borel transforms, it is clear it will follow a completely analogous reasoning to the previous subsection. As made clear for the one-point function in \eqref{eq:int-rep-W1}, but which is generic for all higher-point correlators, one of the key advantages of the nonperturbative topological recursion is that it provides an integral representation for all the $\widehat{W}_{n}^{(1)}$. In the $n=2$ case one now uses \eqref{eq:W2-int} which, upon performing the same steps as we did in the one-point case leading up to \eqref{eq:W11-approx-Borel-resum}, results in
\be
\widehat{W}^{(1)}_2 \left(z_1,z_2; g_{\text{s}} \right) \approx \frac{1}{2 \pi \rmi} \int_{\widetilde{\mathcal{I}}} \dd s\, \textcolor{blue}{\frac{1}{V_{\text{eff}}' \left( x (s) \right)}}\, \textcolor{red}{\frac{1}{x (s)-z_1^{2}}}\, \textcolor{violet}{\frac{1}{x (s)-z_2^{2}}}\, \rme^{-\frac{s}{g_{\text{s}}}}.
\ee
\noindent
Again, the choice of perturbative integration contour leads us to claim that the above integrand may be interpreted as the approximate Borel transform for $\widehat{W}_2^{(0)} \left(z_1,z_2; g_{\text{s}}\right)$. It has exactly the singularity structure that we expect, with poles at the three instanton-action locations which contribute in \eqref{eq:W2-zz-fzzt}. This very same structure emerges in a surprisingly clear way from a Borel--Pad\'e approximant analysis, as shown in figures~\ref{fig:Pade-2} and~\ref{fig:Pade-3}. It is clear in the plots how FZZT singularities may leave the principal-sheet of the Borel plane in turns, hence leading to a more complex competition for large-order dominance. As for the one-point function, these figures clearly illustrate the resonance features of both ZZ and FZZT nonperturbative corrections in the context of JT gravity (further discussed in \cite{gs21, sst23}).

\begin{figure}
\centering
     \begin{subfigure}[h]{0.49\textwidth}
         \includegraphics[width=\textwidth]{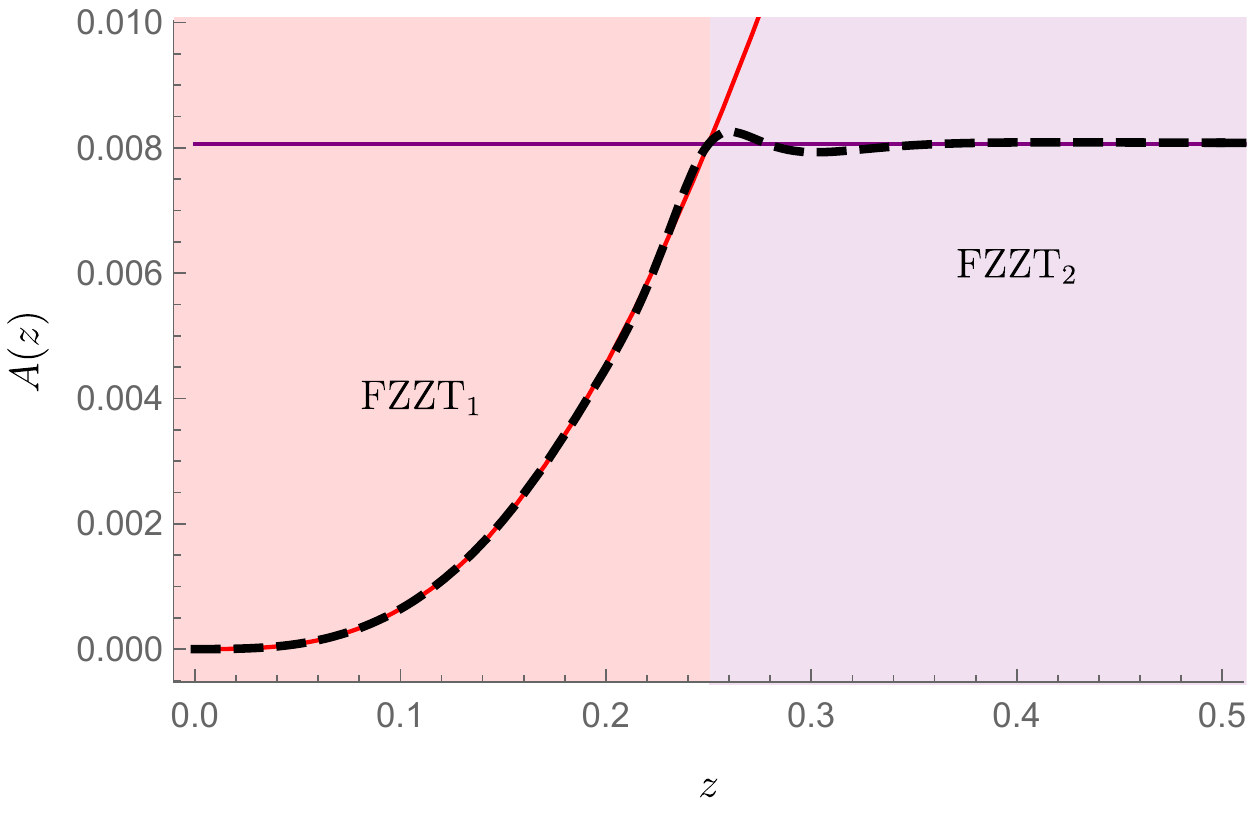}
     \end{subfigure}
     \begin{subfigure}[h]{0.49\textwidth}
         \includegraphics[width=\textwidth]{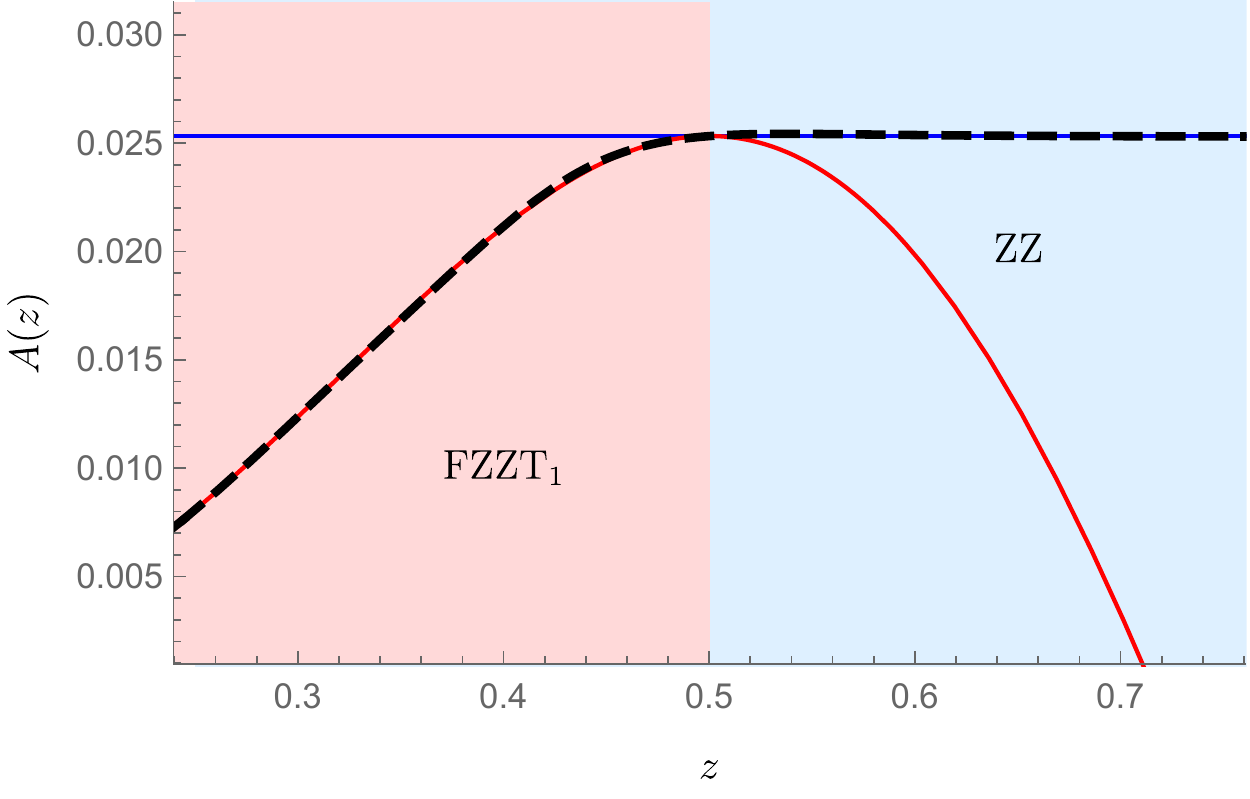}
     \end{subfigure}
\caption{The black-dashed lines are the fourth Richardson transforms of the instanton action sequences via the inverse square root of \eqref{eq:Qg} for $\widehat{W}_2 (z,0.25)$ (left) and $\widehat{W}_2 (z,0.75)$ (right), as a function of $z_1 \equiv z$. In red, purple, and blue, the predicted values associated for the FZZT$_1$, FZZT$_2$ and ZZ contributions, respectively.}
\vspace{3mm}
\label{fig:2point-act}
\end{figure}

\begin{figure}
\centering
     \begin{subfigure}[h]{0.49\textwidth}
         \includegraphics[width=\textwidth]{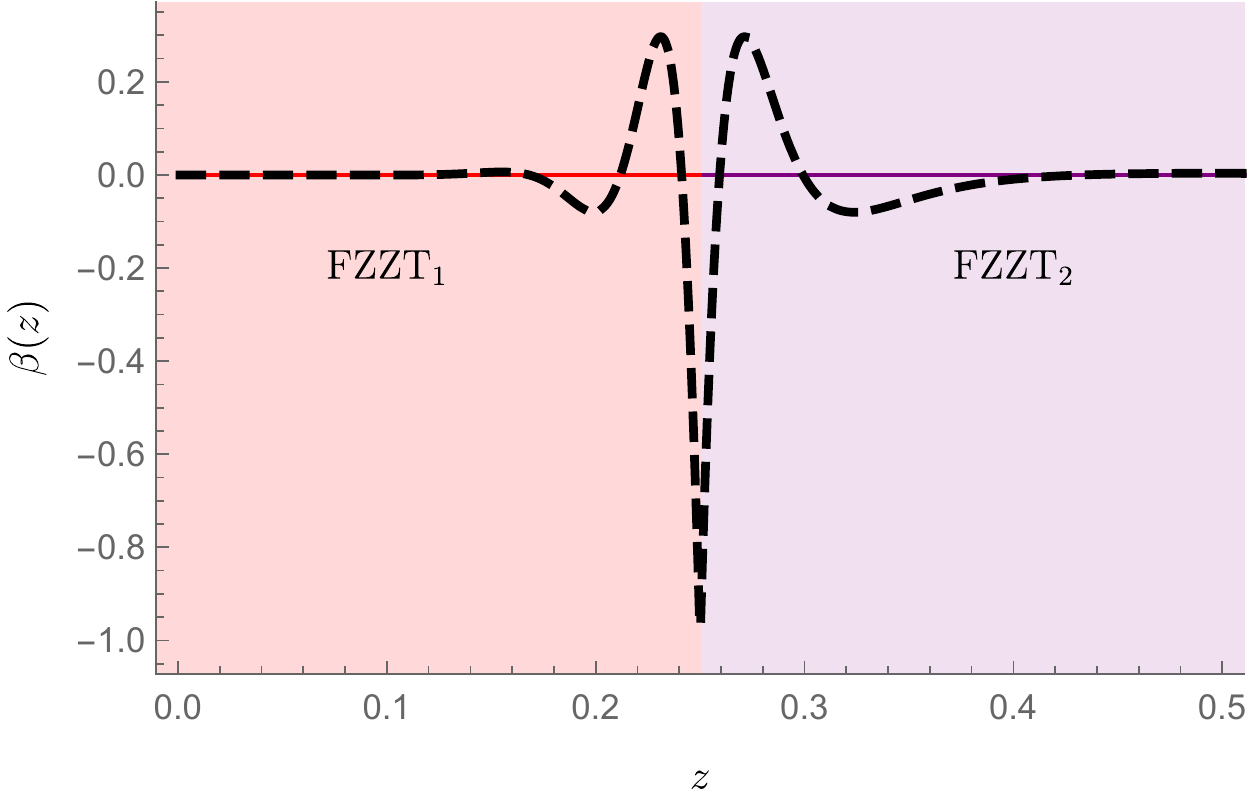}
     \end{subfigure}
     \begin{subfigure}[h]{0.49\textwidth}
         \includegraphics[width=\textwidth]{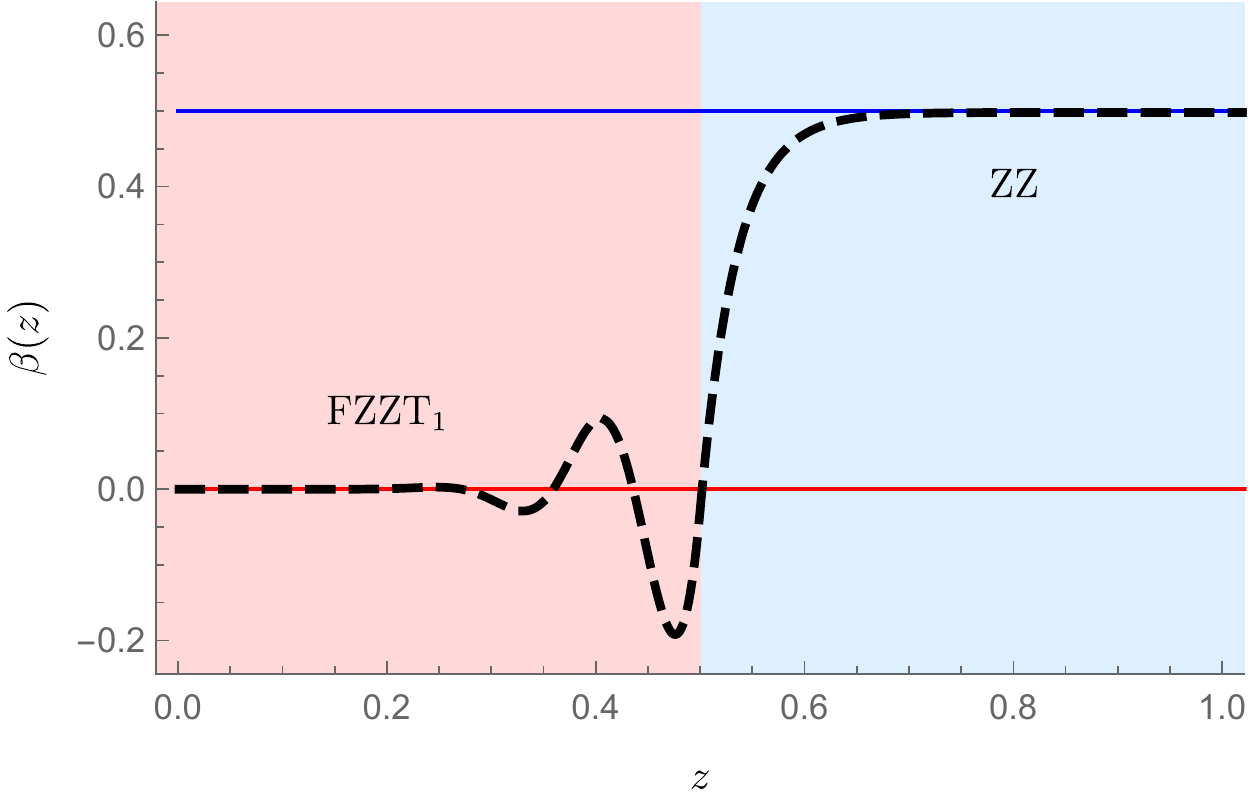}
     \end{subfigure}
\caption{The black-dashed lines are the fourth Richardson transforms of the characteristic exponent \eqref{eq:Bg} sequences for $\widehat{W}_2 (z,0.25)$ (left) and $\widehat{W}_2 (z,0.75)$ (right), as a function of $z_1 \equiv z$.}
\label{fig:2point-beta}
\end{figure}

\begin{figure}
\centering
     \begin{subfigure}[h]{0.49\textwidth}
         \includegraphics[width=\textwidth]{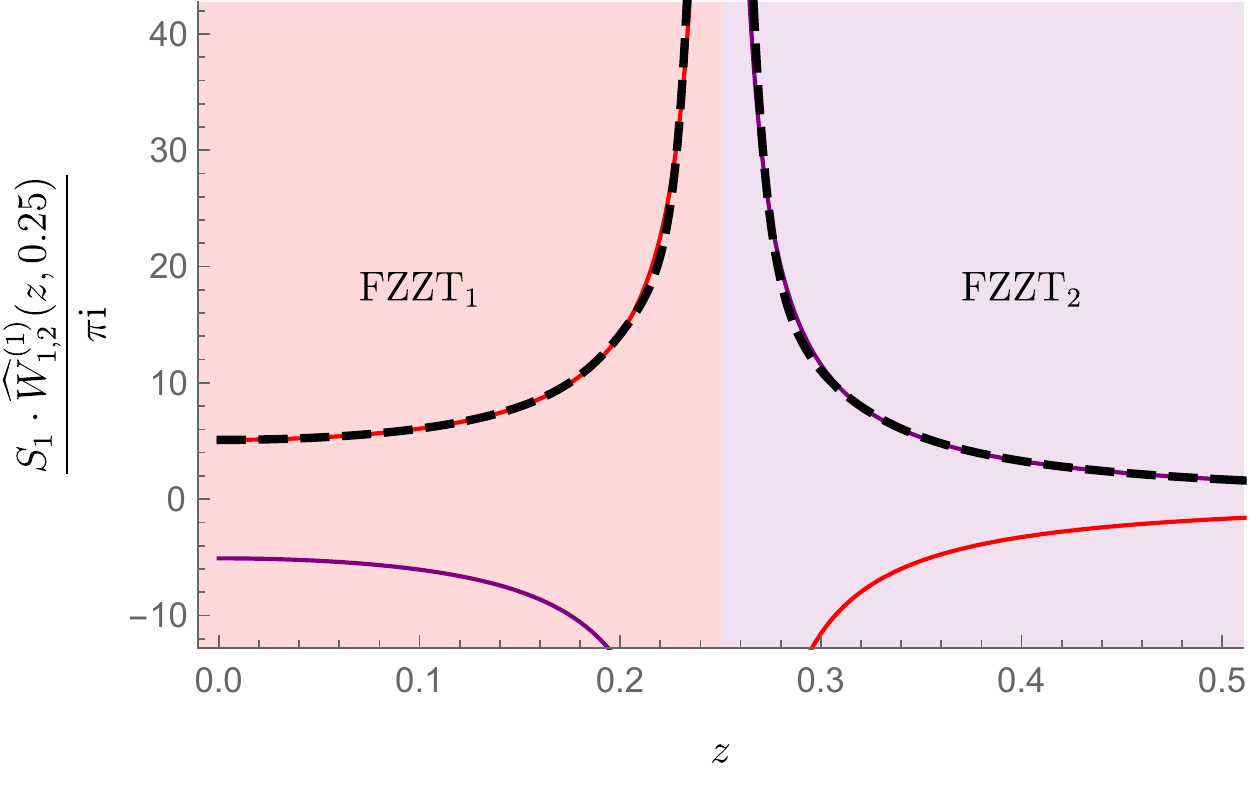}
     \end{subfigure}
     \begin{subfigure}[h]{0.49\textwidth}
         \includegraphics[width=\textwidth]{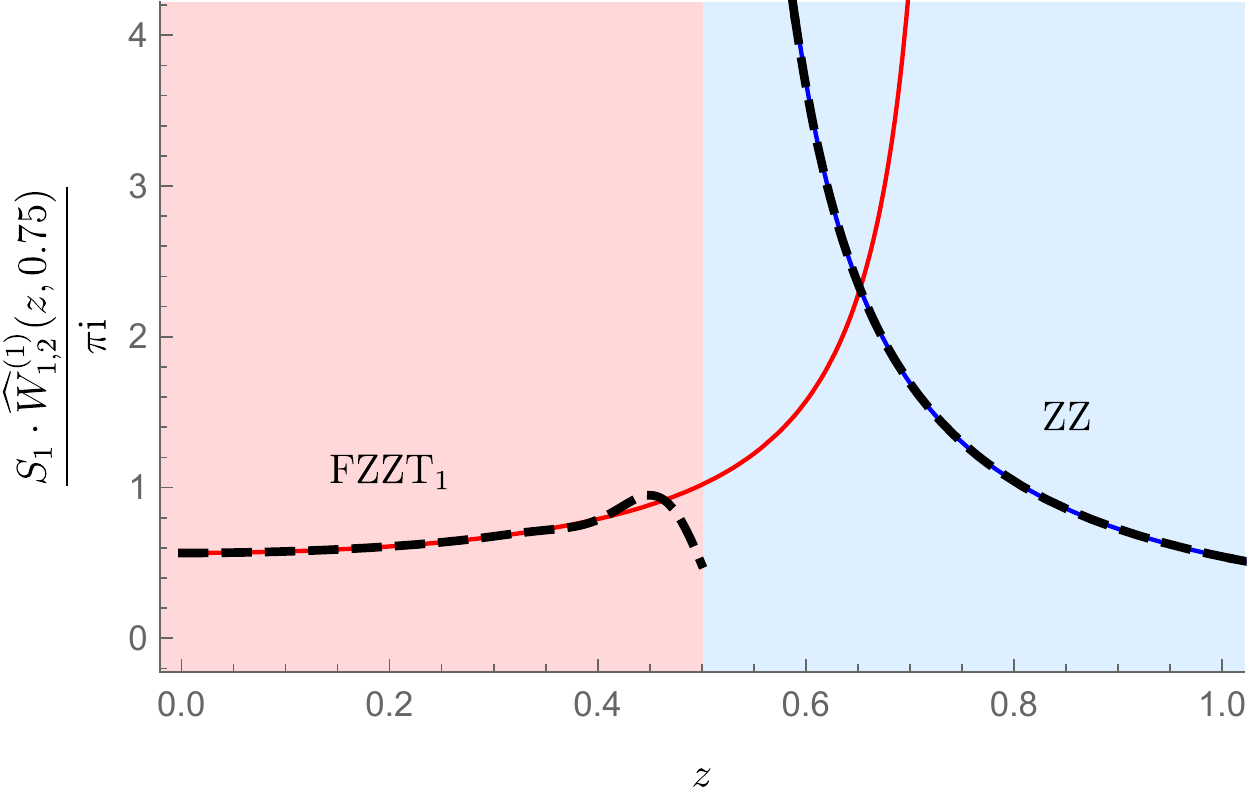}
     \end{subfigure}
\caption{The black-dashed lines are the fourth Richardson transforms of the \eqref{eq:M1g} sequences, determining the one-loop contribution around the one-instanton sector, for $\widehat{W}_2 (z,0.25)$ (left) and $\widehat{W}_2 (z,0.75)$ (right), as a function of varying $z_1 \equiv z$.}
\label{fig:2point-one}
\end{figure}

\begin{figure}
\centering
     \begin{subfigure}[h]{0.49\textwidth}
         \includegraphics[width=\textwidth]{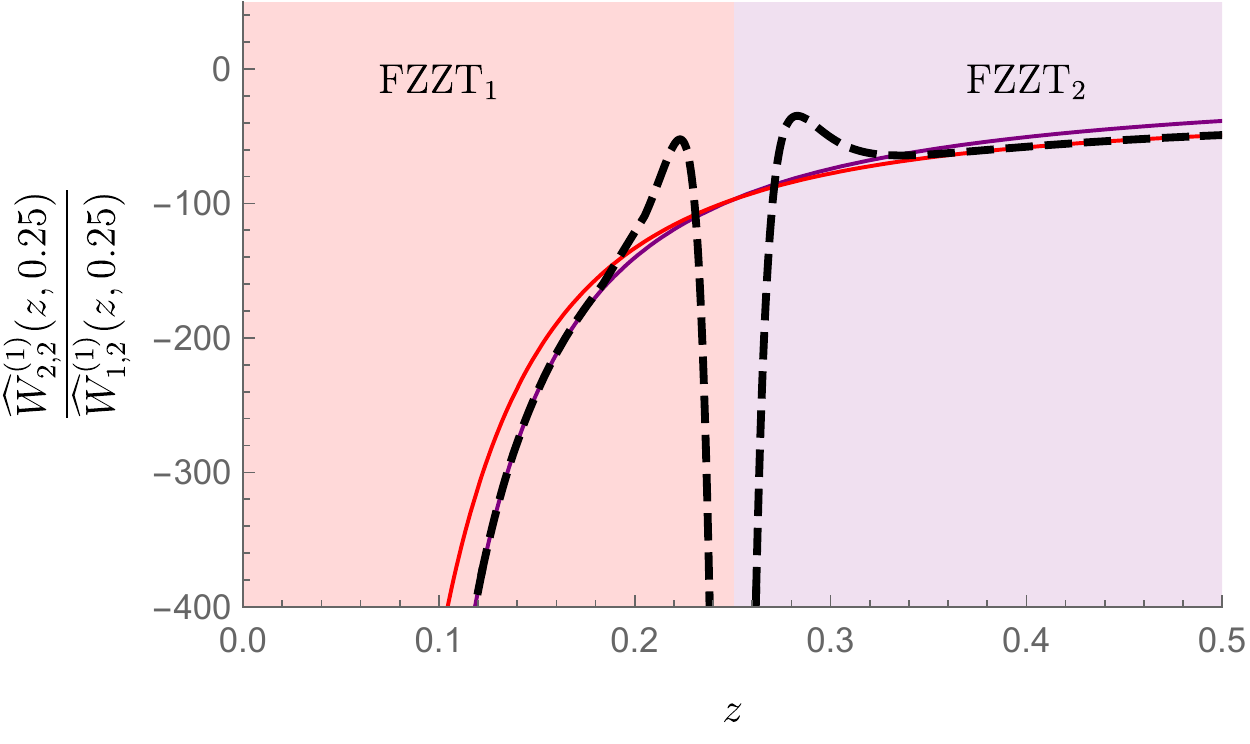}
     \end{subfigure}
     \begin{subfigure}[h]{0.49\textwidth}
         \includegraphics[width=\textwidth]{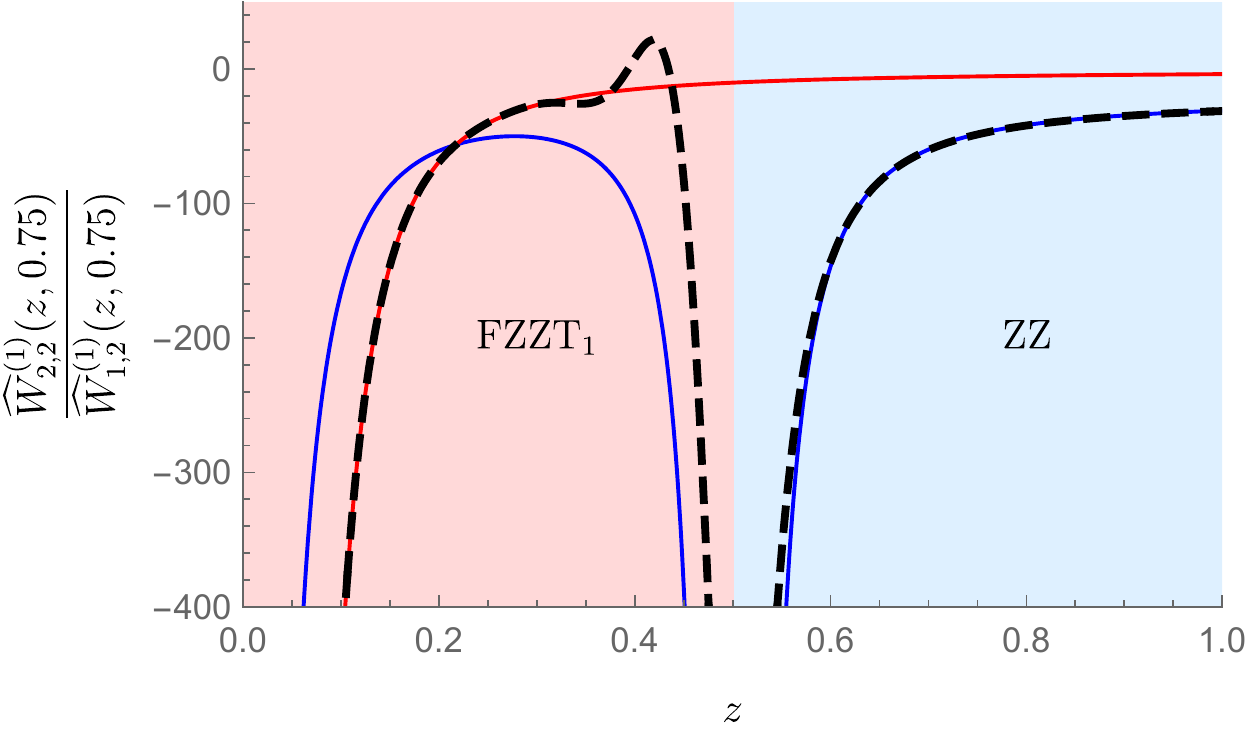}
     \end{subfigure}
\caption{The black-dashed lines are the first (left) and fifth (right) Richardson transforms of the \eqref{eq:M2g} sequences, determining the two-loop contribution around the one-instanton sector, for $\widehat{W}_2 (z,0.25)$ (left) and $\widehat{W}_2 (z,0.75)$ (right), as a function of $z_1 \equiv z$.}
\label{fig:2point-two}
\end{figure}

Large-order asymptotic analyses of the perturbative sequence $\widehat{W}_{g,2}^{(0)} \left( z_1,z_2 \right)$ associated to the two-point function are depicted in figures~\ref{fig:2point-act}, \ref{fig:2point-beta}, \ref{fig:2point-one}, and~\ref{fig:2point-two}. They were carried out at two different fixed values of $z_2$, namely $1/4$ and $3/4$, and for varying $z_1$ argument. When $z_2=1/4$ we observe the two FZZT contributions competing against each other for the leading asymptotics, with the interchange of dominance happening as soon as $z_1>1/4$. When $z_2=3/4$, having left the principal-sheet of the Borel plane the second FZZT contribution is out of the picture. As such, we just observe the remaining competition between the first FZZT contribution and the ZZ contribution, much like in the one-point function large-order behavior---\textit{e.g.}, compare the right-hand plots of the present figures~\ref{fig:2point-act}, \ref{fig:2point-beta}, \ref{fig:2point-one}, and~\ref{fig:2point-two} with the one-point figures~\ref{fig:1point-act}, \ref{fig:1point-one}, and~\ref{fig:1point-two-and-three}.

\section{Resurgent Asymptotics of Weil--Petersson Volumes}
\label{sec:resurgence-volumes}

Having understood the role of ZZ and FZZT nonperturbative corrections in the large-order asymptotic behavior of multi-resolvent correlation functions, we may now come full circle and return to Weil--Petersson volumes, at the basis of the JT gravity construction in \cite{sss19} as described in subsections~\ref{subsec:WPvolumes} and~\ref{subsec:classicalTR}. After all, Laplace transform \cite{eo07b} of the results in the previous section is bound to yield Weil--Petersson results; recall \eqref{eq:fromWtoV} or \eqref{eq:fromVtoW}
\be
\widehat{W}_{g,n} \left( \boldsymbol{z} \right) = \left( \prod_{i=1}^n \NCL_{z_i} \right) \cdot V_{g,n} \left( \boldsymbol{b} \right) \qquad \longleftrightarrow \qquad V_{g,n} \left( \boldsymbol{b} \right) = \left( \prod_{i=1}^n \NCL^{-1}_{b_i} \right) \cdot \widehat{W}_{g,n} \left( \boldsymbol{z} \right).
\ee
\noindent
This is obviously true generically, \textit{i.e.}, given the multi-resolvent correlation functions immediately yields not only the above Weil--Petersson volumes but also \cite{sss19} JT partition-function correlators\footnote{With the integration prescription that the singularity at the origin is avoided to the right (at positive $x_k$).} (via \eqref{eq:PFtoWP})
\be
Z_{g,n} \left( \boldsymbol{\beta} \right) = \frac{\rmi}{4\pi} \prod_{k=1}^n \int_{-\rmi \infty}^{+\rmi \infty} \mathrm{d}x_k\, \frac{\mathrm{e}^{\beta_k x_k}}{\sqrt{x_k}}\, \widehat{W}_{g,n} \left(\sqrt{x_1},\ldots,\sqrt{x_n}\right),
\ee
\noindent
or correlation functions of the spectral density $\rho (E)$ via further Laplace transform as in \eqref{eq:rho-from-Z-LAP}. Herein we focus solely on Weil--Petersson volumes (extension to other observables is straightforward). In particular, the resurgent ZZ/FZZT brane picture we unfolded in the previous section implies that also the large-genus asymptotics of Weil--Petersson volumes will be dictated by two distinct, possibly competing classes of nonperturbative effects. This procedure should in fact be generic towards obtaining the complete large-genus asymptotics of any $V_{g,n} (\boldsymbol{b})$. In parallel with what we did in the previous section~\ref{sec:resurgence-correlators}, let us first see how they emerge in asymptotic formulae, and then test them against explicit large-order behavior.

\subsection{Large Genus Asymptotics of Weil--Petersson Volumes}
\label{subsec:largegenusWP-formulae}

Soon after Mirzakhani found the recursion relation for Weil--Petersson volumes \cite{m07a, m07b}, a study of their large-genus\footnote{See as well the earlier \cite{p92}, focusing on leading-growth asymptotic estimates. Further, it is worth mentioning that besides their large-genus asymptotics, it has also been of interest in the mathematical literature to study the asymptotic growth of Weil--Petersson volumes on the number of boundaries, \textit{e.g.}, \cite{mz99}. It would be interesting to understand if our present analysis may have any extensions along this direction.} asymptotics was also initiated in several papers \cite{z08, m10, mz11}.  One of their main results was the asymptotic expression for the $V_{g,n}$ volumes \cite{mz11}
\be
\label{eq:mirzog-asympt}
V_{g,n} \simeq \frac{\Gamma \left( 2g+n-2 \right)}{\left( \frac{1}{4\pi^2} \right)^{2g+n-3}}\, \frac{c_n^{(0)}}{\sqrt{g}} \left( 1 + \frac{c_n^{(1)}}{g} + \frac{c_n^{(2)}}{g^2} + \cdots  \right),
\ee
\noindent
where $c_n^{(0)}$ was conjectured to be $\frac{1}{\sqrt{\pi}}$, and where the $c_n^{(k)}$ were shown to be polynomials in $n$ of degree $\deg c_n^{(k)} = 2k$, whose order $n^{2k}$ coefficient is $\frac{(-1)^k}{k! \left( 2\pi^2 \right)^k}$. For example, it was explicitly computed \cite{mz11}
\be
\label{eq:mirzog-asympt-coef}
c_n^{(1)} = - \frac{n^2}{2\pi^2} + \left( \frac{5}{2\pi^2} - \frac{1}{4}\right) n - \frac{17}{6\pi^2} + \frac{7}{12}.
\ee
\noindent
Further considering the geodesic boundaries and hence addressing the $V_{g,n} (b_1,\ldots,b_n)$ volumes was later pursued in \cite{mp17}, which proved\footnote{In the specific case of $n=1$ this had been previously conjectured in unpublished work of 2010 by Norman Do (private communications). In the physics literature, this case of $n=1$ was also explicitly computed in \cite{sss19} (up to the one-loop contribution and via a matrix integral approach). Starting from this result in \cite{sss19}, expression \eqref{eq:kimura-asympt} was also conjectured in \cite{k20,k21}.} the leading asymptotics (again with the same conjectural overall constant $c_n^{(0)}$)
\be
\label{eq:kimura-asympt}
V_{g,n} \left( b_1, \ldots, b_n \right) \simeq \frac{\Gamma \left( 2g+n-\frac{5}{2} \right)}{\left( \frac{1}{4\pi^2} \right)^{2g+n-3}}\, c_n^{(0)} \sqrt{2}\, \prod_{k=1}^{n} \CS \left( b_k \right) + \cdots.
\ee
\noindent
Let us also mention that the structure of the large-genus asymptotics of Weil--Petersson volumes was thoroughly studied in \cite{am20}, in terms of sums of products of hyperbolic functions in the $b_i$ variables, times certain polynomials in the $b_i$ (whose existence was also proved). This work further provided an explicit expression for the first subleading correction\footnote{Since this work focuses on the ratio $V_{g,n} \left( b_1, \ldots, b_n \right) / V_{g,n}$, it could also not capture the overall constant $c_n^{(0)}$.} to equation \eqref{eq:kimura-asympt}, which we also reproduce with our methods. In fact, as should be clear, we are now expecting that our present resurgence analysis reproduces these results alongside many others, eventually including all higher ZZ/FZZT instanton corrections. Our starting point will be the generic\footnote{By generic we imply one should think of the obvious generalization of \eqref{eq:largeorder} to $n$-point functions.} multi-resolvent large-order relation \eqref{eq:largeorder}, out from which we will obtain the large-genus asymptotics of Weil--Petersson volumes using the ZZ/FZZT nonperturbative data computed in the previous sections. In doing this, we make the plausible assumption that the generating series of Weil--Petersson volumes is a resurgent asymptotic series. To sum up, the novelty of our result are the nonperturbative techniques we use, which allow us to provide a proof for the value of the overall constant $c_n^{(0)}$ (up to the aforementioned natural assumption of resurgence). Moreover, we give a universal interpretation to the $\CS \left (b_k \right)$ factors of equation \eqref{eq:kimura-asympt}, since they appear as the anti-Laplace transform of the Bergman kernel---which plays a universal role for many different problems that are governed by the topological recursion.

We start by focusing on the ZZ contributions and consider the first (blue) line of equation \eqref{eq:largeorder}. Translating that result to Weil--Petersson volumes $V_{g,n}(b_1,\ldots,b_n)$, plugging-in the ZZ nonperturbative data computed in section~\ref{sec:NPtoprec}, and using the familiar $\beta=\beta^{(1)}-\beta^{(0)}=\frac{5}{2}-n$ from section~\ref{sec:resurgence-correlators}, one immediately obtains the large-genus asymptotics
\bea
\label{eq:largeorderV-ZZ}
V_{g,n} (b_1,\ldots,b_n) &\simeq& \frac{1}{\sqrt{2}\, \pi^\frac{3}{2}} \left(4\pi^2\right)^{2g + n - \frac{5}{2}}\, \Gamma \left( 2g + n - \frac{5}{2} \right) \times \\
&&
\times \left\{ \widetilde{{V}}_{1,n}^{(1)} \left( b_{1}, \ldots, b_n \right) + \frac{1}{4\pi^2}\, \frac{1}{2g+n-\frac{7}{2}}\, \widetilde{V}_{2,n}^{(1)} \left( b_{1}, \ldots, b_{n} \right) + \cdots\right\}. \nonumber
\eea
\noindent
The nonperturbative loop corrections are the ones in section~\ref{sec:NPtoprec}, namely $\widetilde{{V}}_{1,n}^{(1)} (b_{1}, \ldots, b_n)$ in \eqref{eq:sinhLn} and $\widetilde{V}_{2,n}^{(1)} (b_{1}, \ldots, b_{n})$ in \eqref{eq:WP2loop}. At leading order, this reproduces \eqref{eq:kimura-asympt}; but now on top of having \textit{derived} the generic-$n$ Weil--Petersson large-genus asymptotics, our results explicitly produce (in principle) \textit{all} higher-loop subleading\footnote{The \textit{first} subleading correction to the asymptotics was computed in \cite{os19} for $n=1$, and generically in \cite{am20}.} corrections alongside \textit{generic} Stokes data. In contrast with the Weil--Petersson volumes themselves, which are polynomials in the $b_i$, these nonperturbative corrections now also involve hyperbolic functions. As already explained, albeit our method recursively computes arbitrarily higher loops, obtaining more and more precise asymptotics, such formulae quickly become too cumbersome to display---hence why we truncated the series at two-loops. We will next go to higher loops focusing on the Weil--Petersson volumes with $\boldsymbol{b} = \boldsymbol{0}$. But let us first point out that, at this stage, one could suspect that all asymptotic formulae based on \eqref{eq:largeorderV-ZZ} are seemingly \textit{incomplete}, with this asymptotic behavior only being valid for sufficiently small $b_i$. Or, in the words of section~\ref{sec:resurgence-correlators}, with this asymptotic behavior only being valid when ZZ-branes dominate the asymptotics. As we shall see below, this is not exactly the case.

Next, let us use \eqref{eq:largeorderV-ZZ} in order to make contact with the asymptotics of $V_{g,n}$ Weil--Petersson volumes \eqref{eq:mirzog-asympt} obtained in \cite{mz11}, which is done by simply setting all the $b_i$ to zero. At the level of pure ZZ-asymptotics \eqref{eq:largeorderV-ZZ} it immediately follows
\bea
\label{eq:largeorderV2}
V_{g,n} &\simeq& \frac{1}{\sqrt{2}\, \pi^\frac{3}{2}} \left(4\pi^2\right)^{2g + n - \frac{5}{2}}\, \Gamma \left( 2g + n - \frac{5}{2} \right) \times \\
&&
\times \left\{ 1 + \frac{1}{4\pi^2}\, \frac{1}{2g+n-\frac{7}{2}}\, \widetilde{V}_{2,n}^{(1)} + \left(\frac{1}{4\pi^2}\right)^2 \frac{1}{2g+n-\frac{7}{2}}\, \frac{1}{2g+n-\frac{9}{2}}\, \widetilde{V}^{(1)}_{3,n} + \cdots \right\},
\nonumber
\eea
\noindent
with $\widetilde{V}_{2,n}^{(1)}$ computed in \eqref{eq:V12n} and $\widetilde{V}^{(1)}_{3,n}$ given by \eqref{eq:V13n}. We added the three-loop contribution to the asymptotics in order to show how one can systematically compute higher-loop corrections. If we now use
\be
\frac{\Gamma \left( 2g+n-2 \right)}{\sqrt{2g}\, \Gamma \left( 2g+n-\frac{5}{2} \right)} = 1 + \frac{4 n - 11}{16g} - \frac{16 n^2 - 88 n + 119}{512 g^2} + \cdots
\ee
\noindent
it immediately follows
\be
\frac{\Gamma \left( 2g+n-2 \right)}{\sqrt{2g}\, \Gamma \left( 2g+n-\frac{5}{2} \right)} \left( 1 + \frac{c_n^{(1)}}{g} + \cdots \right) \simeq 1 + \frac{1}{4\pi^2}\, \frac{1}{2g+n-\frac{7}{2}}\, \widetilde{V}_{2,n}^{(1)} + \cdots,
\ee
\noindent
which explicitly checks upon expansion that our \eqref{eq:largeorderV2} reproduces both \eqref{eq:mirzog-asympt} and \eqref{eq:mirzog-asympt-coef} up to two-loops. Taking this reasoning one step further, it is straightforward to obtain an expression for $c_n^{(2)}$ in \eqref{eq:mirzog-asympt}, as:
\bea
c_n^{(2)} &=& \frac{1}{576 \pi^4} \left( 72 n^4 - 1152 n^3 + 6504 n^2 - 15168 n + 12104 \right) + \\
&&
+ \frac{1}{576 \pi^2} \left( 204 n^3 - 1608 n^2 + 4032 n - 3160 \right) + \frac{1}{576} \left( 54 n^2 - 261 n + 290 \right). \nonumber
\eea
\noindent
The degree in $n$ of this coefficient is indeed $4 = 2 \times 2$ with leading coefficient $\frac{72}{576 \pi^4} = \frac{1}{8 \pi^4} = \frac{(-1)^2}{2! \left( 2\pi^2 \right)^2}$ as shown in \cite{mz11}. On top of this, we have now further computed Stokes data analytically, rather than numerically conjectured as in \cite{mz11}. It should also be noted that the addition of the above three-loop contribution constitutes an improvement with respect to the large-genus asymptotics of \cite{mz11}, and that we can systematically compute any desired higher-loop corrections.

Let us finally recall the discussion in section~\ref{sec:resurgence-correlators} where, for the $\widehat{W}_{n}$ correlators, we observed a competition between ZZ and FZZT effects in the large-genus asymptotics, depending on the values of the $z_i$. This has now (partially) translated to the above ZZ-dominance regime \eqref{eq:largeorderV-ZZ} occurring when the geodesic boundaries are small. But it further implies that, as we vary the values of the $b_i$, we may eventually find a crossover regime with both ZZ and FZZT contributions to the large-genus asymptotics of Weil--Petersson volumes. Then, at large boundary lengths, \eqref{eq:largeorderV-ZZ} could be replaced by an FZZT-dominance regime, which would be obtained by applying the inverse Laplace transform \eqref{eq:fromVtoW} to the second (red) line of \eqref{eq:largeorder}. Let us understand what exactly happens. Unfortunately, unlike in the ZZ case, at the current stage we can not yet write generic-$n$ FZZT formulae for the large-genus asymptotics of $V_{g,n} (b_1,\ldots,b_n)$. Instead, our present FZZT analysis can only be done for $n=1$ via \eqref{eq:W1-fzzt}, or $n=2$ via \eqref{eq:W2-fzzt-a}, and is further hampered by the inability to perform the Laplace transform analytically on the (red/purple) FZZT contributions for either \eqref{eq:W1asymp} or \eqref{eq:W2asymp}. One may however expand the JT FZZT contributions around their Airy limit\footnote{Let us remark that in this limit the Weil--Petersson spectral curve becomes the Airy spectral curve which generates Witten--Kontsevich intersection numbers \cite{w91, k92}. The large-genus asymptotics of these intersection numbers have been studied in \cite{a20}. Note that the techniques developed in the present work do not immediately apply to the Airy curve, due to the lack of non-trivial (nonperturbative) saddle points. In \cite{egggl23}, this asymptotic regime will be addressed for several enumerative problems using a combination of topological recursion and resurgence.} (where they are exactly solvable) and Laplace transform order by order.

Begin with the FZZT contribution to the large-order behavior of the perturbative one-point function \eqref{eq:W1-pert-exp}, via the second (red) line of \eqref{eq:W1asymp}. Working around the Airy limit \cite{egggl23}, \textit{i.e.}, focusing on the first few leading contributions as $z \to 0$ in the second (red) line of \eqref{eq:W1asymp} amounts to considering $V_{\text{eff}} (z^{2}) \approx \frac{2}{3} z^3 + \cdots$. In this approximation, the inverse Laplace transform \eqref{eq:antiLapl} may be evaluated analytically to yield the first few FZZT contributions to Weil--Petersson large-genus asymptotics as
\be
\label{eq:largeorderV-FZZT-1}
V_{g,1} (b) \simeq \frac{1}{2\pi} \left(\frac{3}{2}\right)^{2g-1} \frac{\Gamma \left(2g-1\right)}{\Gamma \left(6g-2\right)}\, b^{6g-4} \left\{ 1 - \frac{17}{18}\, \frac{1}{2g-2} + \frac{1225}{648}\, \frac{1}{\left(2g-2\right) \left(2g-3\right)} + \cdots \right\}.
\ee
\noindent
In particular, we obtain the first JT perturbation around the Airy limit also addressed in \cite{sss19}, now alongside a few subleading corrections. A completely analogous calculation yields the FZZT contribution to the large-order behavior of the perturbative two-point function, now via the second (red) and third (purple) lines of \eqref{eq:W2asymp}. Again working around the Airy limit \cite{egggl23}, and performing the inverse Laplace transform \eqref{eq:antiLapl} analytically, now yields new FZZT contributions to Weil--Petersson large-genus asymptotics as
\bea
\label{eq:largeorderV-FZZT-2}
V_{g,2} (b_1,b_2) &\simeq& \frac{1}{2\pi} \left( \frac{3}{2} \right)^{2g} \frac{\Gamma \left(2g\right)}{\Gamma \left(6g+1\right)} \left\{ \frac{\left(b_1+b_2\right)^{6g} - \left(b_1-b_2\right)^{6g}}{b_1 b_2} - \right. \\
&&
\left. - \frac{4}{3}\, \frac{1}{2g-1} \left( \frac{17}{24}\, \frac{\left(b_1+b_2\right)^{6g} - \left(b_1-b_2\right)^{6g}}{b_1 b_2} - 6g \left( b_1^{6g-2} + b_2^{6g-2} \right) \right) +\cdots \right\}. \nonumber
\eea
\noindent
Interestingly enough, the Weil--Petersson FZZT-dominated asymptotics are no longer of the standard resurgent type as in, \textit{e.g.}, \eqref{eq:largeorder}---in particular they are now factorially \textit{suppressed} rather than factorially \textit{enhanced}, hence extremely subtle to scrutinize from under the large-order asymptotics. This follows directly from the above inverse Laplace transforms: in the Airy limit of either \eqref{eq:W1asymp} or \eqref{eq:W2asymp} the FZZT action is $\propto z^3$, implying one has to inverse-Laplace transform a term $\propto z^{-6g}$, which then yields the the factorial suppression $\propto 1 / \Gamma \left(6g\right)$ explicitly appearing in \eqref{eq:largeorderV-FZZT-1} and \eqref{eq:largeorderV-FZZT-2}. In other words, except in the strict large boundary limit, the Weil--Petersson asymptotics is actually never purely FZZT dominated. As already explained in \cite{sss19}, this factorial suppression can only be overcome in the $b_i \gg g$ regime; and in the large-genus asymptotics, for all finite (even possibly very large) boundary lengths, one never really finds clear FZZT effects, with the \eqref{eq:largeorderV-ZZ} asymptotics still remaining valid. These effects are only manifest when in the $b_i \gg g$ regime, which corresponds to the Airy limit. As a final comment, it would be very interesting to extend the present FZZT calculation to arbitrary $n$, possibly building upon \cite{os21}.

\subsection{Large-Order Asymptotic Analysis of Weil--Petersson Volumes}
\label{subsec:largegenusWP-tests}

\begin{figure}
\centering
     \begin{subfigure}[h]{0.48\textwidth}
         \centering
         \includegraphics[width=\textwidth]{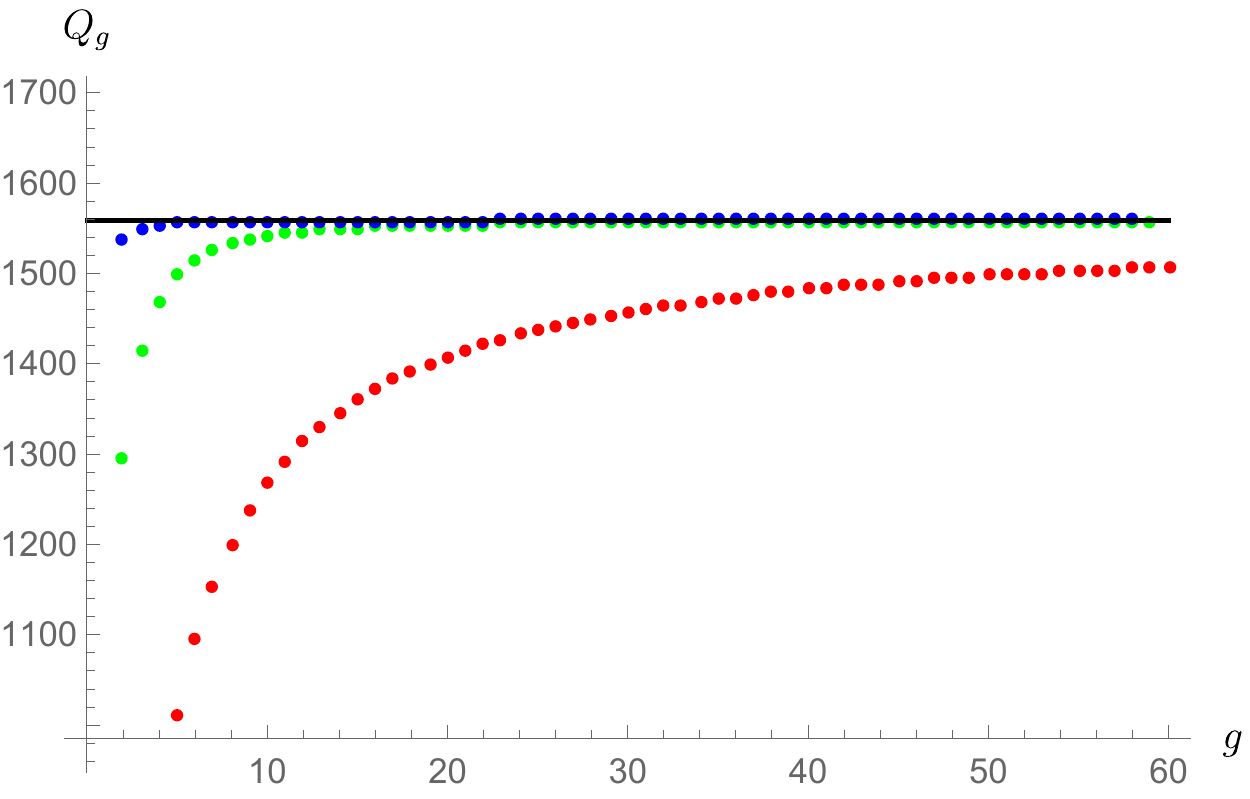}
     \end{subfigure}
\hspace{2mm}
     \begin{subfigure}[h]{0.48\textwidth}
         \centering
         \includegraphics[width=\textwidth]{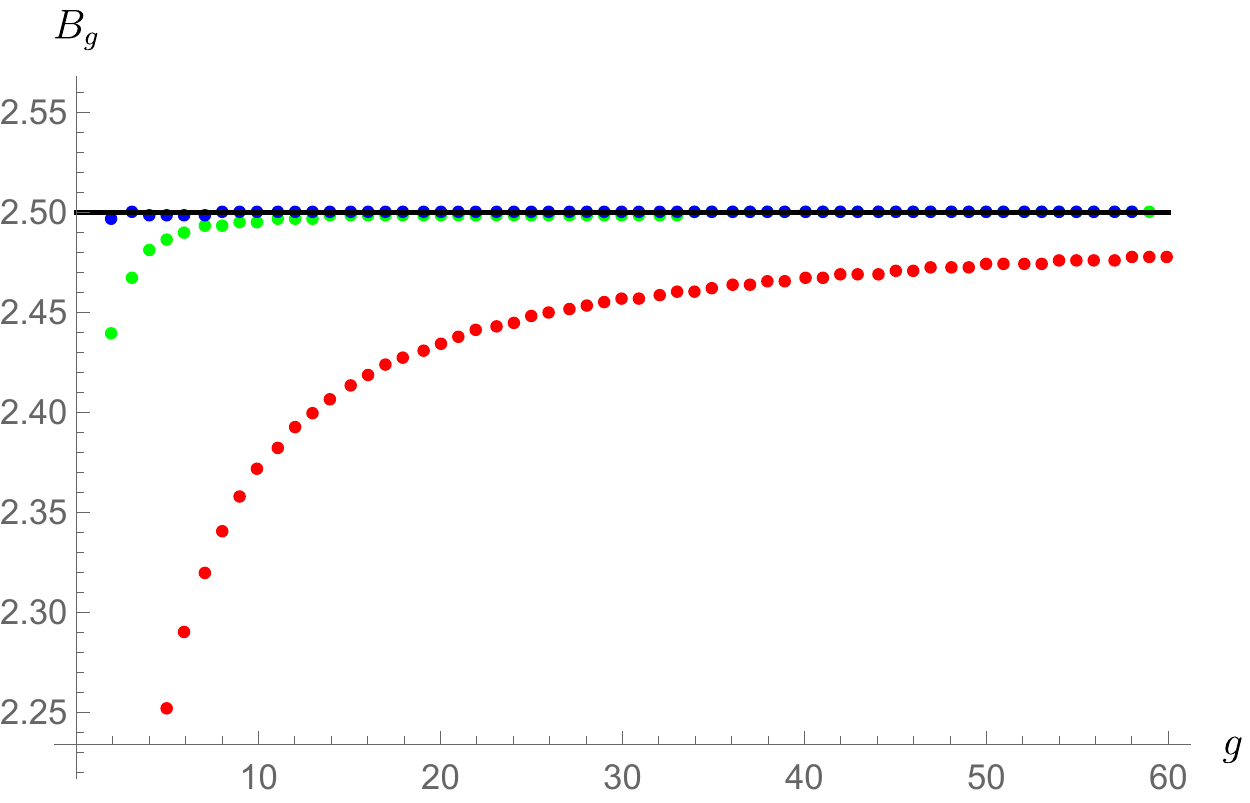}
     \end{subfigure}
     \begin{subfigure}[h]{0.48\textwidth}
         \centering
         \includegraphics[width=\textwidth]{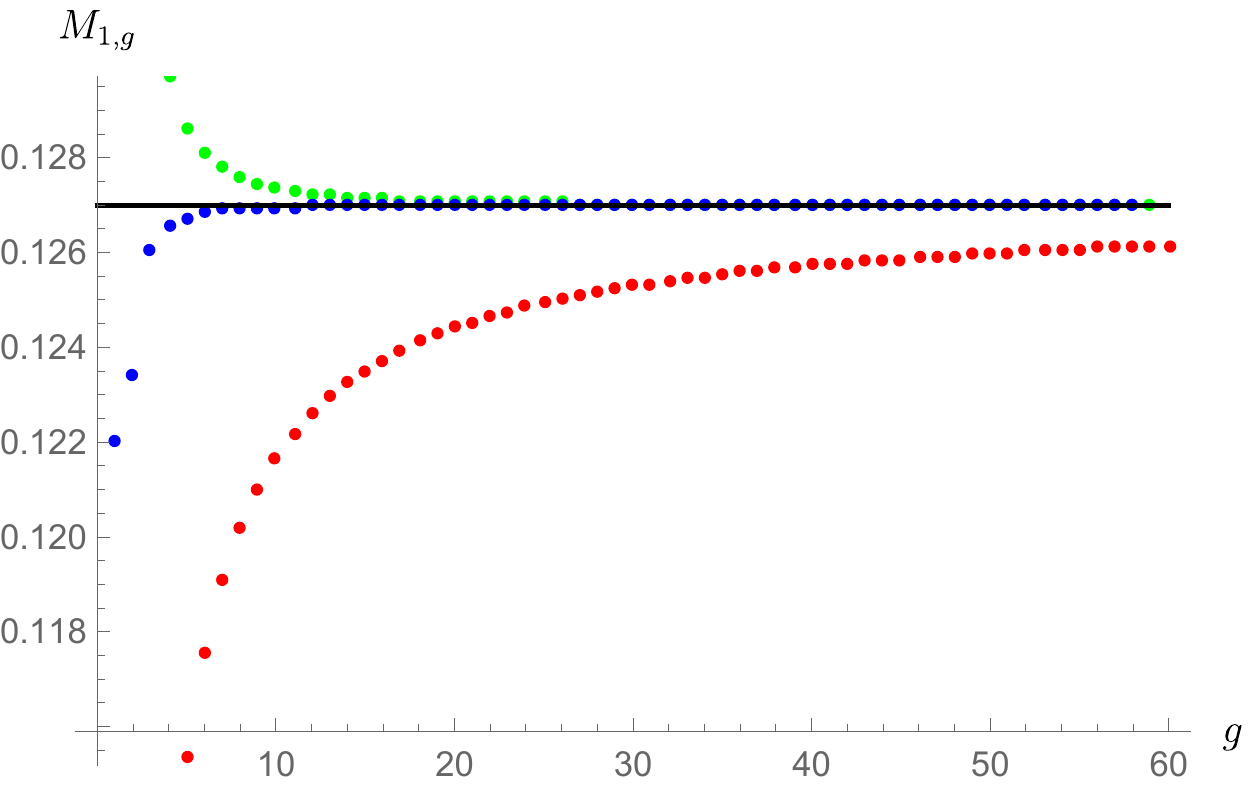}
     \end{subfigure}
\hspace{2mm}
     \begin{subfigure}[h]{0.48\textwidth}
         \centering
         \includegraphics[width=\textwidth]{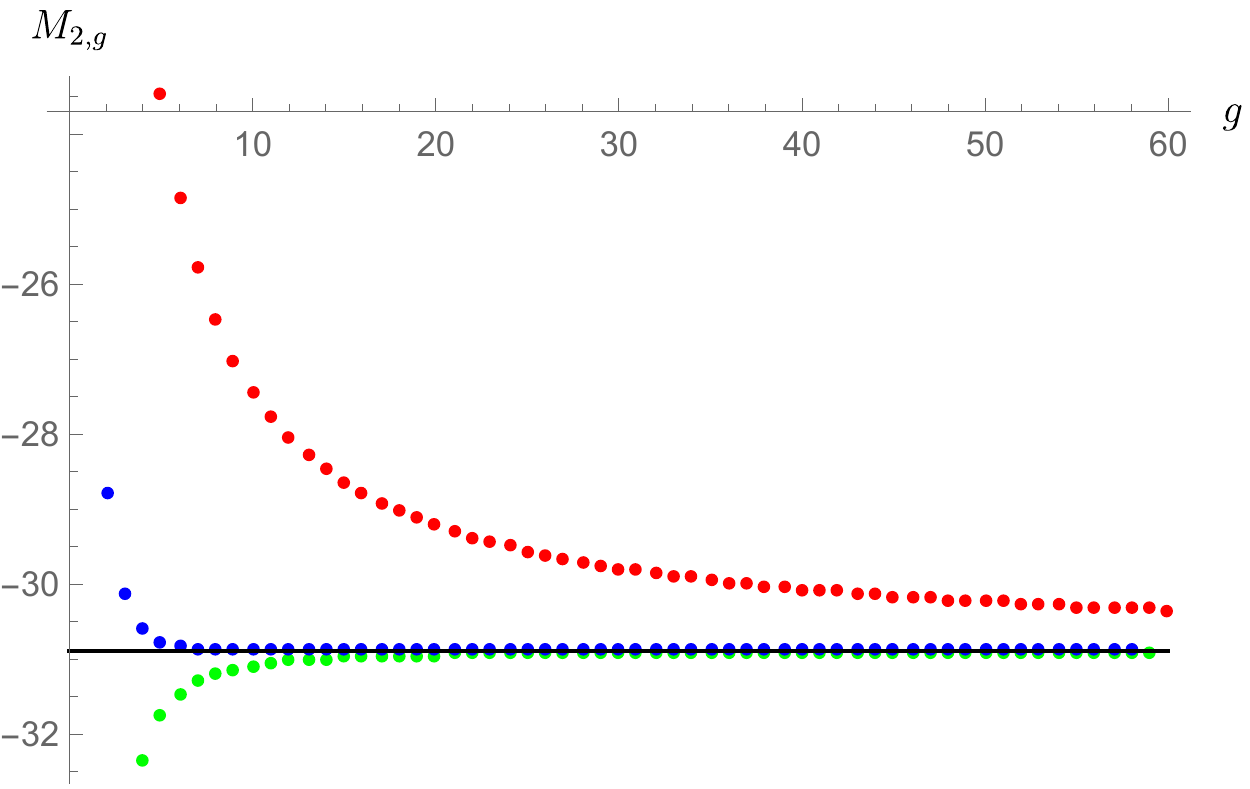}
     \end{subfigure}
     \begin{subfigure}[h]{0.48\textwidth}
         \centering
         \includegraphics[width=\textwidth]{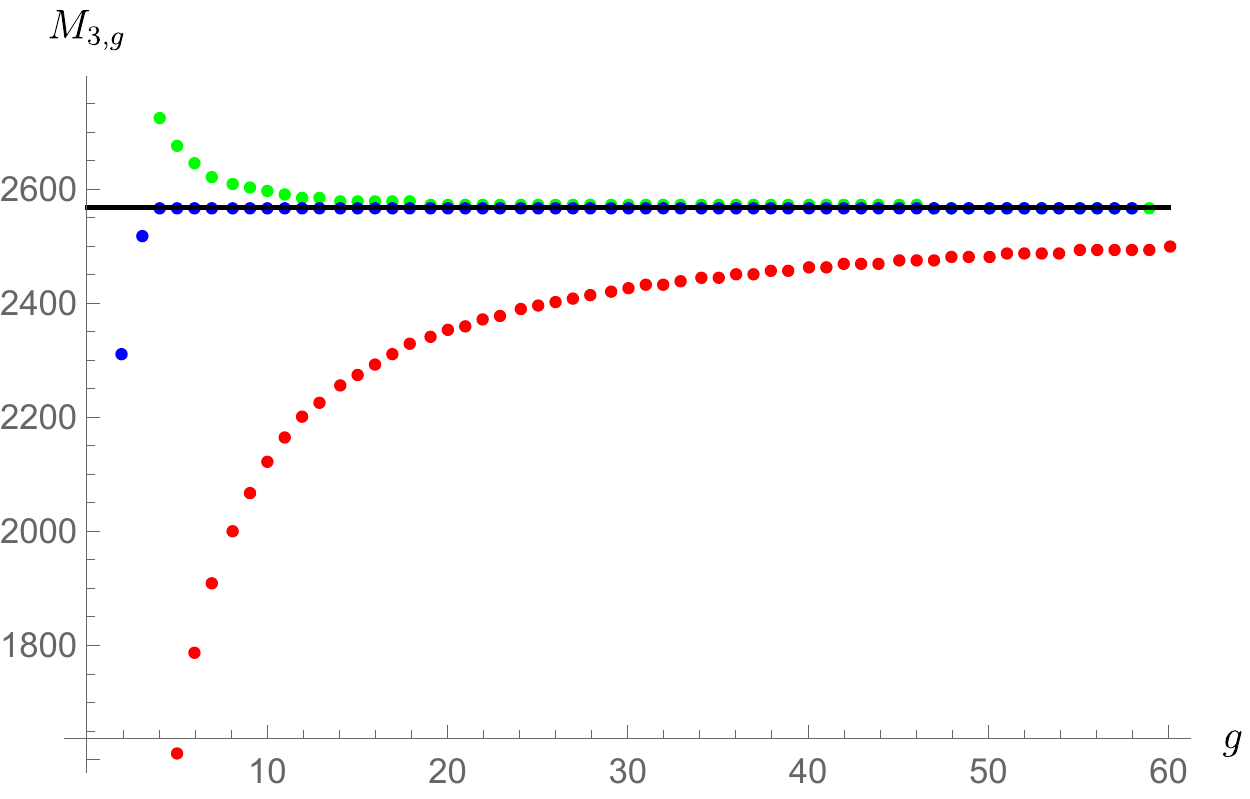}
     \end{subfigure}
\hspace{2mm}
     \begin{subfigure}[h]{0.48\textwidth}
         \centering
         \includegraphics[width=\textwidth]{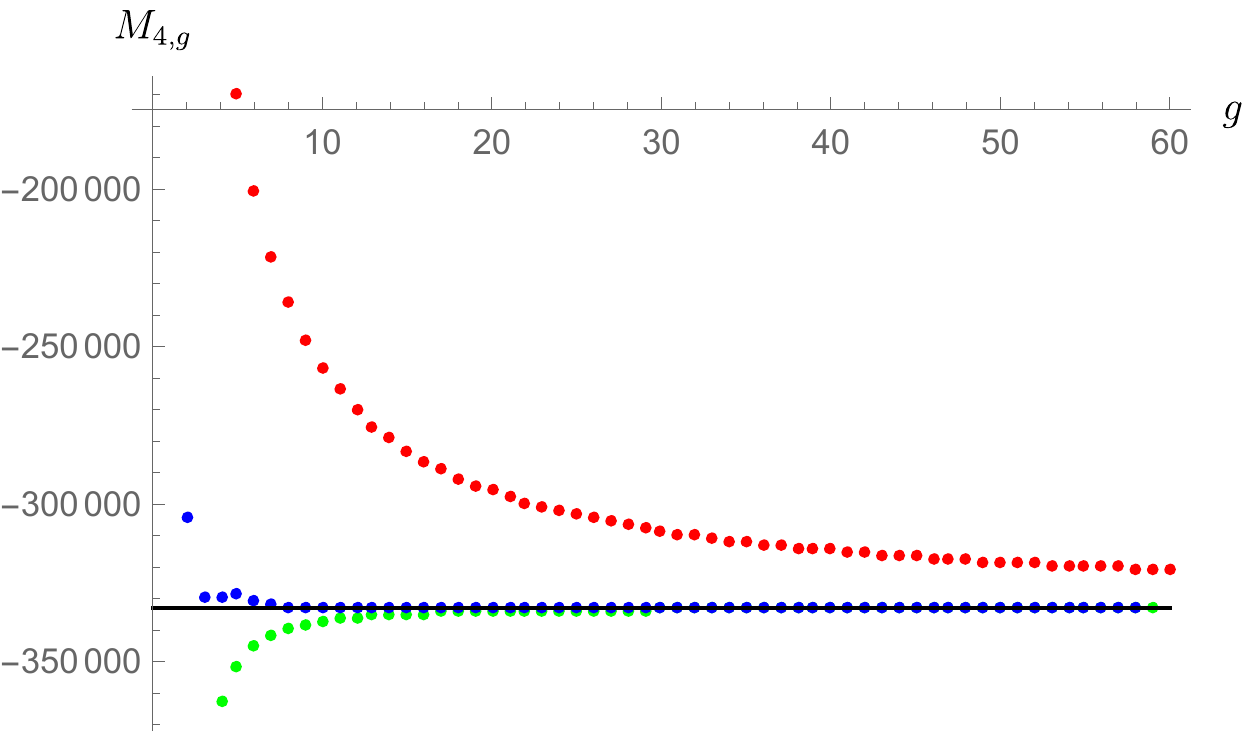}
     \end{subfigure}

     \begin{subfigure}[h]{0.48\textwidth}
         \centering
         \includegraphics[width=\textwidth]{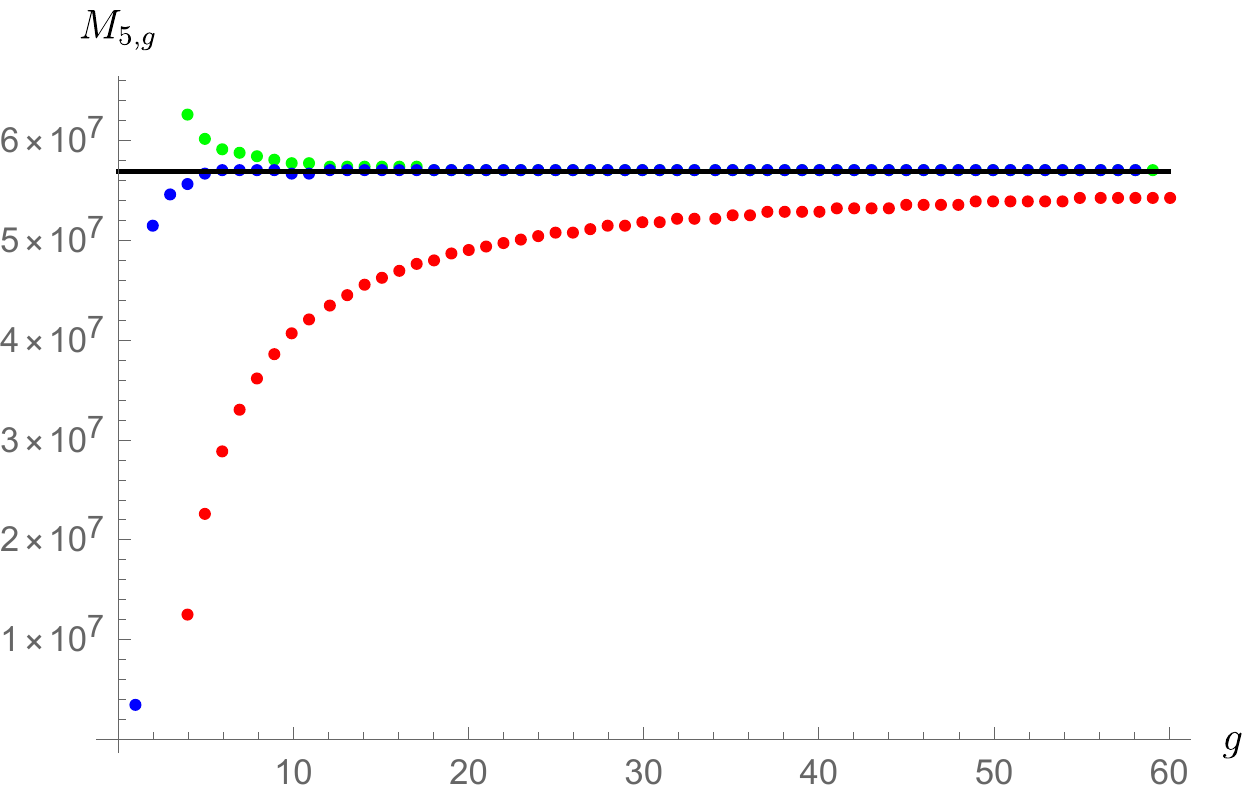}
     \end{subfigure}
\hspace{2mm}
     \begin{subfigure}[h]{0.48\textwidth}
         \centering
         \includegraphics[width=\textwidth]{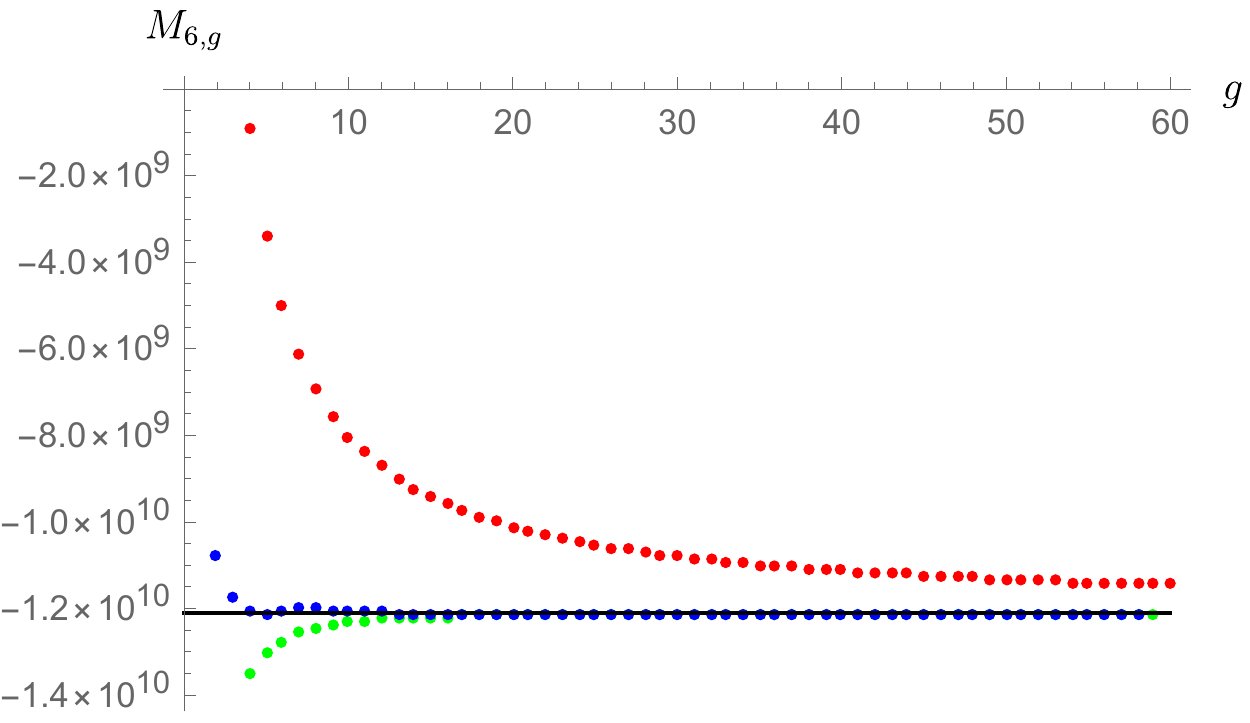}
     \end{subfigure}
     \caption{The sequences $Q_g$, $B_g$, $M_{\ell,g}$ ($\ell=1\ldots6$) for the $V_{g,0}$. In red, the sequences themselves; in green and blue the first and second Richardson transforms, respectively. The black lines correspond to the predicted values. In all cases the relative error is of the order of $\sim 10^{-6}$.}
\label{fig:JT-test1}
\end{figure}

\begin{figure}
\centering
     \begin{subfigure}[h]{0.48\textwidth}
         \centering
         \includegraphics[width=\textwidth]{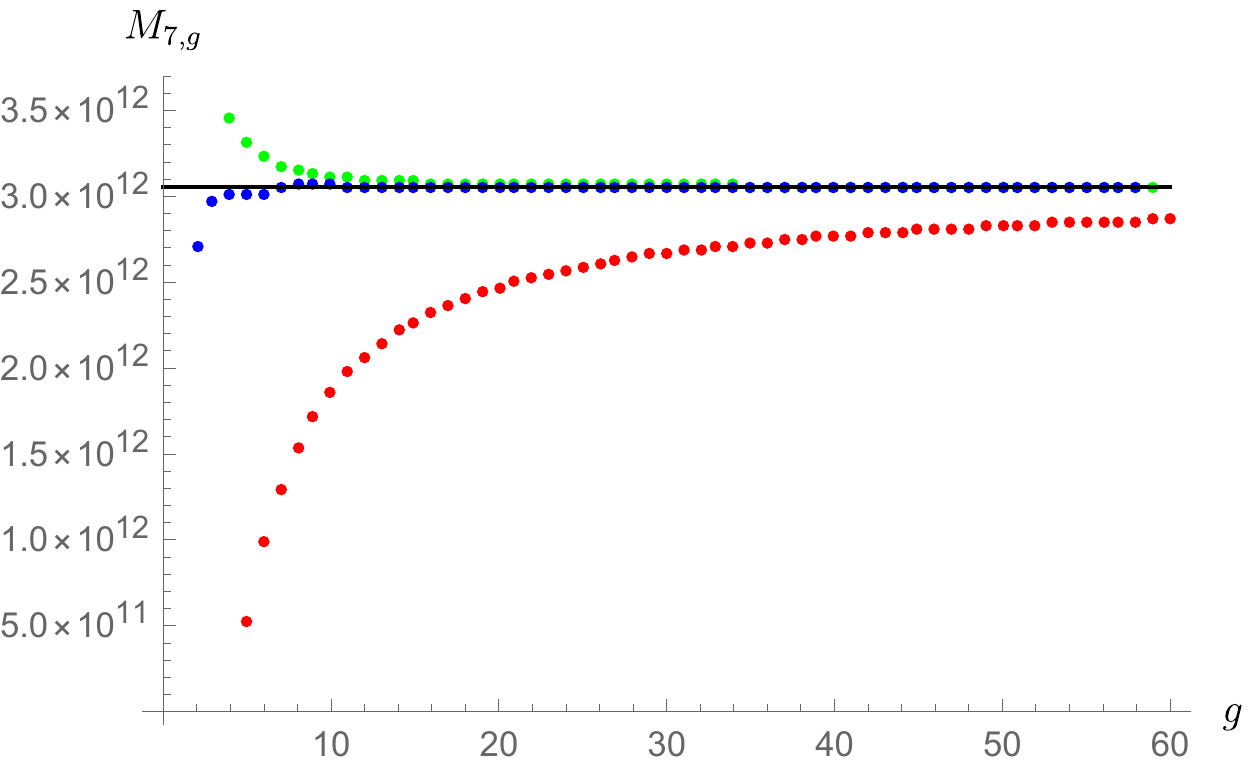}
     \end{subfigure}
\hspace{2mm}
     \begin{subfigure}[h]{0.48\textwidth}
         \centering
         \includegraphics[width=\textwidth]{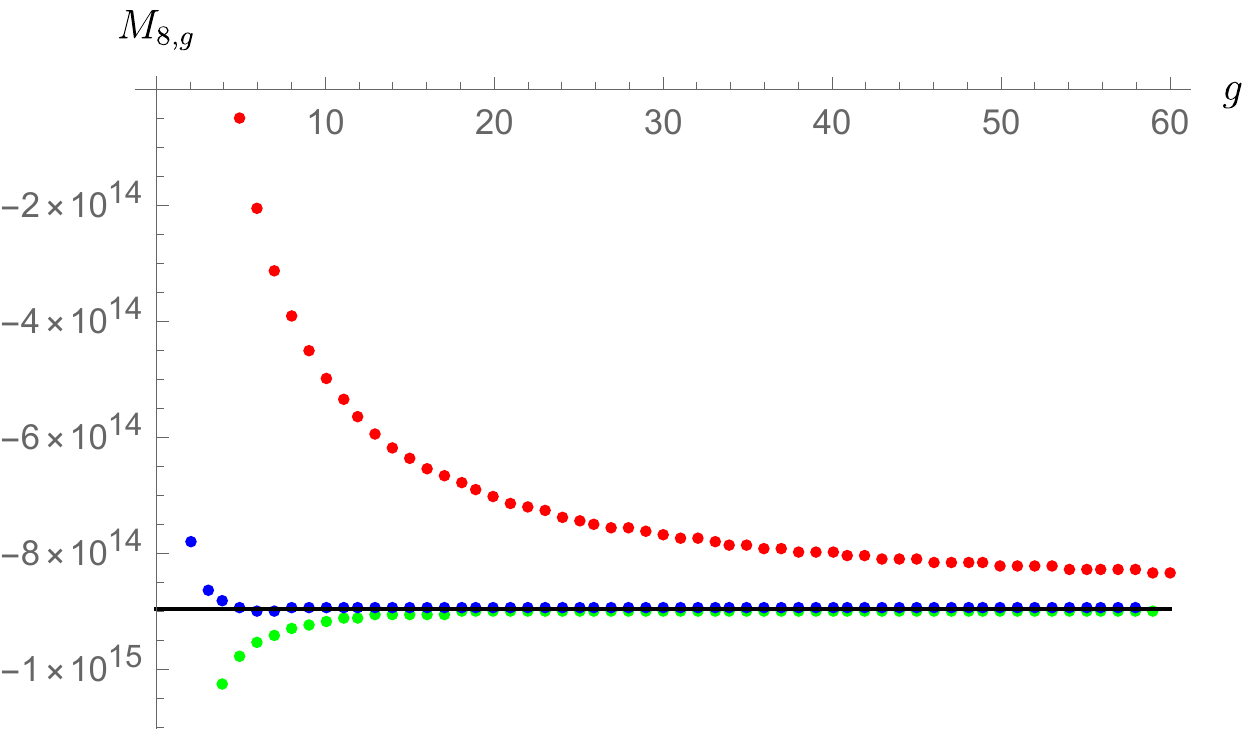}
     \end{subfigure}
     \begin{subfigure}[h]{0.48\textwidth}
         \centering
         \includegraphics[width=\textwidth]{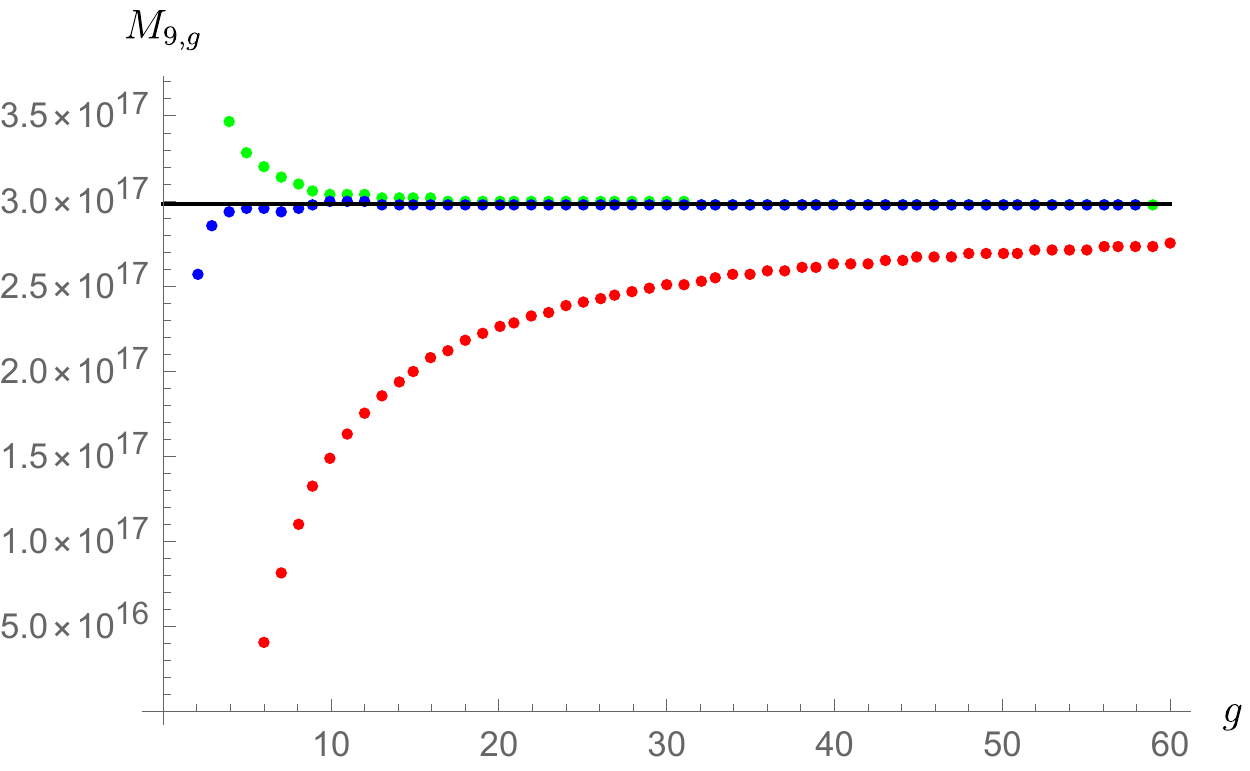}
     \end{subfigure}
\hspace{2mm}
     \begin{subfigure}[h]{0.48\textwidth}
         \centering
         \includegraphics[width=\textwidth]{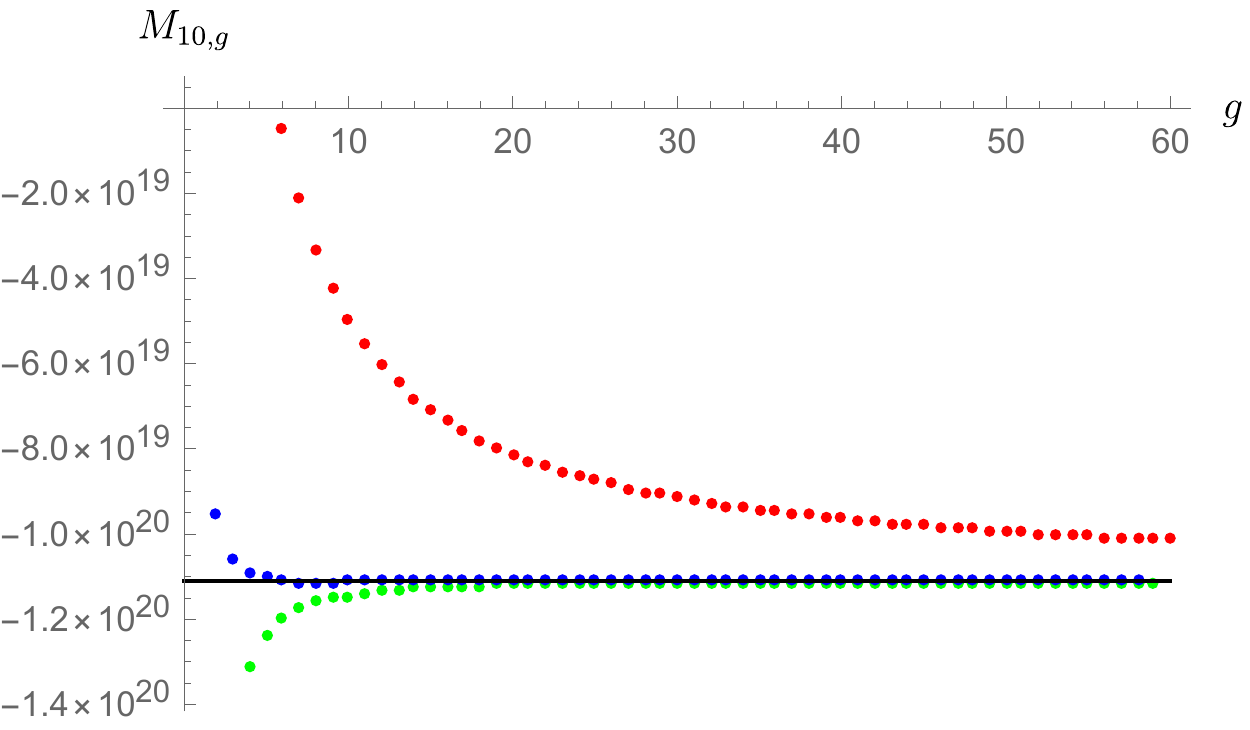}
     \end{subfigure}

     \begin{subfigure}[h]{0.48\textwidth}
         \centering
         \includegraphics[width=\textwidth]{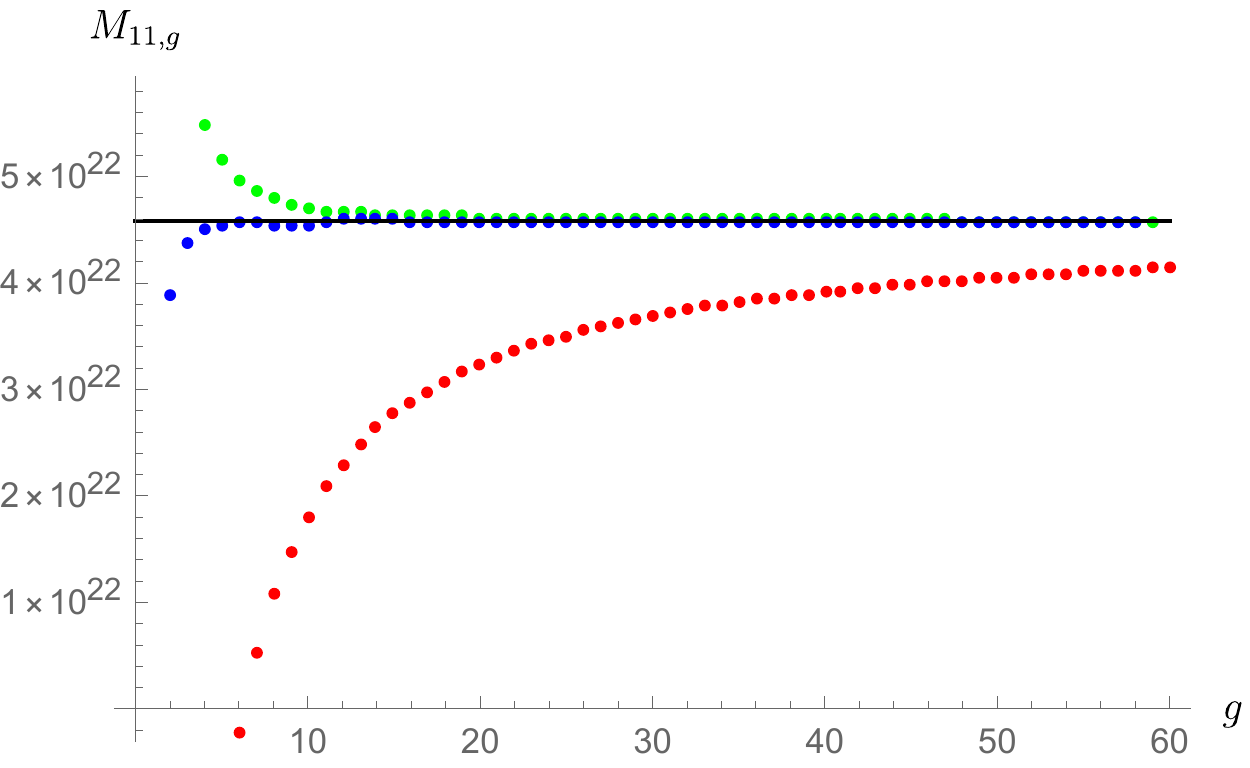}
     \end{subfigure}
\hspace{2mm}
     \begin{subfigure}[h]{0.48\textwidth}
         \centering
         \includegraphics[width=\textwidth]{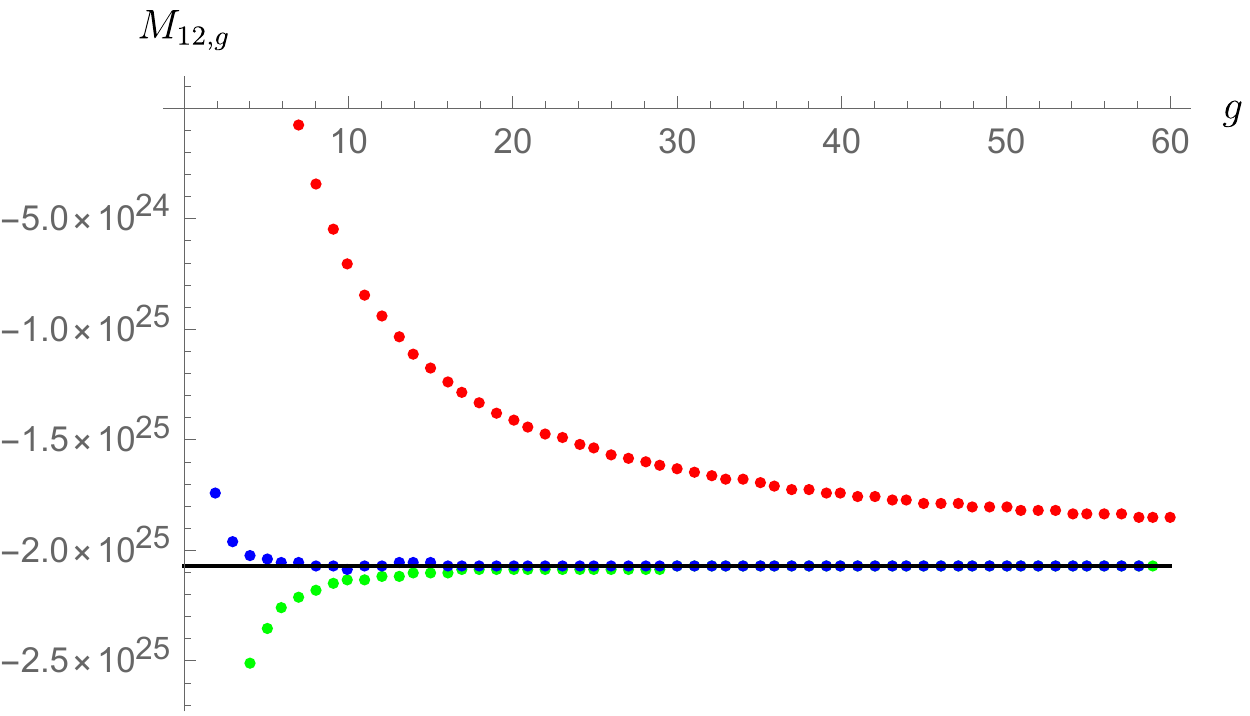}
     \end{subfigure}
     \caption{The sequences $M_{\ell,g}$ ($\ell=7\ldots12$) for the $V_{g,0}$. In red, the sequences themselves; in green and blue the first and second Richardson transforms, respectively. The black lines correspond to the predicted values. In all cases the relative error is of the order of $\sim 10^{-6}$.}
\label{fig:JT-test2}
\end{figure}

As in the previous section~\ref{sec:resurgence-correlators}, the above considerations can be applied to the large-genus asymptotic behavior of Weil--Petersson volumes, in order to provide numerical checks of the nonperturbative data we have obtained analytically. For this purpose, we computed the first $60$ $V_{g,0}$ volumes using the Zograf algorithm \cite{z07,z08}, and used them to construct the above defined sequences $Q_g$ in \eqref{eq:Qg}, $B_g$ in \eqref{eq:Bg}, and all the $M_{\ell,g}$, building upon \eqref{eq:M1g}-\eqref{eq:M2g}, with loop order $\ell$ from $1$ to $12$. These allow us to test our analytic predictions for the instanton action $A$, characteristic exponent $\beta$, and the Stokes and $\ell$-loop contributions to the one-instanton configuration up to the twelve-loop contribution. The resulting numerical checks, which are plotted in figures~\ref{fig:JT-test1} and~\ref{fig:JT-test2}, turn out to be very convincing, with a relative error never exceeding the order of $\sim 10^{-6}$.

\acknowledgments
We would like to thank
Salvatore~Baldino,
Norman~Do,
Maximilian~Schwick,
Peter~Zograf,
for useful discussions, comments and/or correspondence. In particular, we would like to thank Adrien~Ooms for initial collaboration in this work. EGF, PG, and DL would like to thank IHES and IPhT for hospitality during part of the work. RS would like to thank CERN-TH for extended hospitality, where parts of this work were conducted. BE was supported by the European Research Council (ERC) grant described below and by the grant host institutes IHES and IPhT. EGF was supported by the public grant ``Jacques Hadamard'' as part of the ``Investissement d'Avenir'' project, reference ANR-11-LABX-0056-LMH, LabEx LMH, and received funding from the ERC under the European Union's Horizon 2020 research and innovation programme, grant agreement No.~ERC-2016-STG 716083  ``CombiTop''. PG was supported by the FCT-Portugal grant PTDC/MAT-OUT/28784/2017. DL was supported by the Swiss National Foundation via the Ambizione project ``Resurgent Topological Recursion, Enumerative Geometry and Integrable Hierarchies'' under grant agreement PZ00P2-202123, by the Department of Mathematics, Informatics and Geoscience at the University of Trieste, by the INdAM group GNSAGA, and by the Istituto Nazionale di Fisica Nucleare under the project Mathematical Methods for NonLinear Physics (MMNLP). RS was supported in part by CAMGSD/IST-ID and via the FCT-Portugal grants UIDB/04459/2020, PTDC/MAT-OUT/28784/2017. This paper is partly a result of the ERC-SyG project, Recursive and Exact New Quantum Theory (ReNewQuantum) funded by the ERC under the European Union's Horizon 2020 research and innovation programme, grant agreement 810573.

\newpage

\appendix

\section{Details on the Shifted Partition Function}
\label{app:shift}

One of our main steps builds upon equation \eqref{eq:shiftedZ} (which also plays an important role in \cite{mss22}). In the present appendix we expand on the derivation of this equation, \textit{i.e.}, we give the details on how to compute
\be
\label{eq:init}
\CZ_{N-1}^{(0)} \ev{\det \left( x \textbf{1} - M'\right)^2}_{N-1}^{(0)} \rme^{- \frac{1}{g_{\text{s}}}V(x)},
\ee
\noindent
as the partition function of a spectral curve, shifted by specific generalized cycles. We will first focus on the computation of the correlator $\CZ_{N}^{(0)} \ev{\det \left( x \textbf{1} - M\right)^2}_{N}^{(0)}$, and then perform the shift from $N$ to $N-1$. These steps will be first carried out in the standard case of a finite-cut matrix model, and then the double-scaling limit will be taken.

\paragraph{The Squared Determinant:}

Start by rewriting the expectation value of some power $\uplambda/g_{\text{s}}$ of the determinant insertion as the partition function of some other matrix model with shifted potential,
\be
\CZ_{N}^{(0)} \ev{\det \left( x \textbf{1} - M\right)^{\frac{\uplambda}{g_{\text{s}}}}}_{N}^{(0)} = \int \mathrm{d}M\, \det \left( x \textbf{1} - M\right)^{\frac{\uplambda}{g_{\text{s}}}}\, \mathrm{e}^{-\frac{1}{g_{\text{s}}} \Tr\, V(M)} \equiv \int \mathrm{d}M\, \mathrm{e}^{-\frac{1}{g_{\text{s}}} \Tr\, V_{\uplambda}(M)} \equiv \CZ_{N,\uplambda}^{(0)},
\ee
\noindent
where
\be
\label{eq:lambda-pot}
V_{\uplambda}(M) \equiv V(M) - \uplambda\, \log \left( x \textbf{1} - M\right).
\ee
\noindent
Our goal of course is to compute $\CZ_{N,2 g_{\text{s}}}^{(0)}$, which we may recast as
\be
\label{eq:ZN2}
\left. \CZ_{N,2 g_{\text{s}}}^{(0)} \equiv \widehat{\mathcal{O}}_{\uplambda+2 g_{\text{s}}} \cdot  \CZ_{N,\uplambda}^{(0)} \right|_{\uplambda=0},
\ee
\noindent
where we introduced an operator which shifts $\uplambda$ by $2g_{\text{s}}$, immediately defined as
\be
\label{eq:Olambda2}
\widehat{\mathcal{O}}_{\uplambda+2 g_{\text{s}}} \equiv \exp \left( 2 g_{\text{s}}\, \frac{\partial}{\partial \uplambda}\right) = \sum_{k=0}^{+\infty} \frac{1}{k!} \left( 2 g_{\text{s}}\, \frac{\partial}{\partial \uplambda} \right)^k.
\ee

Next, express the $\uplambda$-deformed partition function in terms of its associated free energy
\be
\mathcal{F}^{(0)}_{N,\uplambda} \simeq \sum_{g=0}^{+\infty} \widetilde{\omega}_{g,0}\, g_{\text{s}}^{2g-2},
\ee
\noindent
where the $\widetilde{\omega}_{g,n}$ denote the multi-differential invariants in the $\uplambda$-deformed model, in contrast with the $\omega_{g,n}$ invariants of the original model in \eqref{eq:TRomega}. It follows
\be
\label{eq:OZNlambda}
\widehat{\mathcal{O}}_{\uplambda+2 g_{\text{s}}} \cdot \CZ_{N,\uplambda}^{(0)} = \widehat{\mathcal{O}}_{\uplambda+2 g_{\text{s}}} \cdot \rme^{\mathcal{F}^{(0)}_{N,\uplambda}} = \exp \left( \widehat{\mathcal{O}}_{\uplambda+2 g_{\text{s}}} \cdot \mathcal{F}^{(0)}_{N,\uplambda} \right).
\ee
\noindent
Now, using \eqref{eq:lambda-pot}, in the $\uplambda$-deformed matrix model we have\footnote{We are herein working in the variable $x$, where $\infty_{\pm}$ in $x$ correspond to $\pm\infty$ in $z$ illustrated in figure~\ref{fig:sketch}.}
\be
\label{eq:lambda-free}
\left. \frac{\partial}{\partial \uplambda} \mathcal{F}^{(0)}_{N,\uplambda} \right|_{\uplambda=0} = \frac{1}{g_{\text{s}}} \ev{ \Tr\, \log \left( x \textbf{1} - M\right)} = \frac{1}{g_{\text{s}}}\, \mathsf{R}\hspace{-6pt}\int_{\infty_+}^{x} \mathrm{d}x'\, \ev{\Tr\, \frac{1}{x'-M}},
\ee
\noindent
where the expectation values are taken in the undeformed $N\times N$ matrix model. We further introduced the notation $\mathsf{R}\hspace{-5pt}\int$ to denote the regularized integral defined as
\be
\mathsf{R}\hspace{-6pt}\int_{\infty_+}^{x} \mathrm{d}x'\, f(x') \equiv \int_{\infty_+}^{x} \mathrm{d}x' \left( f(x')-\frac{1}{x'}\right) + \log x.
\ee
\noindent
For the rest of this appendix, all integrals are assumed to be regularized in this way. Note that this regularization plays no role and can therefore be dropped in the final result, since the final integration contour does not pass through $\infty_{\pm}$ (recall figure~\ref{fig:sketch}). As such, in general we have
\bea
\label{eq:lambda-main}
\left. \frac{\partial}{\partial \uplambda} \ev{\Tr\, \frac{1}{x_1-M} \cdots \Tr\, \frac{1}{x_n-M}}' \right|_{\uplambda=0} &=& \frac{1}{g_{\text{s}}} \ev{\Tr\, \log \left( x \textbf{1} - M\right) \Tr\, \frac{1}{x_1-M} \cdots \Tr\, \frac{1}{x_n-M}} = \nonumber \\
&&
\hspace{-50pt}
= \frac{1}{g_{\text{s}}} \int_{\infty_+}^{x} \mathrm{d}x'\, \ev{\Tr\, \frac{1}{x'-M}\, \Tr\, \frac{1}{x_1-M} \cdots \Tr\, \frac{1}{x_n-M}},
\eea
\noindent
where the ``primed'' notation on the left-hand side denotes the expectation value taken in the $\uplambda$-deformed matrix model. Matching powers of $g_{\text{s}}$ on both sides of \eqref{eq:lambda-free} and \eqref{eq:lambda-main}, and making use of equations \eqref{eq:Rn-multi}-\eqref{eq:Rn-multi2} and \eqref{eq:fromwtoW}-\eqref{eq:fromhattonohat}, one obtains a relation between the $\widetilde{\omega}_{g,n}$ and the $\omega_{g,n}$ as
\be
\label{eq:lambda-der}
\left. \frac{\partial}{\partial \uplambda }\widetilde{\omega}_{g,n} \right|_{\uplambda=0} = \left. \int_{\infty_+}^{x} \widetilde{\omega}_{g,n+1} \right|_{\uplambda=0} = \int_{\infty_+}^{x}\omega_{g,n+1}, \qquad \text{ for } \qquad (n,g)\neq (0,0),
\ee
\noindent
and\footnote{Here we make use of \eqref{eq:TR-init} and of the usual matrix model relation $y(x) = V'(x) - 2 W_{0,1}(x)$.}
\be
\left. \frac{\partial}{\partial \uplambda} \widetilde{\omega}_{0,0 }\right|_{\uplambda=0} = \int_{\infty_+}^{x} \omega_{0,1} + \frac{1}{2}V(x).
\ee

Combining all of the above yields:
\be
\label{eq:Z-sqed-det}
\CZ_{N,2 g_{\text{s}}}^{(0)} = \exp \left( \frac{1}{g_{\text{s}}} V(x) + \sum_{g,n\geq 0} \frac{2^{n}}{n!}\, g_{\text{s}}^{2g+n-2} \overbrace{\int_{\infty_+}^x \cdots \int_{\infty_+}^x}^{n} \omega_{g,n} \right) = \rme^{\frac{V(x)}{g_{\text{s}}}}\, \CZ \left( g_{\text{s}}^{-1}\mathscr{S} + 2\gamma_{\infty_+ \rightarrow x} \right).
\ee

\paragraph{The $N$ to $N-1$ Shift:}

Taking into account the $N\rightarrow N-1$ shift which sends $\CZ_{N,2 g_{\text{s}}}^{(0)}$ into \eqref{eq:init} goes along a very similar line of reasoning. The shift in the number of eigenvalues can be seen as an equivalent shift in the 't~Hooft coupling $t = g_{\text{s}} N$, from $t$ to $t-g_{\text{s}}$. Then, analogously to what we did earlier in \eqref{eq:ZN2}-\eqref{eq:Olambda2}, we can introduce a $t$-shifting operator
\be
\widetilde{\mathcal{O}}_{t-g_{\text{s}}} \equiv \exp \left( - g_{\text{s}}\, \frac{\partial}{\partial t}\right) = \sum_{k=0}^{+\infty} \frac{1}{k!} \left( - g_{\text{s}}\, \frac{\partial}{\partial t} \right)^k,
\ee
\noindent
which computes what we want, that is,
\be
\label{eq:ZN-1}
\CZ_{N-1}^{(0)} \equiv \widetilde{\mathcal{O}}_{t-g_{\text{s}}} \cdot \CZ_{N}^{(0)}.
\ee
\noindent
We also have the analogue of \eqref{eq:OZNlambda},
\be
\label{eq:OZNt-gs}
\widetilde{\mathcal{O}}_{t-g_{\text{s}}} \cdot \CZ_{N}^{(0)} = \exp \left( \widetilde{\mathcal{O}}_{t-g_{\text{s}}} \cdot \mathcal{F}^{(0)}_{N} \right).
\ee

Next, we will use the property that
\be
\label{eq:t-derivative}
\frac{\partial}{\partial t} \omega_{g,n} = - \int_{\infty_-}^{\infty_+} \omega_{g,n+1}.
\ee
\noindent
To prove this equality, start with its matrix model equivalent via \eqref{eq:fromwtoW}-\eqref{eq:fromhattonohat}
\be
g_{\text{s}}\, \frac{\partial}{\partial t} \widehat{W}_n = - \int_{\infty_-}^{\infty_+} \widehat{W}_{n+1}.
\ee
\noindent
This may be explicitly evaluated out of multi-resolvent definition \eqref{eq:Rn-multi}, by simply going back to the matrix integral \eqref{eq:ZN} and trading the string coupling $g_{\text{s}}$ by the 't~Hooft coupling $t = g_{\text{s}} N$. Then
\bea
\label{eq:fstep}
\frac{\partial}{\partial t} \widehat{W}_{n} \left( x_1, \ldots, x_n \right) &=& \frac{N}{t^2} \ev{\Tr\, V(M) \Tr\, \frac{1}{x_1-M} \cdots \Tr\, \frac{1}{x_n-M}}_{(\text{c})} = \\
&=& - \frac{1}{t g_{\text{s}}}\, \underset{x \to \infty_+}{\operatorname{Res}} V(x)\, \widehat{W}_{n+1} \left( x; x_1, \ldots, x_n \right), \nonumber
\eea
\noindent
where in the second equality we used
\be
\underset{x \to \infty_+}{\operatorname{Res}} V(x)\, \Tr\, \frac{1}{x-M} = \underset{x \to \infty_+}{\operatorname{Res}} V(x)\, \Tr\, \frac{1}{x} \sum_{k\geq 0} \left(\frac{M}{x}\right)^k = - \Tr\, V(M).
\ee
\noindent
Next, introduce the primitive of $\widehat{W}_{n+1}$
\be
\Phi_{n+1} \left( x; x_1, \ldots, x_n \right) = \int_{\infty_+}^x \rmd x'\, \widehat{W}_{n+1} \left( x'; x_1, \ldots, x_n \right) + C,
\ee
\noindent
where regularization is only needed for $\widehat{W}_1$ (for all others the behavior at $\infty$ is $\sim x'^{-2}$). It is convenient to simplify the calculation by choosing the integration constant $C$ such that $\Phi_{n+1}$ is odd under the involution $\iota(z)=-z$ exchanging the two sheets of the spectral curve. This implies:
\be
C = \frac{1}{2} \int_{\infty_-}^{\infty_+} \rmd x'\, \widehat{W}_{n+1} \left( x'; x_1, \ldots, x_n \right).
\ee
\noindent
This primitive allows for evaluation of the residue integral in \eqref{eq:fstep} via integration by parts as
\bea
t g_{\text{s}}\, \frac{\partial}{\partial t} W_{n} \left( x_1, \ldots, x_n \right) &=& \ev{\Tr\, V(M) \Tr\, \frac{1}{x_1-M} \cdots \Tr\, \frac{1}{x_n-M}}_{(\text{c})} = \\
&=& \underset{x \to \infty_+}{\operatorname{Res}} V'(x)\, \Phi_{n+1} \left( x; x_1, \ldots, x_n \right). \nonumber
\eea
\noindent
Now we remove the potential function using $V'(x)= y(x) + 2W_{0,1}(x)$, and the fact that $y(x)$ is an odd function under the involution changing sheets. Since also $\Phi_{n+1}$ is odd, then the product is even; hence
\be
\underset{x \to \infty_+}{\operatorname{Res}} y(x)\, \Phi_{n+1} \left( x; x_1, \ldots, x_n \right) = 0.
\ee
\noindent
This implies
\be
t g_{\text{s}}\, \frac{\partial}{\partial t} \widehat{W}_{n} \left( x_1, \ldots, x_n \right) = 2 \underset{x \to \infty_+}{\operatorname{Res}} W_{0,1}(x)\, \Phi_{n+1} \left( x; x_1, \ldots, x_n \right).
\ee
\noindent
Finally, since at $\infty_+$ we have $\Phi_{n+1} \to C$ and $W_{0,1} \sim t/x$ we find what we wanted to show,
\be
g_{\text{s}} \frac{\partial}{\partial t} \widehat{W}_n \left( x_1, \ldots, x_n \right) = - 2 C = -  \int_{\infty_-}^{\infty_+} \rmd x'\, \widehat{W}_{n+1} \left(x'; x_1, \ldots, x_n \right).
\ee

With all these in hand, we can finally use \eqref{eq:ZN-1}-\eqref{eq:OZNt-gs}-\eqref{eq:t-derivative} to write
\be
\label{eq:Z-N-1-shift}
\CZ_{N-1}^{(0)} = \exp \left( \sum_{g,n\geq 0} \frac{1}{n!}\, g_{\text{s}}^{2g-2+n} \overbrace{\int_{\infty_-}^{\infty_+} \cdots \int_{\infty_-}^{\infty_+}}^{n} \omega_{g,n} \right) = \CZ \left( g_{\text{s}}^{-1}\mathscr{S} + \gamma_{\infty_- \rightarrow \infty_+} \right).
\ee

Combining the two operations, \eqref{eq:Z-sqed-det} and \eqref{eq:Z-N-1-shift}, we obtain our wanted result for \eqref{eq:init}, which is \eqref{eq:shiftedZ} in the main body of the text,
\be
\label{eq:shiftedZ-appendix}
\CZ_{N-1}^{(0)} \ev{\det \left( x \textbf{1} - M'\right)^2}_{N-1}^{(0)} \rme^{- \frac{1}{g_{\text{s}}} V(x)} = \CZ \left( g_{\text{s}}^{-1}\mathscr{S} + 2\gamma_{\infty_+ \rightarrow x} + \gamma_{\infty_- \rightarrow \infty_+} \right).
\ee

\paragraph{The Double-Scaling Limit:}

All we have left to do is to address the double-scaling limit in which the JT-gravity spectral-curve is expressed. Starting from a finite-cut curve such as in \eqref{eq:finite-cut-SC}, introduce a uniformization variable $\upzeta$ on the Riemann sphere as
\bea
\label{eq:finite-cut-x}
x(\upzeta) &=& \frac{a+b}{2} + \frac{a-b}{4} \left(\upzeta+\frac{1}{\upzeta}\right), \\
\label{eq:finite-cut-y}
y(\upzeta) &=& M \left(x(\upzeta)\right) \frac{a-b}{4} \left(\upzeta-\frac{1}{\upzeta}\right).
\eea
\noindent
The endpoints of the cut, $a$ and $b$, are mapped to $\upzeta=1$ and $\upzeta=-1$, respectively. The double-scaling limit zooming-in on the endpoint $a$ is then obtained by expanding around $\upzeta=1$. This is simply done introducing the new uniformization variable $z$
\be
\label{eq:newvar}
\upzeta = 1 + \delta z,
\ee
\noindent
and performing a Taylor expansion around $\delta=0$. Equation \eqref{eq:finite-cut-x} then becomes
\be
x(z) = a + \frac{1}{4} \left( a-b \right) \delta^2 z^2 + \cdots,
\ee
\noindent
\textit{i.e.}, up to an uninfluential shift of $x$ by $a$, we obtain $x \propto z^2$, which is the appropriate relation for double-scaled models. Furthermore, the points $\infty_+$ and $\infty_-$ in the finite-cut matrix model correspond, in the double-sheeted $x$-plane, to the values $0$ and $\infty$ of the uniformization variable $\upzeta$. These two points in the $\upzeta$-plane are related to each other via the involution $\upzeta \mapsto \frac{1}{\upzeta}$ which exchanges the two sheets in the $x$-plane. In the double-scaling limit, and upon the introduction of the new uniformization variable $z$, the points $\infty_+$ and $\infty_-$ are mapped to $+\infty$ and $-\infty$ in the $z$-plane now connected to each other via the new involution which sends $z \mapsto -z$. 

The concern one might have with the double-scaling limit---in the derivation of the above \eqref{eq:shiftedZ-appendix}---is that, at first sight, integrals appearing in \eqref{eq:t-derivative} might appear ill-defined in this limit. In order to illustrate how to carefully take the double-scaling limit in this context, let us explicitly evaluate it in \eqref{eq:t-derivative}, for $(g,n)$=$(0,2)$. For a generic one-cut matrix model with spectral curve given by \eqref{eq:finite-cut-SC}, we have \cite{ajm90}
\bea
W_{0,2} (x_1,x_2) &=& \frac{1}{2\left(x_1-x_2\right)^2} \left( \frac{x_1 x_2 - \frac{1}{2} \left(x_1+x_2\right) \left(a+b\right) + ab}{\sqrt{\left(x_1 - a\right) \left(x_1 - b\right) \left(x_2 - a\right) \left(x_2 - b\right)}} - 1 \right), \\
W_{0,3} (x_1,x_2,x_3) &=& \frac{1}{8\sqrt{\left(x_1 - a\right) \left(x_1 - b\right) \left(x_2 - a\right) \left(x_2 - b\right) \left(x_3 - a\right) \left(x_3 - b\right)}} \times \\
&&
\times \left(\frac{a-b}{M(a)}\, \frac{1}{\left(x_1 - a\right) \left(x_2 - a\right) \left(x_3 - a\right)} + \frac{b-a}{M(b)}\, \frac{1}{\left(x_1 - b\right) \left(x_2 - b\right) \left(x_3 - b\right)} \right). \nonumber
\eea
\noindent
The derivatives with respect to the 't~Hooft coupling are then taken using \cite{msw07}
\be
\frac{\partial a}{\partial t} = \frac{4}{a-b}\, \frac{1}{M(a)}, \qquad \frac{\partial b}{\partial t} = \frac{4}{b-a}\, \frac{1}{M(b)}.
\ee
\noindent
Ensemble, we therefore get
\bea
\frac{\partial}{\partial t} W_{0,2} (x_1,x_2) &=& \frac{1}{2 \sqrt{\left(x_1 - a\right) \left(x_1 - b\right) \left(x_2 - a\right) \left(x_2 - b\right)}} \times \\
&&
\times \left(\frac{1}{M(a) \left(x_1 - a\right) \left(x_2 - a\right)} + \frac{1}{M(b) \left(x_1 - b\right) \left(x_2 - b\right)} \right). \nonumber
\eea
\noindent
Next, recasting everything in terms of the uniformization variable $\upzeta$ and multi-differentials $\omega_{g,n}$, we obtain\footnote{Note that the derivative with respect to $t$ is taken at fixed $x_i$.}
\bea
\frac{\partial}{\partial t} \omega_{0,2} (\upzeta_1,\upzeta_2) &=& \frac{8}{\left(a-b\right)^2}  \left( \frac{1}{M(a) \left(\upzeta_1-1\right)^2 \left(\upzeta_2-1\right)^2} + \frac{1}{M(b) \left(\upzeta_1+1\right)^2 \left(\upzeta_2+1\right)^2} \right) \rmd \upzeta_1 \rmd \upzeta_2, \\ 
\omega_{0,3} (\upzeta_1,\upzeta_2,\upzeta_3) &=& \frac{8}{\left(a-b\right)^2 \left(\upzeta_1^2-1\right)^2 \left(\upzeta_2^2-1\right)^2 \left(\upzeta_3^2-1\right)^2} \times \\
&&
\times \left( \frac{\left(\upzeta_1+1\right)^2 \left(\upzeta_2+1\right)^2 \left(\upzeta_3+1\right)^2}{M(a)} + \frac{\left(\upzeta_1-1\right)^2 \left(\upzeta_2-1\right)^2 \left(\upzeta_3-1\right)^2}{M(b)} \right) \rmd \upzeta_1 \rmd \upzeta_2 \rmd \upzeta_3. \nonumber
\eea
\noindent
Integrating $\omega_{0,3}$ with respect to the third variable from $0$ to $\infty$, we see that indeed \eqref{eq:t-derivative} is satisfied as
\be
\label{eq:t-derivative-ex}
\int_0^{+\infty} \omega_{0,3} (\upzeta_1,\upzeta_2,\bullet) = - \frac{\partial}{\partial t} \omega_{0,2} (\upzeta_1,\upzeta_2).
\ee

One may now explicitly address the double-scaling limit of \eqref{eq:t-derivative-ex} by first performing the change of uniformization variable \eqref{eq:newvar}, namely,
\be
z_i = \frac{1}{\delta} \left( \upzeta_i - 1 \right),
\ee
\noindent
and keeping only the leading piece in the $\delta\rightarrow 0$ limit. One hence obtains
\be
\label{eq:double-scaling-w03}
\int_0^{+\infty}\omega_{0,3} (\upzeta_1,\upzeta_2,\bullet) = \delta^3 \int_{-1/\delta}^{+\infty} \omega_{0,3} \left( 1+\delta z_1, 1+\delta z_2, 1+\delta\bullet \right) = - \frac{1}{\delta^2}\, \frac{8}{M(a) \left(a-b\right)^2 z_1^2 z_2^2} + \cdots,
\ee
\noindent
which, together with
\be
\frac{\partial}{\partial t} \omega_{0,2} (\upzeta_1,\upzeta_2) = \delta^2\, \frac{\partial}{\partial t} \omega_{0,2} \left( 1+\delta z_1, 1+\delta z_2 \right) = \frac{1}{\delta^2}\, \frac{8}{M(a) \left(a-b\right)^2 z_1^2 z_2^2} + \cdots,
\ee
\noindent
precisely satisfies \eqref{eq:t-derivative}, now in the double-scaling limit. Notice how, in order to keep track of the correct powers of $\delta$, we could not directly take the integral from $-\infty$ to $+\infty$ in \eqref{eq:double-scaling-w03}. It is also worth pointing out that all these precautions become irrelevant once we perform the composition of cycles \eqref{eq:cyclecomp}, as all integrals no longer pass through $\infty_+$ or $\infty_-$.

\section{Comments on Correlators and the KdV Hierarchy}
\label{app:corr_func}

In the main body of the paper we addressed the resurgent properties of a large class of correlation functions, the multi-resolvent correlators \eqref{eq:Rn-multi}, which, of course, are the generating functions for multi-trace correlation functions \eqref{eq:Rn-multi-trace}. In spite of the clear advantage of working with multi-resolvents, as they are the prime natural objects of the topological-recursion construct as reviewed in subsection~\ref{subsec:classicalTR}, it is also the case that there other classes of interesting operators in generic two-dimensional quantum and topological gravity, and minimal string theory. For instance, a natural basis of minimal string operators includes ``tachyon'' ghost-number-one vertex-operators alongside ghost-number-zero ground-ring operators \cite{ss03} (see \cite{km20} for comments on the JT extension). Another interesting class of matrix-model operators are the double-scaled $k$-th multicritical-model operators $\mathsf{O}_{\ell}$, with $\ell \in \BN_0$, given by \cite{gm90b, bdss90}
\be
\label{eq:O-multicrit-op}
\mathsf{O}_{\ell} = \mathsf{H}^{\ell+\frac{1}{2}},
\ee
\noindent
where $\mathsf{H}$ is the Gel'fand--Dikii hamiltonian operator \cite{gd75} (see \eqref{eq:FZZT-schro-eq}). In particular, string equations are nonlinear ordinary differential equations (ODE) for the \textit{two-point function} of the lowest-dimension operator in the $k$-th multicritical theory, \textit{i.e.}, $u = \ev{\mathsf{O}_{0} \mathsf{O}_{0}}$ (see below). It is well-known in the literature, \textit{e.g.}, \cite{gikm10, asv11, sv13, gs21}, that $u$ only receives ZZ-brane nonperturbative corrections. What we illustrate in this appendix is how generic correlation functions of the operators \eqref{eq:O-multicrit-op} in two-dimensional quantum gravity are fully determined by the two-point function $\ev{\mathsf{O}_{0} \mathsf{O}_{0}}$, hence also \textit{only} receive ZZ-brane corrections and \textit{no} FZZT-brane contributions (unlike for the multi-resolvent correlation functions discussed in the text). In other words, the resurgent structure of the specific-heat two-point function dictates (via the Gel'fand--Dikii KdV potentials \cite{gd75}) the full resurgent structure of the $\mathsf{O}_{\ell}$ $n$-point correlation functions (see \cite{gs21} as well).

Generic correlation functions of the $\mathsf{O}_{\ell}$ operators---at \textit{leading} order---were computed in \cite{gm90b} as\footnote{This expression corrects the prefactor of (4.19) in \cite{gm90b} (but where the extra overall minus is from conventions).}
\be
\label{eq:leading-correlation-fcts}
\ev{\mathsf{O}_{\ell_1} \cdots \mathsf{O}_{\ell_p}} = \frac{1}{k}\, \frac{\Gamma \left( \frac{1}{k} \left( \sum_{i=1}^p \ell_i + 1 \right) \right)}{\Gamma \left( \frac{1}{k} \left( \sum_{i=1}^p \ell_i + 1 \right) - p + 3 \right)}\, \kappa^{2 - p + \frac{1}{k} \left( \sum_{i=1}^p \ell_i + 1 \right)} + \cdots,
\ee
\noindent
where, recall, $\kappa$ is the double-scaled variable. Exact (all-orders) correlation functions may then be computed using the techniques in \cite{bdss90}---which amount to a clever use of the higher-order KdV flows in \cite{gd75}. Building upon the one-point functions, one iteratively finds \cite{bdss90} (the $\NCO_{\ell}$ will relate to the $\mathsf{O}_{\ell}$ momentarily)
\be
\frac{\partial^2}{\partial \kappa^2} \ev{\NCO_{\ell_1} \NCO_{\ell_2} \cdots \NCO_{\ell_p}} = \xi_{\ell_p+1} \cdots \xi_{\ell_2+1} \cdot \xi_{\ell_1+1} \cdot u.
\ee
\noindent
Herein, $u (\kappa)$ is the solution to the string equation (for double-scaled models, a nonlinear ODE in $\kappa$) and $\xi_{\ell}$ the mutually commuting KdV vector fields generating KdV flows \cite{gd75, gs21}, \textit{e.g.},
\be
\label{eq:KdV_flow}
\xi_{\ell} \cdot u = \frac{\partial}{\partial \kappa} R_{\ell} \left[ u \right].
\ee
\noindent
where\footnote{Again, we follow in parallel with section~3.1 of \cite{gs21}, to where we refer the reader for further details.} the $R_{\ell} \left[ u \right]$ are the Gel'fand--Dikii KdV potentials \cite{gd75}. In this way, the one-point functions are absolutely straightforward as
\be
\label{eq:one-pt-fct}
\frac{\partial}{\partial \kappa} \ev{\NCO_{\ell}} = R_{\ell+1}.
\ee
\noindent
Comparison of one-point functions in \eqref{eq:leading-correlation-fcts} and \eqref{eq:one-pt-fct}, via the spherical content of the Gel'fand--Dikii KdV potentials (in the conventions of \cite{gs21}) yields the normalization of operators in both computations as
\be
\NCO_{\ell} = (-1)^{\ell+1}\, \frac{\left( 2\ell+1 \right)!!}{2^{\ell+2}\, \ell!}\, \mathsf{O}_{\ell}.
\ee
\noindent
In particular, one may rewrite the (exact) one-point functions as
\be
\ev{\mathsf{O}_{\ell}} = (-1)^{\ell+1}\, \frac{2^{\ell+2}\, \ell!}{\left( 2\ell+1 \right)!!}\, \int \rmd \kappa\, R_{\ell+1}.
\ee
\noindent
Recursively using \eqref{eq:KdV_flow} one may further compute other (exact) correlation functions, \textit{e.g.},
\bea
\langle \underbrace{\mathsf{O}_{0} \cdots \mathsf{O}_{0}}_{n} \mathsf{O}_{\ell} \rangle &=& \frac{(-1)^{\ell+1}\, 2^{\ell+2}\, \ell!}{\left( 2\ell+1 \right)!!}\, R_{\ell+1}^{(n-1)}, \\
\langle \underbrace{\mathsf{O}_{0} \cdots \mathsf{O}_{0}}_{n} \mathsf{O}_{1} \mathsf{O}_{\ell} \rangle &=& \frac{(-1)^{\ell}\, 2^{\ell+4}\, \ell!}{\left( 2\ell+1 \right)!!} \left( R_{1} R_{\ell+1}' + \frac{1}{3 \cdot 2^3} R_{\ell+1}''' \right)^{(n-1)}, \\
\langle \underbrace{\mathsf{O}_{0} \cdots \mathsf{O}_{0}}_{n} \mathsf{O}_{1} \mathsf{O}_{1} \mathsf{O}_{\ell} \rangle &=& \frac{(-1)^{\ell+1}\, 2^{\ell+6}\, \ell!}{3 \left( 2\ell+1 \right)!!} \left( R_2 R_{\ell+1}' + \frac{1}{8} R_1 R_{\ell+1}''' + \left\{ \frac{1}{8} R_1 R_{\ell+1}'' + \frac{1}{3 \cdot 2^6} R_{\ell+1}'''' \right\}' \right)^{(n)}, \\
\langle \underbrace{\mathsf{O}_{0} \cdots \mathsf{O}_{0}}_{n} \mathsf{O}_{2} \mathsf{O}_{\ell} \rangle &=& \frac{(-1)^{\ell+1}\, 2^{\ell+6}\, \ell!}{3 \left( 2\ell+1 \right)!!} \left( R_{2} R_{\ell+1}' + \left\{ \frac{1}{4} R_{1} R_{\ell+1}'' + \frac{1}{5 \cdot 2^5} R_{\ell+1}'''' \right\}' \right)^{(n-1)}.
\eea
\noindent
From the first formula above it immediately follows $\ev{\mathsf{O}_{0} \mathsf{O}_{0}} = u$ as earlier advertised (again, in the conventions of \cite{gs21}). Due to integrations, these expressions depend on new data $P_{mn}$, on top of the Gel'fand--Dikii KdV potentials, which we now specify. They follow from a result in \cite{gd75}, which establishes the generic existence of polynomials $P_{mn} \left[ u \right]$, in $u$ and its derivatives, satisfying
\be
R_{m} R_{n}' = \frac{\rmd}{\rmd \kappa} P_{mn}.
\ee
\noindent
One can compute the first few:
\bea
P_{21} &=& - \frac{1}{64} \left( u^3 - \frac{1}{2} \left( u' \right)^2 \right), \\
P_{31} &=& \frac{1}{256} \left( \frac{5}{2} u^4 - 5 u \left( u' \right)^2 - \frac{1}{2} \left( u'' \right)^2 + u' u''' \right).
\eea

It is now rather explicit how all these correlation functions are dictated by the Gel'fand--Dikii KdV potentials, which themselves are known functionals of the specific-heat two-point-function solution to the string equation, $u (\kappa) = \ev{\mathsf{O}_{0} \mathsf{O}_{0}}$. Hence, if the resurgent transseries structure and properties of $u(\kappa)$ are known (\textit{e.g.}, \cite{gikm10, asv11, sv13, gs21, bssv22}), then immediately so will be the resurgence properties of all above class of $n$-point correlators.

\newpage

\bibliographystyle{plain}

\end{document}